\documentclass[10pt, american]{article}
\newif\ifdraft \global\drafttrue
\def\production{\global\draftfalse}
\production
\usepackage[T1]{fontenc}
\usepackage[latin1]{inputenc}
\usepackage{latexsym}
\usepackage{amsmath}
\usepackage{babel}
\usepackage{bbm}
\usepackage{amssymb}
\usepackage{amsfonts}
\usepackage{times}
\usepackage{theorem}
\usepackage{epsfig}
\usepackage{enumerate}
\usepackage{combelow}
\usepackage[colorlinks=true]{hyperref}
\usepackage{rotating}
\ifdraft
\usepackage{showkeys}
\fi

\numberwithin{equation}{section}

\newcounter{smallarabics}
\newenvironment{arabicenumerate}
{\begin{list}{{\normalfont\textrm{(\arabic{smallarabics})}}}
  {\usecounter{smallarabics}\setlength{\itemindent}{0cm}
   \setlength{\leftmargin}{5ex}\setlength{\labelwidth}{4ex}
   \setlength{\topsep}{0.75\parsep}\setlength{\partopsep}{0ex}
   \setlength{\itemsep}{0ex}}}
{\end{list}}

\newcounter{smallroman}

\newcommand{\ben}{\begin{arabicenumerate}}  
\newcommand{\een}{\end{arabicenumerate}}  
 
\newtheorem{theoreme}{Theorem }[section]
\newtheorem{proposition}[theoreme]{Proposition}
\newtheorem{lemma}[theoreme]{Lemma}
\newtheorem{definition}[theoreme]{Definition}

\def\rr{{\mathbb R}}
\def\zz{{\mathbb Z}}
\def\cc{{\mathbb C}}
\def\nn{{\mathbb N}}

\def\textsl{{}}

\def\Im{{\rm Im}\,}
\def\Re{{\rm Re}\,}
\def\ch{{\mathfrak{h}}}

\def\slim{\mathop{\mbox{s-lim}}}

\def\c0inf{C_0^\infty}
\def\bep{\begin{proposition}}
\def\eep{\end{proposition}}
\def\s{{\rm s}}
\def\proof{\noindent {\bf Proof.}\ \ }

\def\cH{{\cal  H}}

\def\cR{{\cal R}}
\def\cA{{\cal A}}

 \def\cB{{\cal B}}

\def\i{{\rm i}}
\newcommand{\beq}{\begin{equation}}
\newcommand{\eeq}{\end{equation}}
\newcommand{\bear}[1]{\begin{array}{#1}}
\newcommand{\ear}{\end{array}}

\def\sp{{\hat e}}

\newcommand{\CAR}{\mathrm{CAR}}

\newcommand{\e}{\mathrm{e}}
\renewcommand{\i}{\mathrm{i}}

\renewcommand{\d}{\mathrm{d}}

\newcommand{\vertleq}[0]{\begin{turn}{-90} $\le$\end{turn}}

\def\qed{\hfill$\Box$\medskip}

\def\cA{{\cal A}}
\def\cQ{{\cal Q}}
\def\cK{{\cal K}}

\def\bel{\begin{lemma}}
\def\eel{\end{lemma}}
\def\bet{\begin{theoreme}}
\def\eet{\end{theoreme}}
\def\bed{\begin{definition}}
\def\eed{\end{definition}}
\def\bar{\overline}
\def\ubar{\underline}

\def\12{\frac{1}{2}}

\def\e{{\rm e}}

\def\d{{\rm d}}
\def\Ran{{\rm Ran}\,}

\def\cH{{\cal H}}
\def\cO{{\cal O}}

\def\Ker{{\rm Ker}\,}
\def\Dom{{\rm Dom}\,}

\def\sp{{\rm sp}}
\def\cS{{\cal S}}
\def\cR{{\cal R}}

\def\Ent{{\rm Ent}}

\def\fh{{\mathfrak h}}

\def\fM{{\mathfrak M}}

\def\s{{\rm s}}

\def\Tr{{\rm Tr}}

\begin{document}

\title{Quantum Hypothesis Testing\\ and\\ Non-Equilibrium Statistical Mechanics}
\author{  V. Jak\v{s}i\'c$^{1}$,  Y. Ogata$^{2}$,  C.-A. Pillet$^{3}$, R. Seiringer$^{1}$
\\ \\ 
\\ \\
$^1$Department of Mathematics and Statistics\\ 
McGill University\\
805 Sherbrooke Street West \\
Montreal,  QC,  H3A 2K6, Canada
\\ \\
$^2$Department of Mathematical Sciences\\
University of Tokyo\\
Komaba, Tokyo, 153-8914, Japan
\\ \\
$^3$Centre de Physique Th\'eorique -- FRUMAM\\
CNRS -- Universit\'es de Provence, de la M\'editerran\'e et du Sud Toulon-Var\\
B.P. 20132, 83957 La Garde, France
}
\def\today{}
\maketitle

\vskip 5cm
{\bf\noindent Abstract.}
We extend the mathematical theory of quantum hypothesis testing to the general $W^\ast$-algebraic 
setting and explore its relation with recent developments in non-equilibrium quantum statistical 
mechanics. In particular, we relate the large deviation principle for the full counting statistics of 
entropy flow to quantum hypothesis testing of the arrow of time.
\thispagestyle{empty}
\eject
\tableofcontents
\eject

\section{Introduction}
Starting with the works \cite{JP1, Ru1, Pi}, the mathematical theory of non-equilibrium quantum
statistical mechanics has developed rapidly in recent years \cite{Ab, AF, AP, AJPP1, AJPP2, 
FMS, FMU, JKP, JOP1, JOP2, JOP3, JOP4, JP2, JP3, JP4, MMS1, MMS2, MO, Na, Og1, Og2, Ro, 
Ru2, Tas, TM1, TM2}. The current research efforts are centered around the theory of entropic 
fluctuations (see \cite{JOPP1, JOPP2}) and it is these developments that will concern us here. 

Since Shannon's rediscovery of Gibbs-Boltzmann entropy  there has  been a close interplay 
between  information theory and  statistical mechanics. One of the deepest links is provided by 
the theory of large deviations \cite{DZ, El}. We refer the reader to \cite{Me} for a  beautiful 
and easily accessible account of this interplay. In this context, it was natural to try to interpret
recent results in non-equilibrium statistical mechanics in terms of quantum information theory.

The link  can be roughly summarized as follows\footnote{Needless to say, all the notions discussed 
in this paragraph will be defined later in the paper.}.
Consider the large deviation principle  for the full counting statistics for the repeated quantum 
measurement of the energy/entropy flow over the time interval $[0, t]$ \cite{LL, Ro, JOPP1}. Let 
$I(\theta)$ be the rate function and $e(s)$ its  Legendre transform.  Let $\hat e (s)$ be the  
Chernoff error exponent  in the quantum hypothesis testing of the arrow of time, {\sl i.e.}, of the 
family of states $\{(\omega_{t/2}, \omega_{-t/2})\}_{t>0}$, where $\omega_{\pm t/2}$ is the state 
of our quantum  system at the time $\pm t/2$. Then $e(s)=\hat e(s)$. In this paper we prove this
result and elaborate on the relation between quantum hypothesis testing and non-equilibrium 
statistical mechanics. 

Hypothesis testing has a long tradition in theoretical and applied statistics \cite{Pe, Ch, LR}. During 
the last decade many results of classical hypothesis testing have been extended to the quantum 
domain 
\cite{ACM,ANS,BDK1,BDK2,BDK3,Ha1,Ha2,HMO1,HMO2,HMO3,Ka,Mo,MHO,NS1,NS2,OH,ON}.
The culmination of these efforts was the proof of a long standing conjecture regarding the quantum 
Chernoff bound \cite{ACM, ANS}. The following trace inequality of \cite{ACM,ANS} played a key 
role in the resolution of this conjecture. 
\bep\label{paris-hot}
Let $A> 0$ and $B>0$ be  matrices on $\cc^n$. Then 
\begin{equation}
\frac{1}{2}\left(\Tr \,A + \Tr\, B- \Tr \, |A-B|\right)\leq \Tr \,A^{1-s} B^{s}
\label{ACM}
\end{equation}
holds for any $s\in[0,1]$.
\eep
\proof (Communicated by N. Ozawa, unpublished).  For $X$ self-adjoint, $X_{\pm}$ denotes its 
positive/negative part. Decomposing $A-B = (A-B)_+ - (A-B)_-$ one gets 
$$
\frac{1}{2}\left(\Tr \,A + \Tr\, B- \Tr \, |A-B|\right)=\Tr\, A - \Tr\, (A-B)_+,
$$
and \eqref{ACM} is equivalent to 
\begin{equation}
\Tr \, A - \Tr\, B^sA^{1-s}\leq \Tr\, (A-B)_+.
\label{inequality-oz}
\end{equation}
Note that $ B+ (A-B)_+\geq B$  and $B + (A-B)_+=A + (A-B)_- \geq A$.  Since, for $s\in[0,1]$,
the function $x\mapsto x^s$ is operator monotone ({\sl i.e.,} $X\le Y\Rightarrow X^s\le Y^s$ for any
positive matrices $X$, $Y$), we can write 
\begin{align*}
\Tr\,A -\Tr \, B^s A^{1-s}=\Tr(A^s-B^s)A^{1-s}&\leq \Tr\, ((B + (A-B)_+)^s - B^s)A^{1-s}\\[2mm]
&\leq \Tr\, ((B + (A-B)_+)^s -B^s)(B + (A-B)_+)^{1-s}\\[2mm]
&=\Tr\, B + \Tr\, (A-B)_+ - \Tr \, B^s(B+ (A-B)_+)^{1-s}\\[2mm]
&\leq \Tr \,(A-B)_+.
\end{align*}
\nopagebreak[4]
\qed
\pagebreak[1]

We have singled out this result for the following reason. $W^\ast$-algebras and modular theory 
provide a natural general mathematical framework for quantum hypothesis testing.  For example, 
Inequality \eqref{inequality-oz} can be formulated in $W^\ast$-algebraic language as 
\begin{equation}
\frac{1}{2}(\omega(1) +\nu(1)-\|\omega-\nu\|)
\leq (\Omega_\omega|\Delta_{\nu|\omega}^{s}\Omega_\omega)
\label{ozzie}
\end{equation}
for $0\leq s \leq 1$, where $\omega$ and $\nu$ are faithful, normal, positive linear functionals on 
a $W^\ast$ algebra $\fM$ in standard form, $\Omega_\omega$ is the vector representative of 
$\omega$ in the natural cone and $\Delta_{\nu|\omega}$ is the relative modular operator. 
The extension of quantum hypothesis testing to $W^\ast$-algebras  was hindered by the fact 
that the original proof \cite{ACM, ANS} of the inequality \eqref{inequality-oz} could not be 
extended/generalized to a proof of \eqref{ozzie}. Ozawa's proof, however, can and  the inequality 
\eqref{ozzie} was recently proven in \cite{Og3}.  An alternative proof, which  links \eqref{ozzie} to 
Araki-Masuda theory of non-commutative $L^p$-spaces, is given in Section~\ref{sec-q-prelim}.

In Section~\ref{sec-walgebra} we extend  the mathematical theory of quantum hypothesis testing to  
the $W^\ast$-algebraic setting  and prove the Chernoff bound, the Hoeffding bound, and Stein's 
Lemma. We develop a model independent axiomatic approach to quantum hypothesis testing which
clarifies its mathematical structure and reduces the study of concrete models to the verification of 
the proposed axioms. We emphasize that apart from Inequality \eqref{ozzie}, whose proof is 
technically involved and has no classical counterpart, the proofs follow essentially  line by line the 
classical arguments. The verification of the large deviation axioms that underline the proofs leads to a 
novel class of analytic problems in quantum statistical mechanics. 

To make the paper and our main points accessible to a reader without
prior knowledge of modular theory, we describe in
Sections~\ref{sec-approxfin} and \ref{sec-fullep} quantum hypothesis
testing, non-equilibrium quantum statistical mechanics, and their
relation in the context of finite quantum systems. Typical examples
the reader should keep in mind are a quantum spin system or a Fermi
gas confined to a finite part $\Lambda$ of some infinite lattice
$L$. Needless to say, the thermodynamic limit $\Lambda \rightarrow L$
has to be taken before the large time limit $t\rightarrow \infty$. The
reader not familiar with (or not interested in) the algebraic theory
may directly proceed to Section~\ref{sec-openqs} after reading
Sections~\ref{sec-approxfin} and \ref{sec-fullep}.

For reasons of space we have not attempted to prove quantum hypothesis testing results
under the most general conditions possible. In particular, we  shall only discuss hypothesis testing
of faithful states (this restriction is inconsequential as far as statistical mechanics is concerned). 
The extensions to non-faithful states follow typically by  straightforward limiting arguments 
(see \cite{Og3} for an example).

This work is of a review nature. Our goal is to point to a surprising link between two directions of 
research which were largely unaware of each other, in a hope that they both may benefit from this 
connection. We shall discuss here only the single parameter full counting statistics for 
entropy flow and its relation to binary hypothesis testing. The multi-parameter full counting statistics 
describing the energy flow between different parts of the system is a well understood object
but its relation with quantum hypothesis testing is unclear and would presumably involve multiple 
quantum state discrimination which is poorly understood at the moment (the proposal of \cite{NS1}
appears unsuitable in this context). 

We also point out that for obvious space reasons this paper is not a review of either quantum statistical 
mechanics or quantum hypothesis testing, but of the link between the two of them. The reader who 
wishes to learn more about these topics individually may consult \cite{AJPP1, JOPP1, ANS}.

The paper is organized as follows. In Section~\ref{sec-prologue} we review the results of Large 
Deviation Theory that we will need. Since these results are not stated/proven in this form in the 
classical references \cite{DZ, El}, we provide  proofs for the reader's convenience. In 
Section~\ref{sec-approxfin} we review the existing results in quantum hypothesis testing of 
finite quantum systems. The non-equilibrium statistical mechanics of finite quantum systems is 
described in Section~\ref{sec-fullep}. Its relation with quantum hypothesis testing is discussed  in 
Section~\ref{sec-arrow}. In Section~\ref{sec-algebrapre} we review the results of modular theory 
that we need and give a new proof of the key preliminary inequality needed to prove  \eqref{ozzie} 
(Proposition \ref{araki-multi}). Sections~\ref{sec-walgebra} and \ref{sec-NECSM} are devoted to 
quantum hypothesis testing and non-equilibrium statistical mechanics of infinitely extended quantum 
systems described by $W^\ast$-algebras and $W^\ast$-dynamical systems. Finally, in 
Section~\ref{sec-openqs} we describe several physical models for which the existence of the large 
deviation functionals have been proven and to which the results described in this paper apply. 

\bigskip
{\noindent\bf Acknowledgments.} The research  of V.J. and R.S. was partly supported by NSERC. 
The research of Y.O. was supported by JSPS Grant-in-Aid for Young Scientists (B), 
Hayashi Memorial Foundation for Female Natural Scientists, Sumitomo Foundation, and 
Inoue Foundation. The research of C.-A.P. was partly supported by ANR (grant 09-BLAN-0098).
A part of this paper was written during the stay of the authors at the University of Cergy-Pontoise. 
We wish to thank L.~Bruneau, V.~Georgescu and F.~ Germinet for making these visits possible and 
for their hospitality.

\section{Prologue}
\label{sec-prologue}

\subsection{Fenchel-Legendre transform}
\label{sec-FLT}

Let $[a,b]$ be a finite closed interval in $\rr$ and $e: [a,b] \rightarrow  \rr$  a convex continuous
function. Convexity implies  that for every $s\in ]a,b[$ the limits 
\[
D^{\pm}e(s) =\lim_{h\downarrow 0}\frac{e(s\pm h)- e(s)}{\pm h}
\]
exist and are finite. Moreover, $D^{-}e(s)\leq D^{+}e(s)$, $D^+e(s) \leq D^{-}e(t)$ for $s<t$, and 
$D^-e(s)=D^+e(s)$ outside a countable set  in $]a,b[$. $D^+e(a)$ and $D^{-}e(b)$ also exist, 
although they may not be finite. If $e^\prime(s)$ exists for all $s\in ]a, b[$, then the mean value
theorem implies that $e^\prime(s)$ is continuous on $]a, b[$ and that 
\[
\lim_{s \downarrow a}e^\prime(s)= D^+e(a), \qquad \lim_{s\uparrow b} e^\prime(s)= D^-e(b).
\]

We set $e(s)=\infty$ for $s\not \in [a,b]$. Then $e(s)$ is a lower semi-continuous convex function 
on $\rr$. The subdifferential of $e(s)$, $\partial e(s)$,  is   defined by 
\[
\partial e(s)=
\begin{cases}
]-\infty, D^+e(a)]& \text{if }s=a;\\[1mm]
[D^-e(s), D^+e(s)]& \text{if }s\in ]a, b[;\\[1mm]
[D^-e(b), \infty[& \text{if }s=b;\\[1mm]
\emptyset&\text{if }s\not\in [a,b].
\end{cases}
\]
The function 
\begin{equation}
\varphi (\theta)= \sup_{s\in[a,b]}(\theta s - e(s))=\sup_{s\in\rr}(\theta s -e(s))
\label{local-fenchel}
\end{equation}
is called the  Fenchel-Legendre transform of $e(s)$. $\varphi(\theta)$ is finite and convex (hence 
continuous) on $\rr$.  Obviously, $a\geq 0\Rightarrow\varphi(\theta)$ is increasing, and
$b \leq 0\Rightarrow\varphi(\theta)$ is decreasing. The subdifferential  of $\varphi(\theta)$ is  
$\partial \varphi(\theta)=[D^-\varphi(\theta), D^+\varphi(\theta)]$. The basic properties of the pair 
$(e, \varphi)$ are summarized in:
\bet \label{prop-fenchel}
\begin{enumerate}[{\rm (1)}] 
\item $s\theta \leq e(s) +\varphi(\theta)$.
\item $s\theta =e(s) +\varphi(\theta)$ $\Leftrightarrow$ $\theta \in \partial e(s)$.
\item $\theta \in \partial e(s)$ $\Leftrightarrow$  $s\in \partial \varphi(\theta)$.
\item $e(s)=\sup_{\theta \in \rr}(s\theta - \varphi(\theta))$.
\item If $0\in ]a, b[$, then $\varphi(\theta)$ is decreasing on $]-\infty, D^-e(0)]$, increasing on 
$[D^+e(0), \infty[$, $\varphi(\theta)=-e(0)$ for $\theta \in \partial e(0)$, and $\varphi(\theta)>-e(0)$
for $\theta \not\in \partial e(0)$. 
\item
\[\varphi(\theta)=
\begin{cases}
a\theta-e(a)& \text{if }\,\theta\leq D^+e(a);\\
b\theta-e(b) &\text{if }\,\theta\geq D^-e(b).
\end{cases}
\]
\end{enumerate}
\eet
The proofs of  these results are simple and can be found in \cite{El, JOPP1}.

The function 
\begin{equation}
\hat \varphi(\theta)=\sup_{s\in[a,b]}(\theta (s-b) - e(s))= \varphi(\theta)-b\theta
\label{local-hat}
\end{equation}
will also play an important role in the sequel. Its properties are easily deduced from the properties of
$\varphi(\theta)$. In particular, $\hat \varphi$ is  convex, continuous and decreasing.
It is a one-to-one map from $]-\infty,D^-e(b)]$ to $[-e(b),\infty[$. We denote by $\hat\varphi^{-1}$
the inverse map.

Suppose now that  $a=0$ and $b=1$. For $r\in \rr$ let
\begin{equation}
\psi(r)=
\begin{cases}
-\varphi(\hat \varphi^{-1}(r))& \text{if }r\geq-e(1);\\
-\infty &\text{otherwise}.
\end{cases}
\label{def-br}
\end{equation}

\bep\label{var-b-1}
\begin{enumerate}[{\rm (1)}] 
\item \begin{equation}
\psi(r)=-\sup_{s\in[0,1[}\frac{-sr -e(s)}{1-s}.
\label{var-b}
\end{equation}
\item $\psi(r)$ is concave, increasing,  finite for $r> -e(1)$, and $\psi(-e(1))=e(1)-D^-e(1)$. 
\item $\psi(r)$ is continuous on $]-e(1), \infty[$ and upper-semicontinuous on $\rr$. 
\end{enumerate}
\eep
The proof of this proposition is elementary and we will omit it.

\subsection{Large deviation bounds}
\label{sec-LDB}

In this paper we shall make use of some results of Large Deviation Theory, and in particular of 
a suitable variant of the G\"artner-Ellis theorem \cite{DZ, El}. In this subsection we describe and 
prove the results we will need. 

Let ${\cal I}\subset \rr_+$ be an unbounded  index set, $(M_t, {\cal F}_t, P_t)_{t\in {\cal I}}$ a 
family of measure spaces, and $X_t: M_t\rightarrow \rr$  a family of measurable functions. We 
assume that the measures $P_t$ are finite for all $t$. The functions
\[
\rr\ni s\mapsto e_t(s) =\log \int_{M_t}\e^{ sX_t}\d P_t,
\]
are convex (by H\"older's inequality) and take values in $]-\infty, \infty]$. We need

\newcommand{\LD}{{\hyperref[LD]{{\rm(LD)}}}}
\begin{quote}\label{LD}
{\bf Assumption (LD).} For some weight function $\mathcal{I}\ni t\mapsto w_t>0$ such
that $\lim_{t\to\infty}w_t=\infty$ the limit
\[
e(s)=\lim_{t\rightarrow \infty}\frac{1}{ w_t}e_t(s),
\]
exists and is finite for $s\in [a, b]$. Moreover, the function $[a,b]\ni s\mapsto e(s)$ is continuous.
\end{quote}

Until the end of this section we shall assume that (LD) holds and set $e(s)=\infty$ for $s\not\in [a,b]$.
The Legendre-Fenchel transform $\varphi(\theta)$ of $e(s)$ is defined by \eqref{local-fenchel}.
\bep\label{LDP-up}
\begin{enumerate}[{\rm (1)}] 
\item Suppose that $0\in [a,b[$. Then 
\begin{equation}
\limsup_{t\rightarrow \infty}\frac{1}{w_t}\log P_t(\{ x\in M_t\,|\, X_t(x) >\theta w_t\})\leq 
\begin{cases}
-\varphi(\theta)&\text{if }\,\theta\geq  D^+e(0);\\
e(0)&\text{if }\,\theta <D^+e(0).
\end{cases}
\label{up-b1}
\end{equation}
\item Suppose that $0\in ]a,b]$. Then 
\begin{equation}
\limsup_{t\rightarrow \infty}\frac{1}{w_t}\log P_t(\{ x\in M_t\,|\, X_t(x) <\theta w_t\})\leq 
\begin{cases}
-\varphi(\theta)&\text{if }\,\theta  \leq  D^-e(0);\\
e(0) &\text{if }\,\theta >D^-e(0).
\end{cases}
\label{up-b2}
\end{equation}
\end{enumerate}
\eep
\proof
We shall prove (1); (2) follows from (1) applied to $-X_t$. For any $s\in]0,b]$, the Chebyshev
inequality yields
\[
P_t(\{ x\in M_t\,|\, X_t(x) >\theta w_t\})
=P_t(\{ x\in M_t\,|\, \e^{sX_t(x)} >\e^{s\theta w_t}\})
\leq \e^{-s\theta w_t}\int_{M_t}\e^{sX_t}\d P_t,
\]
and hence
\[
\limsup_{t\rightarrow \infty}\frac{1}{w_t}\log P_t(\{ x\in M_t\,|\, X_t(x) >\theta w_t\})
\leq-\sup_{s\in[0,b]}(s\theta -e(s)).
\]
Since 
\[
\sup_{s\in[0,b]}(s\theta -e(s))=
\begin{cases}
 \varphi(\theta)&\text{if }\,\theta\geq  D^+e(0);\\
-e(0)&\text{if }\,\theta <D^+e(0),
\end{cases}
\]
the statement follows.
\qed

\bep\label{LDP-basic}
Suppose that  $0\in ]a, b[$, $e(0)\leq 0$, and that  $e(s)$ is differentiable at $s=0$. Then 
for any $\delta >0$ there is $\gamma >0$ such that for $t$ large enough
\[
P_t(\{ x\in M_t\, |\, |w_t^{-1}X_t(x) - e^\prime(0)|\geq \delta\})\leq  \e^{-\gamma w_t}.
\]
\eep
\proof
Since $\varphi(e^\prime(0))=-e(0)$, Theorem \ref{prop-fenchel} (5) implies that 
$\varphi(\theta)>\varphi(e^\prime(0))\geq 0$ for $\theta\not=e^\prime(0)$. 
Proposition \ref{LDP-up} implies
\[
\limsup_{t\rightarrow \infty}\frac{1}{w_t}\log P_t(\{ x\in M_t\,|\, |w_t^{-1}X_t(x)-e^\prime(0)|\geq \delta\})
\leq -\min\{\varphi(e^\prime(0)+\delta), \varphi(e^\prime(0)-\delta)\},
\]
and the statement follows.
\qed

\bep\label{GE-THM}
 Suppose that   $e(s)$ is differentiable on $]a,b[$.  Then for $\theta \in ]D^+ e(a), D^-e(b)[$,                         
\begin{equation}
\liminf_{t\rightarrow \infty}\frac{1}{w_t}\log P_t(\{ x\in M_t\,|\, X_t(x)>\theta w_t)\}\geq -\varphi(\theta).
\label{lim-LPD}
\end{equation}
\eep
\proof 
Let $\theta \in ]D^+e(a), D^-e (b)[$ be given and let $\alpha >\theta$ and $\epsilon >0$ be such that 
$\theta < \alpha -\epsilon <\alpha <\alpha +\epsilon< D^-e(b)$. Let $s_\alpha\in ]a, b[$ be such that 
$e^\prime(s_\alpha)=\alpha$ (so $\varphi(\alpha)=\alpha s_\alpha - e(s_\alpha)$). Let 
\[
\d\hat P_t =\e^{-e_t(s_\alpha)}\e^{s_\alpha X_t}\d P_t.
\]
Then $\hat P_t$ is a probability measure on $(M_t, {\cal F}_t)$ and 
\begin{equation}
\begin{split}
P_{t}(\{ x\in M_t\,|\, X_t(x)>\theta w_t\})
&\geq P_t(\{ x\in M_t\,|\,w_t^{-1}X_t(x)\in [\alpha-\epsilon, \alpha +\epsilon]\})\\[3mm]
&=\e^{e_t(s_\alpha)}\int_{\{ w_t^{-1}X_t \in [\alpha -\epsilon, \alpha +\epsilon]\}} 
\e^{-s_\alpha X_t}\d\hat P_t\\[3mm]
&\geq \e^{e_t(s_\alpha)-s_\alpha  \alpha w_t -|s_\alpha| w_t\epsilon}
\hat P_t(\{ x\in M_t\,|\, w_t^{-1}X_t \in [\alpha -\epsilon, \alpha +\epsilon]\}).
\end{split}
\label{LDP-long}
\end{equation}
Now, if $\hat e_t(s)=\log \int_{M_t}\e^{s X_t}\d\hat P_t$, then
 $\hat e_t(s)=e_t(s + s_\alpha)-e_t(s_\alpha)$ and so for $s\in [a-s_\alpha, b-s_\alpha]$,  
\[
\lim_{t\rightarrow \infty}\frac{1}{w_t}\hat e_t(s)=e(s+s_\alpha)-e(s_\alpha).
\]
Since $\hat e^\prime(0)=e^\prime(s_\alpha)=\alpha$, it follows from Proposition \ref{LDP-basic} that 
\[
\lim_{t\rightarrow \infty}\frac{1}{w_t}\log 
\hat P_t(\{ x\in M_t\,|\, w_t^{-1}X_t(x) \in [\alpha -\epsilon, \alpha +\epsilon]\})=0,
\]
and \eqref{LDP-long} yields 
\[
\liminf_{t\rightarrow \infty}\frac{1}{w_t}\log P_t(\{ x\in M_t\,|\, X_t(x) >\theta w_t\})
\geq -s_\alpha \alpha +e(s_\alpha) -|s_\alpha|\epsilon = 
-\varphi(\alpha)-|s_\alpha|\epsilon.
\]
The statement follows by taking first  $\epsilon \downarrow 0$ and then $\alpha \downarrow \theta$. 
\qed

The following local version of the G\"artner-Ellis theorem is a consequence of Propositions
\ref{LDP-up} and \ref{GE-THM}.

\bet\label{GEoneD}
If $e(s)$ is differentiable on $]a,b[$ and $0 \in ]a, b[$ then, for any open set 
${\mathbb J}\subset ]D^+e(a), D^- e(b)[$, 
\[
\lim_{t\rightarrow \infty}\frac{1}{w_t}\log P_t(\{ x\in M_t\,|\,w_t^{-1} X_t(x) \in {\mathbb J}\})
= -\inf_{\theta \in {\mathbb J}}\varphi(\theta).
\]
\eet
\proof
{\em Lower bound.} For any $\theta\in\mathbb J$ and $\delta>0$ such that
$]\theta-\delta,\theta+\delta[\subset\mathbb J$ one has
$$
P_t(\{ x\in M_t\,|\,w_t^{-1} X_t(x) \in {\mathbb J}\})
\ge P_t(\{ x\in M_t\,|\,w_t^{-1} X_t(x) \in]\theta-\delta,\theta+\delta[\}),
$$
and it follows from Proposition \ref{GE-THM} that
$$
\liminf_{t\to\infty}\frac1{w_t}\log P_t(\{ x\in M_t\,|\,w_t^{-1} X_t(x) \in {\mathbb J}\})
\ge -\varphi(\theta-\delta).
$$
Letting $\delta\downarrow0$ and optimizing over $\theta\in\mathbb J$, we obtain
\begin{equation}
\liminf_{t\to\infty}\frac1{w_t}\log P_t(\{ x\in M_t\,|\,w_t^{-1} X_t(x) \in {\mathbb J}\})
\ge -\inf_{\theta\in\mathbb J}\varphi(\theta).
\label{oneDlower}
\end{equation}

\medskip\noindent{\em Upper bound.}  
By Part (5) of Theorem
\ref{prop-fenchel}, we have $\varphi(\theta)=-e(0)$ for $\theta=e'(0)$ and $\varphi(\theta)>-e(0)$
otherwise. Hence, if $e'(0)\in\bar{\mathbb J}$ (the closure of $\mathbb J$), then
$$
\limsup_{t\to\infty}\frac1{w_t}\log P_t(\{ x\in M_t\,|\,w_t^{-1} X_t(x) \in {\mathbb J}\})\le e(0)
=-\inf_{\theta\in\mathbb J}\varphi(\theta).
$$
In the case $e'(0)\not\in\bar{\mathbb J}$, there exist $\alpha,\beta\in\bar{\mathbb J}$ such
that $e'(0)\in]\alpha,\beta[\subset\rr\setminus\bar{\mathbb J}$. It follows that
\begin{align*}
P_t(&\{ x\in M_t\,|\,w_t^{-1} X_t(x) \in {\mathbb J}\})\\
&\le P_t(\{ x\in M_t\,|\,w_t^{-1} X_t(x)<\alpha\})
+P_t(\{ x\in M_t\,|\,w_t^{-1} X_t(x)>\beta\})\\
&\le 2\max\left(P_t(\{ x\in M_t\,|\,w_t^{-1} X_t(x)<\alpha\}),
P_t(\{ x\in M_t\,|\,w_t^{-1} X_t(x)>\beta\})\right),
\end{align*}
and Proposition \ref{LDP-up} yields
\begin{align*}
\limsup_{t\to\infty}\frac1{w_t}\log P_t(\{ x\in M_t\,|\,w_t^{-1} X_t(x) \in {\mathbb J}\})\le
-\min(\varphi(\alpha),\varphi(\beta)).
\end{align*}
Finally, by Part (5) of Proposition \ref{prop-fenchel}, one has
$$
\inf_{\theta\in\mathbb J}\varphi(\theta)=\min(\varphi(\alpha),\varphi(\beta)),
$$
and therefore
\begin{equation}
\limsup_{t\to\infty}\frac1{w_t}\log P_t(\{ x\in M_t\,|\,w_t^{-1} X_t(x) \in {\mathbb J}\})\le
-\inf_{\theta\in\mathbb J}\varphi(\theta)
\label{oneDupper}
\end{equation}
holds for any $\mathbb J\subset ]D^+e(a), D^- e(b)[$.
The result follows from the bounds \eqref{oneDlower} and \eqref{oneDupper}. 
\qed

\section{Approximately finite quantum hypothesis testing}
\label{sec-approxfin}

\subsection{Setup}
\label{sec-setup}

Consider a quantum system $\cQ$ described by a finite dimensional Hilbert space 
$\cH$. We denote by $\cO_\cH$ (or $\cO$ whenever the meaning is clear within the context)
the $\ast$-algebra of all linear operators on $\cH$, equipped with the usual operator norm.
The symbol $1_\cH$ (or simply $1$) stands for the unit of $\cO_\cH$.  The spectrum of $A\in\cO_\cH$ 
is denoted by $\sp(A)$. $\cO_{\cH, {\rm self}}$ (or simply $\cO_{\rm self}$) is the set of all 
self-adjoint elements of $\cO_\cH$. $A\in\cO_{\cH, {\rm self}}$ is positive, written $A\ge0$,
if $\sp(A)\subset[0,\infty[$.
For $A\in\cO_{\cH,{\rm self}}$ and $\lambda\in\sp(A)$,
$P_\lambda(A)$ denotes the spectral projection of $A$. We adopt the shorthand notation
$$
\s_A=\sum_{0\not=\lambda\in\sp(A)}P_\lambda(A),\qquad
A_\pm=\pm\sum_{\lambda\in\sp(A),\pm\lambda> 0} \lambda P_\lambda(A),\qquad
|A|=A_+-A_-.
$$
$\s_A$ is the orthogonal projection on the range of $A$ and $A_\pm$ are the positive/negative parts
of $A$.

A linear functional $\nu:\cO\to\cc$ is called:
\begin{enumerate}[{\rm (1)}]
\item hermitian if $\nu(A)\in\rr$ for all $A\in\cO_{\rm self}$;
\item positive if $\nu(A^\ast A)\ge0$ for all $A\in\cO$;
\item a state if it is positive and normalized by $\nu(1)=1$;
\item faithful if it is positive and such that $\nu(A^\ast A)=0$ implies $A=0$ for any $A\in\cO$.
\end{enumerate}
For any $\nu\in\cO$, $A\mapsto\Tr\,\nu A$ defines a linear functional on $\cO$, {\sl i.e.,}
an element of the dual space $\cO^\ast$. Any linear functional on $\cO$ arises in this way. 
In the following, we shall identify $\nu\in\cO$ with the corresponding linear functional and write 
$\nu(A)=\Tr\,\nu A$. With this identification, the norm of the dual space $\cO^\ast$ is just the 
trace norm on $\cO$, {\sl i.e.,} $\|\nu\|=\Tr\,|\nu|=|\nu|(1)$ with $|\nu|=(\nu^\ast\nu)^{1/2}$.
A functional $\nu$ is hermitian iff $\nu\in\cO_{\rm self}$ and positive iff $\nu\ge0$ as an operator.
If $\nu\ge0$ then it is faithful iff $\sp(\nu)\subset]0,\infty[$, which we denote by $\nu>0$. 
The functional $\nu$ is a state iff the operator $\nu$ is a density matrix, {\sl i.e.,} $\nu\ge0$ and 
$\nu(1)=\Tr\,\nu=1$. If $\nu$ is a positive linear functional then $\s_\nu$ is its support projection, 
the smallest orthogonal projection $P$ such that $\nu(1-P)=0$. In particular 
$\|\nu\|=\nu(1)=\nu(\s_\nu)$ and $\nu$ is faithful iff $\s_\nu=1$. If $\nu$ is a Hermitian linear 
functional then $\nu_\pm$ are positive linear functionals such that $\s_{\nu_+}\s_{\nu_-}=0$,
$\nu=\nu_+-\nu_-$ (the Jordan decomposition of $\nu$) and $\|\nu\|=\nu_+(1)+\nu_-(1)$.

Let $\nu$ and $\omega$ be positive linear functionals. The relative entropy of $\nu$ w.r.t.\;$\omega$
is given by
\begin{equation}
\Ent(\nu|\omega)=\left\{\begin{array}{ll}\Tr \,\nu(\log \omega-\log \nu),
&\text{if }\s_\nu\le\s_\omega,\\[5pt]
-\infty&\text{otherwise.}
\end{array}\right.
\label{FinRelEntDef}
\end{equation}
R\'enyi's relative entropy of $\nu$ w.r.t.\;$\omega$ is the convex function
$$
[0,1]\ni s\mapsto\Ent_s(\nu|\omega)=\left\{\begin{array}{ll}
\log\Tr \,\nu^s\omega^{1-s},
&\text{if }\s_\nu\s_\omega\not=0,\\[5pt]
-\infty&\text{otherwise,}
\end{array}\right.
$$
which clearly satisfies $\Ent_s(\nu|\omega)=\Ent_{1-s}(\omega|\nu)$.
If $\s_\nu\le\s_\omega$ then this function has a real analytic extension to $s\in]0,\infty[$ and  
\begin{equation}
\left.\frac{\d\ }{\d s}\Ent_s(\nu|\omega)\right|_{s=1}=-\Ent(\nu|\omega).
\label{DRenyi}
\end{equation}
The observables of the quantum system $\cQ$ are described by elements of $\cO_{\rm self}$ and
its physical states are states on $\cO$, {\sl i.e.,} density matrices.
The possible outcomes of a measurement of $A$ are the  eigenvalues $a\in\sp(A)$. If 
the system is in the state $\nu$, then the probability to observe $a$ is $\nu(P_a(A))$.
In particular, the expectation value of $A$ is $\nu(A)$.

The setup of quantum hypothesis testing is a direct generalization of the corresponding setup 
in classical statistics \cite{LR}. Let $\nu\not=\omega$ be two faithful\footnote{In this paper, for
simplicity of the exposition, we shall only consider faithful states. We note, however, that most 
results extend to the general case by a straightforward limiting argument.} 
states such that one of the following two competing hypotheses holds:

\begin{quote} Hypothesis I :  $\cQ$ is in the state $\omega$;  
\end{quote} 
\begin{quote} Hypothesis II  :  $\cQ$ is in the state $\nu$.
\end{quote} 

A {\em test} is a projection $T\in\cO_{\rm self}$ and the result of a measurement of the corresponding 
observable is either $0$ or $1$. The purpose  of a test is to discriminate between the two hypotheses. 
Given the outcome of the measurement of $T$ one chooses Hypothesis~I or~II. More precisely, if the 
outcome is $1$ one accepts~I and rejects~II. Otherwise, if the outcome is $0$,  one  accepts~II and 
rejects~I. To a given test $T$ one can associate two kinds of errors. A type-I error occurs when 
the system is in the state $\omega$ but the outcome of the test is $0$. The conditional probability of
such an error, given that the state of the system is $\omega$, is $\omega(1-T)$. If the system is
in the state $\nu$ and the outcome of the test is $1$, we get a type-II error, with conditional
probability $\nu(T)$.

Assuming that Bayesian probabilities can be assigned to the states $\omega$ and $\nu$, {\sl i.e.,} that
the state of the system is $\omega$ with probability $p\in ]0, 1[$ and $\nu$ with probability $1-p$,
the total error probability  is equal to 
$$
p\,\omega(1 -T)+(1-p)\,\nu(T),
$$
which we wish to minimize over $T$. It is convenient to absorb the scalar factors into $\nu$ and 
$\omega$ and consider the quantities 
\begin{equation}
\begin{split}
D(\nu,\omega,T)&=\nu(T)+\omega(1-T),\\[3mm]
D (\nu,\omega)&=\inf _{T}D(\nu,\omega,T),
\end{split}
\label{cergy}
\end{equation}
for given faithful linear functionals $\nu,\omega$ on $\cO$. Note that $D(\nu,\omega)$
is unitary invariant, {\sl i.e.,}
$$
D(U\nu U^\ast,U\omega U^\ast)=D(\nu,\omega),
$$
for any unitary $U\in\cO$. Since $D(\nu,\omega,T)=D(\omega,\nu,1-T)$, one also has
$D(\nu,\omega)=D(\omega,\nu)$.

The following result is known as the Quantum Neyman-Pearson Lemma \cite{ANS}.
\bep\label{prop-np}
\[
D(\nu, \omega)
=D(\nu, \omega, \s_{(\omega-\nu)_+})
= \frac{1}{2}(\omega(1) +\nu(1) -\Tr\, |\omega-\nu|).
\]
\eep
\proof 
For any test $T$, 
\begin{align*}
\nu(T) + \omega (1-T)&=\omega(1) -(\omega-\nu)(T)\geq \omega(1) -(\omega-\nu)_+(T)\\[3mm]
&\geq \omega(1)-(\omega-\nu)_+(1)=\frac{1}{2}(\omega(1) +\nu(1)- \Tr\, |\omega-\nu|).
\end{align*}
On the other hand,
\[
D(\nu, \omega, \s_{(\omega-\nu)_+})
=\omega(1)-(\omega-\nu)_+(1)=\frac{1}{2}(\omega(1) +\nu(1) - \Tr\,|\omega-\nu|).
\]
\qed

In the literature one often considers generalized tests defined as $T\in {\cal O}_{\rm self}$ satisfying 
$0\leq T\leq 1$. Proposition \ref{prop-np} holds with the same proof if $D(\nu,\omega)$ is defined by
taking the infimum in \eqref{cergy} over all generalized tests. The same remark applies to all other 
results discussed in this paper.

\subsection{Bounds}
\label{finQHTBounds}
In this section we discuss lower and upper bounds on the minimal error probability $D(\nu,\omega)$ 
that will play a key role in the sequel. These bounds are most easily described in terms of modular 
operators which act on the complex vector space $\cO$ equipped with the Hilbert-Schmidt inner 
product 
\[
(A|B)=\Tr\, A^\ast B.
\]
Operators acting on this Hilbert space are sometimes called {\sl superoperators} in the
physics literature.

The relative modular operator $\Delta_{\nu|\omega}$ associated with two faithful linear 
functionals $\nu$, $\omega$ on $\cO$ is defined by
\[
\cO\ni A\mapsto\Delta_{\nu|\omega}A=\nu A\omega^{-1}.
\]
As a linear operator on $\cO$, $\Delta_{\nu|\omega}$ is positive. Its spectrum consists of the 
eigenvalues $\lambda/\mu$, $\lambda\in\sp(\nu)$, $\mu\in\sp(\omega)$. The corresponding spectral 
projections are the maps $A\mapsto P_\lambda(\nu)AP_\mu(\omega)$.

Set $\Omega_\omega=\omega^{1/2}$ and let $\mu_{\nu|\omega}$ be the 
spectral measure for $-\log\Delta_{\nu|\omega}$ and $\Omega_\omega$. Then one has
\[
(\Omega_\omega|\Delta_{\nu|\omega}^s\Omega_\omega)
=\int\e^{-sx} \d\mu_{\nu|\omega}(x)= {\rm Tr}\,\nu^s\omega^{1-s},
\] 
and 
\[
(\Omega_\omega|\Delta_{\nu|\omega}(1+\Delta_{\nu|\omega})^{-1}\Omega_\omega)
=\int\frac{ \d\mu_{\nu|\omega}(x)}{1+\e^{x}}
=\sum_{(\lambda,\mu)\in\sp(\nu)\times\sp(\omega)}
\frac{\Tr P_\lambda(\nu)P_\mu(\omega)}{\lambda^{-1}+\mu^{-1}}.
\]
The main advantage of writing these quantities in terms of modular operators is that
the following Proposition carries over without change to the infinite dimensional case
(see Theorem \ref{q-NP-Lemma}).

\bep\label{prologue-bounds}
\begin{enumerate}[{\rm (1)}] 
\item Upper bound: for  any $s\in [0,1]$,
\[
D(\nu,\omega) \leq(\Omega_\omega|\Delta_{\nu|\omega}^s\Omega_\omega).
\]
\item Lower bound:
\[
D(\nu,\omega)
\geq(\Omega_\omega|\Delta_{\nu|\omega}(1+\Delta_{\nu|\omega})^{-1}\Omega_\omega).
\]
\end{enumerate}
\eep

\proof  
Part (1) follows from Proposition \ref{paris-hot} and we only need to prove Part (2). 
For any test $T$, one has
\begin{gather*}
\nu(T)=\sum_{\lambda\in\sp(\nu)}\lambda\Tr\,TP_\lambda(\nu)T
=\sum_{(\lambda,\mu)\in\sp(\nu)\times\sp(\omega)}
\lambda\mu^{-1}\omega(TP_\lambda(\nu)T P_\mu(\omega)),\\[5pt]
\omega(1-T)=\sum_{(\lambda,\mu)\in\sp(\nu)\times\sp(\omega)}
\omega((1-T)P_\lambda(\nu)(1-T)P_\mu(\omega)).
\end{gather*}
Since, for $\kappa\ge0$,
$$
\kappa TP_\lambda(\nu)T+ (1-T)P_\lambda(\nu)(1-T)=\frac{\kappa}{1+\kappa}P_\lambda(\nu) 
+ \frac{1}{1+\kappa}(1-(1+\kappa)T)P_\lambda(\nu)(1-(1+\kappa)T)
\geq \frac{\kappa}{1+\kappa}P_\lambda(\nu),
$$
we derive
\[
\nu(T) +\omega(1-T)\ge\sum_{(\lambda,\mu)\in\sp(\nu)\times\sp(\omega)}
\frac{\lambda/\mu}{1+\lambda/\mu}
\omega(P_\lambda(\nu)P_\mu(\omega))=
(\Omega_\omega|\Delta_{\nu|\omega}(1+\Delta_{\nu|\omega})^{-1}\Omega_\omega).
\]
\qed

\subsection{Asymptotic hypothesis testing}
\label{sect-qht-results}

After these preliminaries, we turn to asymptotic hypothesis testing. For each $n \in \nn$, 
let ${\cal Q}_n$ be a quantum system described by the finite dimensional Hilbert spaces $\cH_n$ 
and let $(\nu_n, \omega_n)$ be a pair of faithful linear functionals on $\cO_{\cH_n}$. 
Let $w_n>0$ be given {\em weights} such that $\lim_{n\to\infty} w_n=\infty$. Error exponents 
of the Chernoff type associated to $(\nu_n, \omega_n, w_n)$ are  defined  by 
\[
\bar D=\limsup_{n \rightarrow \infty}\frac{1}{w_n} \log D(\nu_n, \omega_n),
\]
\[
\ubar D=\liminf_{n \rightarrow \infty}\frac{1}{w_n}\log  D(\nu_n, \omega_n).
\]

An immediate consequence of Proposition \ref{prologue-bounds} (1) is: 
\bep\label{bac}
\[
\bar D\leq\inf_{s\in[0,1]}\limsup_{n\to\infty}\frac{1}{w_n}\Ent_s(\nu_n|\omega_n).
\]
\eep
Lower bounds in asymptotic hypothesis testing are intimately linked with the theory of large 
deviations and to discuss them we need 

\newcommand{\AFone}{{\hyperref[AF1]{{\rm(AF1)}}}}
\begin{quote}\label{AF1}
{\bf Assumption (AF1).} The limit 
\[
e(s)= \lim_{n \rightarrow \infty}\frac{1}{w_n}\Ent_s(\nu_n|\omega_n)
\]
exists and is finite for $s\in [0,1]$. The function $s\mapsto e(s)$ is continuous on $[0,1]$,  
differentiable on $]0,1[$, and $D^+e(0)<D^-e(1)$. 
\end{quote}

Note that the limiting function $s\mapsto e(s)$ is convex on $[0,1]$.

The following result is known as the Chernoff bound.

\bet\label{fin-chernoff}
Suppose that Assumption \AFone{} holds. Then
\[
\ubar D=\bar D=\inf_{s\in [0,1]}e(s).
\]
\eet

\proof
By Proposition \ref{bac}, we only need to prove that 
\[
\ubar D \geq \inf_{s\in [0,1]} e(s).
\]
Applying Chebyshev's inequality to Part (2) of Proposition \ref{prologue-bounds} one easily shows
that
\begin{equation}
\ubar D\geq\liminf_{n\to\infty}\frac1{w_n}\log\mu_{\nu_n|\omega_n}(]-\infty,-\theta w_n[),
\label{hotsaturday}
\end{equation}
for any $\theta\ge0$. Suppose first that $e(0)\leq e(1)$. Then $D^-e(1)>0$ and since
\[
e(s)=\lim_{n\rightarrow \infty}\frac{1}{w_n}\log \int\e^{-sx}\d\mu_{\nu_n|\omega_n}(x),
\]
Proposition \ref{GE-THM} implies  that for  $\theta \in ]D^+e(0), D^-e(1)[$, 
\begin{equation}
\liminf_{n\rightarrow \infty}\frac{1}{w_n}\log \mu_{\nu_n|\omega_n}(]-\infty,-\theta w_n[)
\geq -\varphi(\theta),
\label{tues-sick}
\end{equation}
with
\[
\varphi (\theta) =\sup_{s\in [0,1]}(\theta s - e(s)).
\]
If $0>D^{+}e(0)$ then $0\in]D^+e(0), D^-e(1)[$ and it follows from \eqref{hotsaturday} and 
\eqref{tues-sick} that
$$
\ubar D\geq -\varphi(0)=\inf_{s\in [0,1]}e(s).
$$
If $0 \leq D^{+}e(0)$, then \eqref{hotsaturday} and \eqref{tues-sick} imply that
$\ubar D\geq -\varphi(\theta)$ for any $\theta\in]D^+e(0), D^-e(1)[$ and since $\varphi$ 
is continuous, one has
$$
\ubar D\geq -\varphi(D^+e(0))=e(0)\ge\inf_{s\in[0,1]}e(s).
$$

If $e(0)>e(1)$, one derives the result by exchanging the roles of $\nu_n$ and $\omega_n$, using
the fact that $D(\nu_n, \omega_n)=D(\omega_n, \nu_n)$.
\qed

We now turn to asymmetric hypothesis testing. Until the end of this section  we assume that 
$\nu_n$ and $\omega_n$ are states. The asymmetric hypothesis testing concerns individual 
error probabilities $\omega_n(1-T_n)$ (type-I error)  and $\nu(T_n)$ (type-II error).  
For $r\in \rr$, error exponents of the Hoeffding type are defined  by 
\begin{align*}
\bar B(r)&=\inf_{\{T_n\}}\left\{\limsup_{n \rightarrow \infty}\frac{1}{w_n}\log \omega_n(1-T_n)
\,\, \bigg|\,\, \limsup_{n\rightarrow \infty}\frac{1}{w_n}\log 
\nu_n(T_n)<-r\right\},\\[3mm]
\ubar B(r)&=\inf_{\{T_n\}}\left\{\liminf_{n \rightarrow \infty}\frac{1}{w_n}\log \omega_n(1-T_n)
\,\, \bigg|\,\, \limsup_{n\rightarrow \infty}\frac{1}{w_n}\log 
\nu_n(T_n)<-r\right\},\\[3mm]
B(r)&=\inf_{\{T_n\}}\left\{\lim_{n \rightarrow \infty}\frac{1}{w_n}\log \omega_n(1 -T_n)
\,\, \bigg|\,\, \limsup_{n \rightarrow \infty}\frac{1}{w_n}\log 
\nu_n(T_n)<-r\right\},
\end{align*}
where in the last case the infimum is taken over all families of tests $\{T_n\}$ for which 
$\lim_{n}\frac{1}{w_n}\log \omega_n(1 -T_n)$ exists. These exponents give the best exponential 
convergence rate of type-I error under the exponential convergence constraint on the type-II error. 
An alternative interpretation is in terms of state concentration: as $n\rightarrow \infty$ the states 
$\omega_n$ and $\nu_n$ are concentrating along orthogonal subspaces and the Hoeffding 
exponents quantify the degree of this separation on the exponential scale. In classical statistics 
the Hoeffding exponents can be traced back to \cite{Ho, CL, Bl}. 

The following result is known as the Hoeffding bound \cite{Ha2, Na}:
\bet\label{fin-hoeffding-bound}
Suppose that Assumption \AFone{} holds. Then for all $r\in \rr$, 
\[
\ubar B(r)= \bar B(r)=B(r)=-\sup_{s\in[0,1[}\frac{-sr -e(s)}{1-s}.
\]
\eet

For $\epsilon \in ]0,1[$, error exponents of the Stein type are defined by 
\begin{align*}
&\bar B_\epsilon=\inf_{\{T_n\}}\left\{\limsup_{n \rightarrow \infty}\frac{1}{w_n}\log \omega_n(1-T_n)
\,\, \bigg|\,\,\nu_n(T_n)\leq \epsilon\right\},\\[3mm]
&\ubar B_\epsilon=\inf_{\{T_n\}}\left\{\liminf_{n \rightarrow \infty}\frac{1}{w_n}\log \omega_n(1-T_n)
\,\, \bigg|\,\,\nu_n(T_n)\leq \epsilon\right\},\\[3mm]
&B_\epsilon=\inf_{\{T_n\}}\left\{\lim_{n \rightarrow \infty}\frac{1}{w_n}\log \omega_n(1-T_n)
\,\, \bigg|\,\, \nu_n(T_n)\leq \epsilon\right\},
\end{align*}
where in the last case the infimum is taken over all families of tests $\{T_n\}$ for which
$\lim_{n}\frac{1}{w_n}\log \omega_n(1-T_n)$ exists.  Note that if 
\[
\beta_n(\epsilon)=\inf_{T:\nu_n(T)\leq \epsilon}\omega_n(1 -T),
\]
then 
\[
\liminf_{n\rightarrow \infty} \frac{1}{w_n}\log \beta_n(\epsilon)
=\ubar B_\epsilon, \qquad \limsup_{n\rightarrow \infty}
\frac{1}{w_n}\log \beta_n(\epsilon)=\bar B_\epsilon.
\]
The study of Stein's exponents in the quantum setting goes back to \cite{HP}.
To discuss these exponents  we need 

\newcommand{\AFtwo}{{\hyperref[AF2]{{\rm(AF2)}}}}
\begin{quote}\label{AF2}
{\bf Assumption (AF2).}  For some $\delta >0$,  the limit 
\[
e(s)=\lim_{n \rightarrow \infty}\frac{1}{w_n}\Ent_s(\nu_n|\omega_n),
\]
exists and is finite for all  $s\in [0, 1+\delta[$. The function $s\mapsto e(s)$ is differentiable at $s=1$.
\end{quote}

Relation \eqref{DRenyi}, Assumption \AFtwo{} and convexity imply
\[
e^\prime(1)=-\lim_{n\rightarrow \infty}\frac{1}{w_n}\Ent(\nu_n|\omega_n).
\]
In accordance with the terminology used in non-equilibrium statistical mechanics, we shall call 
\[
\Sigma^+=e^\prime(1)
\]
the {\em entropy production} of the hypothesis testing. The following result is known as Stein's 
Lemma \cite{HP, ON}.
\bet\label{fin-St-lm}
Suppose that Assumptions \AFone{}--\AFtwo{} hold. Then for all $\epsilon \in ]0,1[$, 
\[
\ubar B_\epsilon=\bar B_\epsilon=B_\epsilon=-\Sigma^+.
\]
\eet
{\bf Remark.} This result holds under more general conditions then (AF1) and (AF2), see Section~\ref{sec-SL}. 

Just like the Chernoff bound, the Hoeffding bound and Stein's Lemma 
(Theorems \ref{fin-hoeffding-bound} and \ref{fin-St-lm}) are easy consequences of Proposition 
\ref{prologue-bounds} and the G\"artner-Ellis theorem. To avoid repetitions, we refer the reader to 
Sections~\ref{sec-HB} and \ref{sec-SL} for their proofs in the general  $W^\ast$-algebraic setting. 
As we have already mentioned, given Proposition \ref{prologue-bounds} (and its $W^\ast$-algebraic
generalization), the proofs follow line by line the classical arguments. The  non-trivial aspects of 
non-commutativity  emerge only in the verification of Assumptions \AFone--\AFtwo{} in the context of
concrete quantum statistical models. 

\subsection{Examples}

\subsubsection{Quantum i.i.d.\;states}
Quantum i.i.d.\;states are the simplest (and most widely studied) examples of quantum hypothesis
testing. Asymptotic hypothesis testing for such states can be interpreted in terms of multiple
measurements on independent, statistically equivalent systems.
Let $\cK$ be a finite dimensional Hilbert space, $\nu$ and $\omega$ two faithful linear 
functionals on   $\cO_\cK$, and 
\[
\cH_n=\otimes_{j=1}^n\cK, \qquad \nu_n
=\otimes_{j=1}^n\nu, \qquad \omega_n=\otimes_{j=1}^n \omega.
\]
For $s\in \rr$, one has
\[
\Tr\, \nu_n^s \omega_n^{1-s}=(\Tr \, \nu^s \omega^{1-s})^n
\]
and, taking the weights $w_n=n$, we see that 
$$
e(s)=\lim_{n\rightarrow \infty}\frac{1}{n}\Ent_s(\nu_n|\omega_n)=\Ent_s(\nu|\omega)
=\sum_{(\lambda,\mu)\in\sp(\nu)\times\sp(\omega)}
\lambda^s\mu^{1-s}\Tr \, P_\lambda(\nu)P_\mu(\omega).
$$
The function 
$e(s)$ is real-analytic on $\rr$ and is strictly convex iff $\nu\not=\omega$. 
In particular, Theorems \ref{fin-chernoff}, \ref{fin-hoeffding-bound} and 
\ref{fin-St-lm} hold for quantum i.i.d.\;states. 

\subsubsection{Quantum spin systems}
\label{FinQSS}
Let ${\cal P}$ be the collection of all finite subsets of $\zz^d$. For $X\in {\cal P}$,  $|X|$ denotes the
cardinality of $X$ (the number of  elements of $X$), ${\rm diam}\,X=\max\{|x-y|\,|\,x,y\in X\}$ is the 
diameter of $X$,  and $X+a=\{x+a\,|\,x\in X\}$ is the translate of $X$ by $a\in\zz^d$. 
 
Suppose that a single spin is described by the finite dimensional Hilbert space $\cK$. We attach a copy
 $\cK_x$ of $\cK$ to each site $x\in\zz^d$.  For $X\in{\cal P}$ we define $\cH_X=\otimes_{x\in X}\cK_x$
and $\cO_X=\cO_{\cH_X}$. $\|A\|$ is the usual operator norm of $A\in\cO_X$. 
For $X\subset Y$, the identity $\cH_Y=\cH_X\otimes\cH_{Y\setminus X}$ yields a natural
identification of $\cO_X$ with a $\ast$-subalgebra of $\cO_Y$. For $a\in\zz^d$, we denote by
$T^a:\cK_x\to\cK_{x+a}$ the identity map. $T^a$ extends to a unitary map $\cH_X\to\cH_{X+a}$.

An interaction is a collection $\Phi=\{\Phi_X\}_{X\in\cal P}$ such that
$\Phi_X\in\cO_{X,{\rm self}}$ and $T^a\Phi_{X}T^{-a}=\Phi_{X+a}$ for any $a\in\zz^d$. We set 
\[
|||\Phi||| =\sum_{X\ni0}|X|^{-1}\|\Phi_X\|, \qquad \|\Phi\|=\sum_{X\ni0}\|\Phi_X\|.
\]
The interaction $\Phi$ is finite range if for some $R>0$ and any $X\in\cal P$ such that
${\rm diam}\,X >R$, $\Phi_X=0$.

To a given  box  $\Lambda_n =[-n, n]^d$ in $\zz^d$ one associates the Hamiltonian 
\[
H_{\Lambda_n}(\Phi)=\sum_{X\subset \Lambda_n}\Phi_X.
\]
If $|||\Phi|||<\infty$, then the limit
\[
P(\Phi)=\lim_{n \rightarrow \infty}\frac{1}{|\Lambda_n|}\log \Tr\, \e^{- H_{\Lambda_n}(\Phi)}
\]
exists and is called the pressure of $\Phi$. The bound $|P(\Phi)-P(\Psi)|\leq |||\Phi-\Psi|||$ holds and for 
$s\in [0, 1]$, 
\begin{equation}
P(s\Phi + (1-s)\Psi)\leq s P(\Phi) +(1-s)P(\Psi)
\label{pre-con}
\end{equation}
(see \cite{Si}). 
Two interactions $\Phi$ and $\Psi$ are called physically equivalent (denoted $\Phi \sim \Psi$) if
equality holds in \eqref{pre-con} for all $s\in [0,1]$. For further information about the notion of 
physical equivalence we refer the reader to \cite{Is}.

Let $\Phi$ and $\Psi$ be interactions and $\nu_n$, $\omega_n$  the states  defined by 
\[
\nu_n =\frac{\e^{- H_{\Lambda_n}(\Phi)}}{\Tr\, \e^{-H_{\Lambda_n}(\Phi)}},\qquad 
\omega_n =\frac{\e^{- H_{\Lambda_n}(\Psi)}}{\Tr\, \e^{-H_{\Lambda_n}(\Psi)}}.
\]
Let $w_n=|\Lambda_n|$ be the weights.

\bet\label{spin-d-1}
Suppose that $d=1$ and that $\Phi$ and $\Psi$ are finite range. Then
\begin{enumerate}[{\rm (1)}]
\item The limit 
\[
e(s)= \lim_{n \rightarrow\infty}\frac{1}{|\Lambda_n|}\Ent_s(\nu_n,\omega_n)
\]
exists for all $s$. 
\item $e(s)$ is a real analytic function on $\rr$. 
\item If $\Phi\sim \Psi$, then $e(s)=0$ for all $s$. 
\item If $\Phi \not \sim \Psi$, then $e(s)$ is strictly convex on $\rr$.  
\end{enumerate}
\eet
\proof
Part  (1) is proven in \cite{LRB} (see also \cite{NR}) and  Part (2) in \cite{Og2}.  To prove (3), note
that the Golden-Thompson inequality implies that for all $s\in \rr$,
\[
e(s)\geq  p(s)=P(s\Phi + (1-s)\Psi)-sP(\Phi)- (1-s)P(\Psi).
\]
Since $e(0)=e(1)=0$ and $e(s)\leq 0$ for $s\in [0,1]$, if $\Phi\sim \Psi$ then $e(s)=0$ on $[0,1]$, 
and so,  by  analyticity, $e(s)=0$ for all $s$. To prove (4), note that $p(s)$ is also a real  analytic 
convex function satisfying $p(0)=p(1)=0$. If $\Phi \not \sim \Psi$, then $p(s)$ is not identically 
equal to zero, and so $e(s)$ is also not identically equal to zero. The analyticity then implies that 
$e(s)$ is strictly convex. 
\qed

For $d>1$ the following is known. 

\bet\label{spin-d}
Given $r>0$, there is $\delta >0$ such that, for any two finite-range interactions $\Phi$, $\Psi$
satisfying $\|\Phi\|, \|\Psi\|<\delta$, one has: 
\begin{enumerate}[{\rm (1)}]
\item
\[
e(s)= \lim_{n \rightarrow\infty}\frac{1}{|\Lambda_n|}\Ent_s(\nu_n,\omega_n),
\]
exists for $|s|<r$.
\item $e(s)$ is real analytic on $]-r, r[$.
\item If $\Phi\sim \Psi$, then $e(s)=0$ for $s\in ]-r, r[$.  
\item If $\Phi \not \sim \Psi$, then $e(s)$ is strictly convex $]-r, r[$.  
\end{enumerate}
\eet

{\bf Remark 1.} The existence of $e(s)$ is established in \cite{NR} (see also \cite{LRB}).  The real 
analyticity follows from Proposition 7.10 in \cite{NR}. Parts (3) 
and (4) are proven in the same way as in Theorem~\ref{spin-d-1}.

{\bf Remark 2.} One can get an explicit estimate on $\delta$ in terms of $r$ by combining
Theorem 6.2.4  in \cite{BR2} and Proposition 7.10  in \cite{NR}.

{\bf Remark 3.} Theorem \ref{spin-d} is a high-temperature result.  Although, in general, for $d>1$ 
and  low temperatures one does not expect analyticity (or even differentiability) of $e(s)$, one 
certainly expects that $e(s)$ exists. Remarkably, it is {\em not known} in general whether 
$e(s)$ exists outside the high temperature regime.

\subsubsection{Quasi-free CAR states}
\label{sec-CAR}

Quantum hypothesis testing for translation invariant quasi-free states on the algebra of fermionic
creation/anni\-hila\-tion operators on $\zz^d$ has been studied in \cite{MHO}. For later applications 
we start with a more general setup. 

Let $\fh$ be the single particle Hilbert space and let $\cH=\Gamma_{\rm f}(\fh)$ be the antisymmetric
(fermionic) Fock space over $\fh$. The CAR algebra $\CAR(\fh)$ is the $C^\ast$-algebra generated
by $\{ a^\#(f)\,|\, f \in \fh\}$, where $a^\#(f)$ stands for $a(f)$ (the annihilation operator) or $a^\ast(f)$ 
(the creation  operator). $\varphi(f)=\frac{1}{\sqrt 2}(a(f) + a^\ast(f))$ denotes the field operator. 
Every self-adjoint operator $T$ on $\fh$ satisfying $0\leq T\leq 1$ generates a state 
$\omega_T$ on $\CAR(\fh)$ by 
\[
\omega_T(a^\ast(f_n)\cdots a^\ast(f_1)a(g_1)\cdots a(g_m))=\delta_{nm} 
\det\{ (g_i| Tf_j)\}.
\]
$T$ is called the density operator and $\omega_T$ is the gauge invariant quasi-free state with density 
$T$. This state is completely determined by its two point function 
\[
\omega_T(a^\ast(f)a(g))=(g|Tf).
\]
If $h$ is the one-particle Hamiltonian, then the thermal equilibrium state at inverse temperature 
$\beta>0$ and chemical potential $\mu\in\rr$ of the corresponding ideal Fermi gas is described 
by the gauge invariant quasi-free state with the Fermi-Dirac density
\[
T_{\beta, \mu}= \left( 1 + \e^{\beta (h-\mu)}\right)^{-1}.
\]

Let now $A\not=B$ be two  generators such that $\delta <A,B<1-\delta$ for some $\delta >0$. 
Suppose  that $\dim \fh=\infty$. Quantum hypothesis testing is set with respect to a sequence 
of finite dimensional orthogonal projections $p_n$ on $\fh$ such that ${\rm s}-\lim_n p_n=1$. 
Let $\fh_n =\Ran p_n$, $A_n=p_nAp_n$, $B_n=p_nBp_n$, and let $\nu_n$ and $\omega_n$ 
be quasi-free states on $\CAR(\fh_n)$ generated by $A_n$ and $B_n$. A straightforward
computation yields
\[
\Tr\,\nu_n^s\omega_n^{1-s}
=\det \left[ B_n^{(1-s)/2}A_n^s B_n^{(1-s)/2}
+ (1-A_n)^{s/2}(1-B_n)^{1-s}(1-A_n)^{s/2}\right],
\]
and 
\[
\Ent_s(\nu_n|\omega_n)
=\Tr\, \log\left[ B_n^{(1-s)/2}A_n^s B_n^{(1-s)/2}
+ (1-A_n)^{s/2}(1-B_n)^{1-s}(1-A_n)^{s/2}\right].
\]
The natural choice for weights is $w_n=\dim \fh_n$. At this point  one needs to specify the model further.
The case considered in \cite{MHO}  is $\fh=\ell^2 (\zz^d)$, $\fh_n=\ell^2(\Lambda_n)$, with  $A$ 
and $B$ translation invariant. If ${\mathfrak F}:\fh\rightarrow L^2([0, 2\pi]^d, \d k)$ is the usual Fourier
transform, then ${\mathfrak F}A{\mathfrak F}^{-1}$ and ${\mathfrak F}B{\mathfrak F}^{-1}$ are 
operators of multiplication by bounded measurable functions $A(k)$ and $B(k)$ whose essential 
ranges are contained in $[\delta, 1-\delta]$. An application of Sz\"ego's theorem (see \cite{MHO}
for details) yields 
\bet For all $s\in \rr$, 
\[
e(s)=\lim_{n\rightarrow \infty}\frac{1}{|\Lambda_n|}\Ent_s(\nu_n|\omega_n)
=\int_{[0, 2\pi]^d}\log 
\left[ A(k)^sB(k)^{1-s} + (1-A(k))^{s}(1-B(k))^{1-s}\right]\frac{\d k}{(2\pi)^d},
\]
and the function $e(s)$ is real analytic and strictly convex on $\rr$.
\eet

Hypothesis testing for translation invariant quasi-free CCR states on $\zz^d$ has been studied in 
\cite{Mo}.

\section{Entropy production and full counting statistics}
\label{sec-fullep}

\subsection{Setup}
\label{FCS-finite}
We follow \cite{JOPP1}. Let $\cQ$ be a quantum system described by a finite dimensional Hilbert 
space $\cH$, Hamiltonian $H$, and initial state $\omega>0$. The state evolves in time as
$\omega_t=\e^{-\i t H}\omega\e^{\i t H}$ while the observables evolve as 
$A_t=\e^{\i t H}A\e^{-\i t H}$. Obviously, $\omega_t(A)=\omega(A_t)$.

We define the {\sl entropy observable} of $\cQ$ by $S=-\log \omega$. Note that 
\[
S_t =-\log \omega_{-t}.
\]
The {\sl entropy production observable,} defined by 
\[
\sigma=\left.\frac{\d\ }{\d t} S_t\right|_{t=0} =-\i [H, \log \omega],
\]
is the quantum analog of the phase space contraction rate in classical non-equilibrium statistical 
mechanics \cite{JPR}. In terms of the {\sl relative Hamiltonian}
\[
\ell_{\omega_t|\omega}=\log \omega_t- \log \omega=\int_0^t \sigma_{-u}\d u,
\]
which satisfies the cocycle relation 
\[
\ell_{\omega_{t+s}|\omega}=\ell_{\omega_t|\omega} + (\ell_{\omega_s|\omega})_{-t},
\]
the relative entropy of $\omega_t$ w.r.t.\;$\omega$ (recall Equ. \eqref{FinRelEntDef}) is given by
$$
\Ent(\omega_t|\omega)=-\omega_t(\ell_{\omega_t|\omega}).
$$
Defining the {\sl mean entropy production rate observable} over the time interval $[0, t]$ by
\[
\Sigma^t=\frac{1}{t}(S_t-S)=\frac{1}{t}\int_0^t \sigma_u \d u,
\]
we can write the entropy balance equation
\begin{equation}
\omega(\Sigma^t)=\frac1t\int_0^t\omega(\sigma_u)\d u, = -\frac1t\Ent(\omega_t|\omega)\geq 0,
\label{fin-balance}
\end{equation}
which is a finite time expression of the second law of thermodynamics.

We shall call 
$$
(t,s)\mapsto e_t(s)=\Ent_s(\omega_t|\omega)
$$
the {\sl R\'enyi entropic functional} of the quantum system $\cQ$. Since
$\Ent_s(\omega_{-t}|\omega)=\Ent_s(\omega|\omega_t)=\Ent_{1-s}(\omega_t|\omega)$,
one has
\begin{equation}
e_{-t}(s)=e_t(1-s),
\label{RenyiSym}
\end{equation}
for any $t,s\in\rr$. With the notations of Section~\ref{finQHTBounds}
\begin{equation}
e_t(s)=\log (\Omega_\omega|\Delta_{\omega_t|\omega}^s \Omega_\omega)
=\log\int\e^{-sx}\d\mu_{\omega_t|\omega}(x),
\label{Renyimu}
\end{equation}
and Relation \eqref{DRenyi} yields
\begin{equation}
\left.\frac{\d\ }{\d s}e_t(s)\right|_{s=0}=\Ent(\omega|\omega_t),
\qquad
\left.\frac{\d\ }{\d s}e_t(s)\right|_{s=1} = -\Ent(\omega_t|\omega).
\label{StCyr-late}
\end{equation}
Moreover, Relation \eqref{RenyiSym} translates into
\begin{equation}
\d\mu_{\omega_{-t}|\omega}(x)=\e^x\d\mu_{\omega_t|\omega}(-x).
\label{GenES}
\end{equation}

The physical interpretation of the functional $e_t(s)$ and the spectral measure 
$\mu_{\omega_t|\omega}$ is in terms of the full counting statistics. At time $t=0$, with 
the system in the state $\omega$, one performs a measurement of the entropy observable $S$. 
The possible outcomes of this measurement are eigenvalues of $S$  and $\alpha\in\sp(S)$ is 
observed with  probability $\omega(P_\alpha)$, where $P_\alpha$ is the spectral projection of
$S$ for its eigenvalue $\alpha$. After the measurement, the state of $\cQ$ is 
$$
\frac{\omega P_\alpha}{\omega(P_\alpha)},
$$
and this state evolves over the time interval $[0, t]$ to 
$$
\frac{\e^{-\i tH}\omega P_\alpha \e^{\i tH}}{\omega(P_\alpha)}.
$$
A second measurement of $S$ at time $t$ yields the result $\alpha'\in\sp(S)$ with
probability
$$
\frac{\Tr \left(\e^{-\i t H}\omega P_\alpha \e^{\i tH }P_{\alpha'}\right)}{\omega(P_\alpha)}.
$$
The joint probability distribution of the two measurements is given by
$$
\Tr\left(\e^{-\i tH}\omega P_\alpha\e^{\i t H}P_{\alpha'}\right),
$$
and the probability distribution of the mean rate of change of the entropy, $\phi=(\alpha'-\alpha)/t$, is 
given by 
$$
\d{\mathbb P}_t(\phi)
=\sum_{\alpha,\alpha'}\Tr\left(\e^{-\i tH}\omega P_\alpha\e^{\i t H}P_{\alpha'}\right)
\delta(\phi-(\alpha'-\alpha)/t)\d\phi.
$$
We shall say that the probability measure ${\mathbb P}_t$ is the {\sl Full Counting Statistics} (FCS)
of the system $\cQ$. Since
\[
{\rm Tr}\, \omega_{-t}^{s}\omega^{1-s}= 
{\rm Tr}\, \omega_t^{1-s}\omega^{s}= 
\sum_{\alpha,\alpha'\in\sp(S)}\e^{-s(\alpha'-\alpha)} 
{\rm Tr}\left(\e^{-\i tH}\omega P_\alpha\e^{\i t H}P_{\alpha'}\right)
=\int\e^{-st\phi}\d{\mathbb P}_t(\phi),
\]
we conclude that
$$
e_{-t}(s)=e_{t}(1-s)=\log\int\e^{-st\phi}\d{\mathbb P}_t(\phi).
$$
Comparing this relation with Equ. \eqref{Renyimu} and using Equ. \eqref{GenES} we show that 
\begin{equation}
\d{\mathbb P}_t(\phi)=\e^{t\phi}\d\mu_{\omega_t|\omega}(-t\phi)=\d\mu_{\omega_{-t}|\omega}(t\phi),
\label{QPrelat}
\end{equation}
{\sl i.e.,} that ${\mathbb P}_t$ is the spectral measure of $-\frac1t\log\Delta_{\omega_{-t}|\omega}$.

The expectation and variance of $\phi$ w.r.t.\;${\mathbb P}_t$
are given by
\begin{equation}
\begin{split}
{\mathbb E}_t(\phi)
&=\left.-\frac1t\partial_s e_t(1-s)\right|_{s=0}
=\left.\frac1t\partial_s e_{t}(s)\right|_{s=1}=\omega(\Sigma^t),\\
{\mathbb E}_t(\phi^2)-{\mathbb E}_t(\phi)^2
&=\left.\frac1{t^2}\partial_s^2 e_t(1-s)\right|_{s=0}=\left.\frac1{t^2}\partial_s^2 e_{t}(s)\right|_{s=1}
=\omega({\Sigma^t}^2)-\omega(\Sigma^t)^2.
\end{split}
\label{CPT-hot}
\end{equation}
They coincide with the expectation and variance of $\Sigma^t$ w.r.t.\;$\omega$. However, in general 
such a relation does not hold  for higher order cumulants.

The quantum system $\cQ$ is {\sl time reversal invariant} (TRI) if there exists an orthogonal basis
of $\cH$ for which the matrix representatives of $H$ and $\omega$ are both real. Denoting by
$\theta$ the complex conjugation in such a basis, $\Theta(A)=\theta A\theta$ defines an
anti-linear $\ast$-automorphism of $\cO$ for which\footnote{$\bar\omega$ denotes the anti-linear 
functional $\bar\omega(A)=\overline{\omega(A)}=\omega(A^\ast)$}
$$
\Theta(A_t)=(\Theta(A))_{-t},\qquad\omega(\Theta(A))=\bar\omega(A),
$$
and hence $\omega_t(\Theta(A))=\overline{\omega_{-t}}(A)$ holds for any $A\in\cO$. Since, for any 
states $\rho,\nu$, one has $\Ent_s(\overline{\rho\circ\Theta}|\overline{\nu\circ\Theta})=\Ent_s(\rho|\nu)$, 
we can write
$$
e_t(s)=
\Ent_s(\omega_t|\omega)=\Ent_s(\overline{\omega_t\circ\Theta}|\overline{\omega\circ\Theta})=
\Ent_s(\omega_{-t}|\omega)=e_{-t}(s).
$$
Thus, Equ. \eqref{RenyiSym} and the time-reversal invariance of $\cQ$ implies the transient 
{\sl Evans-Searles symmetry} of its R\'enyi entropic functional
\begin{equation}
e_t(s)=e_t(1-s).
\label{paris-1}
\end{equation}
This relation has the following equivalent reformulations:
\begin{enumerate}[{\rm (1)}]
\item $\log\int\e^{-st\phi}\d{\mathbb P}_t(\phi)= e_t(s)$;
\item $\d{\mathbb P}_t(\phi)=\d\mu_{\omega_t|\omega}(t\phi)$;
\item $\d{\mathbb  P}_t(-\phi)=\e^{-t\phi}\d{\mathbb  P}_t(\phi)$;
\item $\d\mu_{\omega_t|\omega}(-x)=\e^{-x}\d\mu_{\omega_t|\omega}(x)$.
\label{ESequiv}
\end{enumerate}

\subsection{Open systems}
To shed some light on the definitions introduced in the previous section, let us describe them in the 
more concrete setup of open quantum systems. 

Let $\cR_j$, $j=1,\ldots,n$, be quantum systems described by finite dimensional Hilbert spaces 
$\cH_j$ and Hamiltonians $H_j$. We denote by $\cO_j$ the corresponding $\ast$-algebras.
Let $N_j\in\cO_{j,{\rm self}}$, $[H_j, N_j]=0$, be given ``conserved charges". We assume 
that each system $\cR_j$  is in thermal equilibrium at inverse temperature $\beta_j>0$ and chemical 
potential $\mu_j\in\rr$, namely that its initial state is 
\[
\omega_j=\frac{\e^{-\beta_j(H_j-\mu_j N_j)}}{\Tr\, \e^{-\beta_j( H_j-\mu_jN_j)}}.
\]
In our context the systems  $\cR_j$ are individual thermal reservoirs. The Hilbert space,
$\ast$-algebra, Hamiltonian, and initial state of the full reservoir system $\cR=\cR_1+\cdots+\cR_n$
are
\[
\cH_\cR =\otimes_{j=1}^n \cH_j, 
\qquad
\cO_\cR=\otimes_{j=1}^n \cO_j, 
\qquad
H_\cR=\sum_{j=1}^nH_j, 
\qquad 
\omega_\cR=\otimes_{j=1}^n \omega_j
\]
(whenever the meaning is clear within the context we identify $A\in\cO_j$ with the element
$(\otimes_{k\not=j}1_{\cH_k})\otimes A$ of $\cO_\cR$). 

Let $\cS$ be  another quantum system described by the finite dimensional Hilbert space $\cH_\cS$,
$\ast$-algebra $\cO_\cS$ and Hamiltonian $H_\cS$. In our 
context $\cS$ will be a small quantum system coupled to the full reservoir $\cR$. The 
thermodynamic limit  discussed in next section concerns only reservoirs while $\cS$ will remain 
unchanged. A convenient reference state of $\cS$ is the chaotic state 
\[
\omega_\cS =\frac{1_{\cH_\cS}}{\dim\cH_\cS},
\]
but under normal circumstances none of the final results (after the thermodynamic and the large time 
limit are taken) depends on this choice. 

The interaction of $\cS$ with $\cR$ is described by the self-adjoint operator 
\[
V=\sum_{j=1}^n V_j,
\]
where $V_j=V_j^\ast\in\cO_{\cS}\otimes\cO_j$. The Hilbert space, $\ast$-algebra, Hamiltonian,
and initial state of the coupled joint system $\cQ=\cS+\cR$ are
$$
\cH=\cH_\cS\otimes\cH_\cR,
\qquad
\cO=\cO_\cS\otimes\cO_\cR,
\qquad
H=H_\cS+H_\cR+V,
\qquad
\omega=\omega_\cS\otimes\omega_\cR.
$$
The entropy observable of $\cQ$ is
\begin{equation}
S=\sum_{j=1}^n \beta_j(H_j -\mu_jN_j)-\sum_{j=1}^n\log\left(\Tr\,\e^{-\beta_j(H_j -\mu_jN_j)}\right)
-\log(\dim\,\cH_\cS),
\label{entro-open}
\end{equation}
and one easily derives that its entropy production observable is  
\[
\sigma=-\sum_{j=1}^n \beta_j(\Phi_j -\mu _j {\cal J}_j),
\]
where 
\begin{equation} \Phi_j=-\i [H, H_j]=-\i[V, H_j], \qquad {\cal J}_j=-\i [H, N_j]=-\i [V, N_j].
\label{def-flux}
\end{equation}
Since 
\[
\omega(H_{jt})-\omega(H_j)=-\int_0^t\omega(\Phi_{ju})\d u, \qquad 
\omega(N_{jt})-\omega(N_j)=-\int_0^t \omega({\cal J}_{ju})\d u,
\]
the observables $\Phi_j/{\cal J}_j$ describe energy/charge currents out of the reservoir $\cR_j$. 

In the framework of open quantum systems the FCS can be naturally generalized.
Consider the commuting set of observables 
\[
{\bf S}= (\beta_1H_1, \cdots, \beta_n H_n, -\beta_1\mu_1 N_1, \cdots, -\beta_n \mu_n N_n).
\]
Let $P_{\boldsymbol\alpha}$ be the joint  spectral projection of ${\bf S}$ corresponding to the eigenvalue 
${\boldsymbol \alpha}=(\alpha_1,\ldots,\alpha_{2n})\in\sp({\bf S})$. Then 
\[
\Tr\, \e^{-\i t H} \omega P_{\boldsymbol \alpha}\e^{\i t H}P_{{\boldsymbol\alpha^\prime}},
\]
is the  joint probability distribution of the measurement of ${\bf S}$ at time $t=0$ followed by a 
later measurement at time $t$. Let ${\mathbb P}_t({\boldsymbol\varepsilon},{\boldsymbol\nu})$ 
be the induced probability distribution of the vector
\[
({\boldsymbol \varepsilon}, {\boldsymbol \nu}) 
=(\varepsilon_1, \cdots, \varepsilon_n, \nu_1, \cdots, \nu_n)
=({\boldsymbol \alpha}^\prime - {\boldsymbol \alpha})/t,
\] 
which describes the mean rate of change of energy and charge of each reservoir.

Expectation and covariance of (${\boldsymbol\varepsilon},{\boldsymbol\nu})$ w.r.t.\;${\mathbb P}_t$
are given by
\begin{align}
{\mathbb E}_t(\varepsilon_j)&=-\frac{\beta_j}t\int_0^t\omega(\Phi_{js})\d s,\nonumber\\[-6pt]
&\label{FCSexp}\\[-6pt]
{\mathbb E}_t(\nu_j)&=\frac{\beta_j\mu_j}t\int_0^t\omega({\cal J}_{js})\d s,\nonumber
\end{align}
and
\begin{align}
{\mathbb E}_t(\varepsilon_j\varepsilon_k)-{\mathbb E}_t(\varepsilon_j){\mathbb E}_t(\varepsilon_k)
&=\frac{\beta_j\beta_k}{t^2}\int_{0}^t\int_0^t
\omega\left((\Phi_{js}-\omega(\Phi_{js}))(\Phi_{ku}-\omega(\Phi_{ku}))\right)\d s \d u,\nonumber\\[12pt]
{\mathbb E}_t(\nu_j\nu_k)-{\mathbb E}_t(\nu_j){\mathbb E}_t(\nu_k)
&=\frac{\beta_j\mu_j\beta_k\mu_k}{t^2}\int_{0}^t\int_0^t
\omega\left(({\cal J}_{js}-\omega({\cal J}_{js}))({\cal J}_{ku}-\omega({\cal J}_{ku}))\right)\d s \d u,
\label{FCSvar}\\[12pt]
{\mathbb E}_t(\varepsilon_j\nu_k)-{\mathbb E}_t(\varepsilon_j){\mathbb E}_t(\nu_k)
&=-\frac{\beta_j\beta_k\mu_k}{t^2}\int_{0}^t\int_0^t
\omega\left((\Phi_{js}-\omega(\Phi_{js}))({\cal J}_{ku}-\omega({\cal J}_{ku}))\right)\d s \d u.\nonumber
\end{align}
The cumulant generating function is
\[
\rr^{2n}\ni{\boldsymbol s}\mapsto
e_t({\boldsymbol s})=\log \sum_{({\boldsymbol\varepsilon},{\boldsymbol\nu})}
\e^{-t {\boldsymbol s}\cdot({\boldsymbol\varepsilon},{\boldsymbol\nu})}
{\mathbb P}_t({\boldsymbol\varepsilon},{\boldsymbol\nu}).
\]
If the open  quantum system $\cQ$ is TRI, then the fluctuation relation 
\[ 
e_t({\boldsymbol s})= e_t({\boldsymbol 1}- {\boldsymbol s})
\]
holds (${\boldsymbol 1}=(1, \cdots, 1)$). $e_t({\boldsymbol s})$ and ${\mathbb P}_t$ can be related 
to the modular structure (see \cite{JOPP1} for details). However, their relation with quantum 
hypothesis testing is unclear at the moment and we will  restrict ourselves to the FCS
of the entropy observable defined by  \eqref{entro-open}.

The above discussion of open quantum systems needs  to be adjusted if the particle statistics 
(fermionic/bosonic) is  taken into the account. These adjustments are minor and we shall discuss 
them only in the concrete example of the electronic black box model (see Section~\ref{sec-ebb}).

{\bf\noindent Remark.} To justify the name ``entropy'' for the observable $S$, note that
$\omega(S)=-\Tr\,\omega\log\omega$ is the Gibbs-von Neumann entropy of $\omega$.
It is well known that if $\omega$ is a thermal equilibrium state then this quantity coincides
with the Boltzmann entropy of the system.

\subsection{Thermodynamic limit}

The dynamics of a finite (or more generally confined) quantum system being quasi-periodic, the
large time asymptotics of its FCS is trivial. In order to get interesting information
from this asymptotics and to relate it to quantum hypothesis testing it is necessary to further idealize
the system by making it infinitely extended. In this section we briefly discuss the thermodynamic 
(TD) limit of the quantum system $\cQ$.

Let $\cQ_m$, $m=1,2,\ldots\,$,  be a sequence of finite quantum systems described by Hilbert 
spaces $\cH_m$, Hamiltonians $H_m$, and states $\omega_m>0$, with the understanding that 
$m\to\infty$ corresponds to the TD limit (for example, in the case of quantum spin 
systems $\cQ_m$ will be the finite spin system in the box $\Lambda_m=[-m,m]^d$ as discussed
in Section~\ref{FinQSS}). 

We shall indicate the dependence of various objects on $m$ by the subscript ${}_m$, {\sl e.g.,} 
${\mathbb P}_ {mt}$ denotes the FCS of $\cQ_m$ and $e_{mt}(s)$ its
R\'enyi entropic functional.
\[
e_{mt}(1-s)= \log \int\e^{-t s \phi}\d {\mathbb P}_{mt}(\phi)
\]
is the cumulant generating function of  ${\mathbb P}_{mt}$ and 
$\e^{-t\phi}\d {\mathbb P}_{mt}(\phi)=\d\mu_{\omega_{mt}|\omega_m}(-t\phi)$.

\newcommand{\TD}{{\hyperref[TD]{{\rm(TD)}}}}
\begin{quote}\label{TD}
{\bf Assumption (TD).} There is an open interval ${\mathbb I}\supset [0,1]$ such that,
for all $t>0$, the limit 
\[
e_t(s)=\lim _{m \rightarrow \infty } e_{mt}(s)
\]
exists and is finite for all $s \in {\mathbb I}$. W.l.o.g.\;we may assume that ${\mathbb I}$ is 
symmetric around the point $s=1/2$.
\end{quote}

\bep\label{paris-end}
Suppose that Assumption \TD{} holds. Then there exist Borel probability measures ${\mathbb P}_t$
on $\rr$ such that, for all $s$ in the complex strip ${\mathbb S}=\{s\in\cc\,|\, \Re s \in{\mathbb I}\}$, 
\[
\lim _{m\to\infty}\int\e^{-t s\phi}\d {\mathbb P}_{mt}(\phi)
=\int\e^{-ts\phi}\d{\mathbb P}_t(\phi).
\]
Moreover, ${\mathbb P}_{mt}\to{\mathbb P}_t$ and 
$\e^{-t\phi}\d {\mathbb P}_{mt}\to\e^{-t\phi}\d{\mathbb P}_t$ weakly as $m\to\infty$.
\eep

\proof Set $k_{mt}(s)=\int\e^{-st\phi}\d{\mathbb P}_{mt}(\phi)$. The functions 
$s\mapsto k_{mt}(s)$ are entire analytic and for any compact set $K\subset{\mathbb S}$ and any
$s\in K$,
\begin{align*}
|k_{mt}(s)|&\le\int_{-\infty}^0\e^{-\Re(s)t\phi}\d{\mathbb P}_{mt}(\phi)
+\int_0^{\infty}\e^{-\Re(s)t\phi}\d{\mathbb P}_{mt}(\phi)\\
&\le\int_{-\infty}^0\e^{-bt\phi}\d{\mathbb P}_{mt}(\phi)
+\int_0^{\infty}\e^{-at\phi}\d{\mathbb P}_{mt}(\phi)
\le \e^{e_{mt}(b)}+\e^{e_{mt}(a)},
\end{align*}
where $a=\min\Re(K)$ and $b=\max\Re(K)$ belong to $\mathbb I$. It follows that
\[
\sup_{m>0, s\in K} |k_{mt}(s)|<\infty.
\]
Vitali's convergence theorem  implies that the sequence $k_{mt}$ converges uniformly on 
compacts subsets of ${\mathbb S}$ to an analytic function $k_t$. Since the functions 
$\rr\ni u\mapsto k_{mt}(\i u)$ are positive definite, so is $\rr\ni u\mapsto k_t(\i u)$. 
Hence, by Bochner's theorem, there exists a Borel probability measure ${\mathbb P}_t$ on 
$\rr$ such that 
\[
k_t(\i u)=\int\e^{-\i tu \phi}\d {\mathbb P}_t(\phi).
\]
By construction, ${\mathbb P}_{mt}\rightarrow {\mathbb P}_t$ weakly. Analytic continuation yields that 
for $s\in {\mathbb S}$,
\[
\lim_{m\rightarrow\infty}k_{mt}(s)=\int\e^{-st\phi}\d {\mathbb P}_t(\phi).
\]
Replacing $s$ with $1-s$ and repeating the above argument we deduce that 
 $\e^{-t\phi}\d {\mathbb P}_{mt}(\phi)\rightarrow \e^{-t\phi}\d{\mathbb P}_t(\phi)$ weakly.
 \qed

\medskip
Note that
$$
{\mathbb I}\ni s\mapsto e_t(s)=\log\int\e^{-t(1-s)\phi }\d{\mathbb P}_t(\phi)
$$
is convex, $e_t(0)=e_t(1)=0$, $e_t(s) \leq 0$ for $\s\in ]0,1[$, $e_t(s)\geq 0$ for $s\not\in [0,1]$.
Moreover, it follows from Vitali's convergence theorem in the proof of Proposition~\ref{paris-end} 
that the derivatives of $e_{mt}(s)$ converge to the corresponding derivatives of $e_t(s)$.
In particular, Equ.~\eqref{StCyr-late} implies
\begin{equation}
e_t^\prime(1)=-\lim_{m\rightarrow \infty}\Ent(\omega_{mt}|\omega_m)
=\lim_{m\rightarrow \infty}\omega_m(\Sigma_m^t).
\label{StCyr-very-late}
\end{equation}

Suppose now that each quantum system ${\cal Q}_m$ is TRI. The symmetries 
$e_{mt}(s)=e_{mt}(1-s)$ imply the finite-time Evans-Searles symmetry
\begin{equation}
e_t(s)=e_t(1-s),
\end{equation}
which holds for $s\in {\mathbb I}$. This fluctuation relation  has the following equivalent 
reformulations:
\begin{enumerate}[{\rm (1)}]
\item For $s\in {\mathbb I}$, 
\[
e_t(s)=\log \int\e^{-t s\phi }\d {\mathbb P}_t(\phi).
\]
\item 
\[
\d {\mathbb P}_t(-\phi)=\e^{-t \phi}\d{\mathbb P}_t(\phi).
\]
\end{enumerate}

We remark that in all examples we will consider the  limiting quantum dynamical system will actually 
exist and that ${\mathbb P}_t(s)$ and $e_t(s)$ can be expressed, via modular structure, in terms 
of states and dynamics of this limiting infinitely extended system. However, a passage through the 
TD limit is necessary for the physical interpretation of the FCS of infinitely extended systems.

\subsection{Large time limit}
\label{sec-LTT}

We shall now make the

\newcommand{\LT}{{\hyperref[LT]{{\rm(LT)}}}}
\begin{quote}\label{LT}
{\bf  Assumption (LT).} In addition to Assumption \TD{}, the limit 
\[
e(s)=\lim_{t\rightarrow \infty}\frac{1}{t}e_t(s)
\]
exists and is finite for any $s\in {\mathbb I}$ . Moreover, the function  ${\mathbb I}\ni s\mapsto e(s)$
is differentiable.
\end{quote}

As an immediate consequence, we note that the limiting function $e(s)$ inherits the following basic
properties from $e_t(s)$:
\begin{enumerate}[{\rm (1)}]
\item $e(s)$ is convex on ${\mathbb I}$;
\item $e(0)=e(1)=0$;
\item $e(s)\leq 0$ for $s\in [0,1]$, and $e(s)\geq 0$ for $s\not\in [0,1]$.
\item If all the $\cQ_m$'s are TRI, then the Evans-Searles symmetry
\begin{equation}
e(s)=e(1-s)
\label{QESS}
\end{equation}
holds for $s\in{\mathbb I}$.
\end{enumerate}

Together with convexity, the Evans-Searles symmetry implies   
\[
e(1/2)=\inf_{s\in [0,1]} e(s).
\]

The asymptotic mean entropy production rate is defined as the double limit
(recall the entropy balance equation \eqref{fin-balance})
\[
\Sigma^+
=\lim_{t\rightarrow \infty}\lim_{m\rightarrow \infty}\omega_m(\Sigma_m^t)
=-\lim_{t\rightarrow \infty}\lim_{m\rightarrow \infty}\frac1t\Ent(\omega_{mt}|\omega_m).
\]
Equ. \eqref{StCyr-very-late} and convexity (see, {\sl e.g.}, Theorem 25.7 in \cite{R}) imply that
$$
\Sigma^+=e^\prime(1)=\lim_{t\to\infty}{\mathbb E}_t(\phi).
$$
Clearly, $\Sigma^+\geq0$ and under normal conditions $\Sigma^+>0$ (that is, systems out of 
equilibrium under normal conditions are entropy producing). The strict positivity of entropy production 
is a detailed (and often difficult) dynamical question that can be answered only in the context of 
concrete models.

Assumption \LT{} implies that the TD limit FCS ${\mathbb P}_t$ converges weakly to the Dirac
measure $\delta_{ \Sigma^+}$ as $t\to\infty$. The G\"artner-Ellis theorem (more precisely, 
Theorem \ref{GEoneD}) allows us to control the fluctuations of ${\mathbb P}_t$ in this limit. 
Namely, the large deviation principle
\[
\lim_{t \rightarrow \infty}\frac{1}{t}\log {\mathbb P}_t({\mathbb J})=-\inf_{\theta \in {\mathbb J}}
\varphi(\theta)
\]
holds for any open set ${\mathbb J}\subset ]\ubar \theta, \bar \theta[$ with
\begin{gather*}
\varphi(\theta) = \sup_{s\in {\mathbb I}}(-\theta s - e(s)),\\
\ubar \theta =\inf_{s \in {\mathbb I}}e^\prime(s), 
\qquad 
\bar \theta =\sup_{s\in {\mathbb I}}e^\prime(s).
\end{gather*}

Under our current assumptions the functions $s\mapsto e_t(s)$ are analytic in the strip $\mathbb S$. 
Suppose that for some $\epsilon>0$ one has
$$
\sup_{t>1,s\in\cc, |s-1|<\epsilon}\frac{1}{t}|e_t(s)|<\infty.
$$
Vitali's theorem then implies that $e(s)$ is analytic for $|s-1|<\epsilon$ and
\[
\lim_{t\to\infty}t^{k-1}\lim_{m\to\infty} C_{mt}^{(k)}=\partial_s^k e(s)|_{s=1}, 
\]
where $C_{mt}^{(k)}$ is the $k$-th cumulant of ${\mathbb P}_{mt}$. Moreover, Brick's
theorem \cite{Br} implies the Central Limit Theorem: for any Borel set ${\mathbb J}\subset\rr$,
$$
\lim_{t\to\infty}\lim_{m\to\infty}{\mathbb P}_{mt}(\phi-{\mathbb E}_{mt}(\phi)\in t^{-1/2}{\mathbb J})=
\mu_D({\mathbb J}),
$$
where $\mu_D$ denotes the centered Gaussian of variance $D=e''(1)$.

Note that if (\ref{QESS}) holds, then
\[
\varphi(-\theta)=\varphi(\theta)-\theta,\qquad
\bar \theta=-\ubar\theta,\qquad \Sigma^+=-e^\prime(0),
\qquad
D=e''(0).
\]
The material described in this and the previous section belongs to a body of structural results known as
{\sl Quantum Evans-Searles Fluctuation Theorem} \cite{JOPP2}.

\subsection{Testing the arrow of time}
\label{sec-arrow}
 
In this section, we establish a connection between the Evans-Searles Fluctuation Theorem and
hypothesis testing. To this end, we consider the problem of distinguishing the past from the future. 
More precisely, we shall apply the results of Section~\ref{sect-qht-results} to the family of pairs
$\{(\omega_{mt}, \omega_{m(-t)})\}_{t>0}$ and investigate the various error exponents associated 
with them. As in the previous section, the thermodynamic limit $m\rightarrow \infty$ has to be taken 
prior to the limit $t\rightarrow \infty$ in order to achieve significant results.
 
Define the exponents of Chernoff type  by
$$
\ubar D=\liminf_{t\rightarrow \infty}\frac{1}{2t}\liminf_{m\rightarrow \infty}
\log D(\omega_{mt},\omega_{m(-t)}),
\qquad
\bar D=\limsup_{t\rightarrow \infty}\frac{1}{2t}\limsup_{m\rightarrow\infty}
\log D(\omega_{mt},\omega_{m(-t)}),
$$
and, for $r\in \rr$, the exponents of Hoeffding type by 
\begin{align*}
\bar B(r)&=\inf_{\{T_{mt}\}}\left\{\limsup_{t \rightarrow \infty}\frac{1}{2t}
\limsup_{m\rightarrow \infty}\log \omega_{mt}(1-T_{mt})
\,\, \bigg|\,\, \limsup_{t\rightarrow \infty}\frac{1}{2t}
\limsup_{m\rightarrow\infty} \log 
\omega_{m(-t)}(T_{mt})<-r\right\},\\[3mm]
\ubar B(r)&=\inf_{\{T_{mt}\}}\left\{\liminf_{t \rightarrow \infty}\frac{1}{2t}
\liminf_{m\rightarrow \infty}\log \omega_{mt}(1-T_{mt})
\,\, \bigg|\,\, \limsup_{t\rightarrow \infty}\frac{1}{2t}\
\limsup_{m\rightarrow \infty} \log 
\omega_{m(-t)}(T_{mt})<-r\right\},\\[3mm]
B(r)&=\inf_{\{T_{mt}\}}\left\{\lim_{t \rightarrow \infty}\frac{1}{2t}
\lim_{m\rightarrow \infty}\log \omega_{mt}(1-T_{mt})
\,\, \bigg|\,\, \limsup_{t\rightarrow \infty}\frac{1}{2t}\
\limsup_{m\rightarrow \infty} \log 
\omega_{m(-t)}(T_{mt})<-r\right\},
\end{align*}
where in the last case the infimum is taken over all families of tests $\{T_{mt}\}$ for which  
\begin{equation}
\lim_{t \rightarrow \infty}\frac{1}{2t}
\lim_{m\rightarrow \infty}\log \omega_{mt}(1-T_{mt})
\label{TestLim}
\end{equation}
exists. Finally, for $\epsilon\in]0,1[$, define the exponents of Stein type as 
\begin{align*}
\bar B_\epsilon&=\inf_{\{T_{mt}\}}\left\{
\limsup_{t \rightarrow \infty}\frac{1}{2t}\limsup_{m\rightarrow \infty}\log \omega_{mt}(1-T_{mt})
\,\, \bigg|\,\,\omega_{m(-t)}(T_{mt})\leq \epsilon\right\},\\[3mm]
\ubar B_\epsilon&=\inf_{\{T_{mt}\}}\left\{
\liminf_{t \rightarrow \infty}\frac{1}{2t}\liminf_{m\rightarrow\infty}\log \omega_{mt}(1-T_{mt})
\,\, \bigg|\,\,\omega_{m(-t)}(T_{mt})\leq \epsilon\right\},\\[3mm]
B_\epsilon&=\inf_{\{T_{mt}\}}\left\{\lim_{t \rightarrow \infty}\frac{1}{2t}\log \omega_{mt}
(1-T_{mt})\,\, \bigg|\,\, \omega_{m(-t)}(T_{mt})\leq \epsilon\right\},
\end{align*}
where in the last case the infimum is taken over all families of tests $\{T_{mt}\}$ for which  
the limit \eqref{TestLim} exists. 

\bet\label{testing-open}
Suppose that Assumption \LT{} holds and that $0\in]e^\prime(0), e^\prime(1)[$. Then,
\begin{enumerate}[{\rm (1)}] 
\item
\[\ubar D=\bar D=\inf_{s\in [0,1]}e(s).\]
\item For all $r$, 
\[ \ubar B(r)= \bar B(r)=B(r)=-\sup_{0\leq s<1}\frac{-sr -e(s)}{1-s}.\]
\item For all $\epsilon \in ]0, 1[$, 
\[ \ubar B_\epsilon=\bar B_\epsilon=B_\epsilon=-\Sigma^+.\]
\end{enumerate}
\eet

{\bf\noindent Remark 1.}  These results hold under more general conditions then \LT{}. 
We have stated them in the present form for a transparent comparison with the results described
in Section~\ref{sec-LTT}.

{\bf\noindent Remark 2.} If the Evans-Searles symmetry $e(s)=e(1-s)$ holds, then  
$]e^\prime(0), e^\prime(1)[=]-\Sigma^+,\Sigma^+]$ so that the condition
$0 \in ]e^\prime(0), e^\prime(1)[$ is equivalent to
$\Sigma^+>0$. In addition, $\inf_{s\in [0,1]}e(s)=e(1/2)$ in this case.

\medskip
\proof We will again prove only (1) to indicate the strategy of the argument. The proofs of (2) and (3) 
follow by a straightforward adaption of the  general $W^\ast$-algebraic proofs described in 
Section~\ref{sec-walgebra}. 

Proposition \ref{prologue-bounds} (1) implies that for $s\in[0,1]$, 
\[
\log  D(\omega_{mt},\omega_{m(-t)})=\log D(\omega_{m2t},\omega_m)\leq e_{m2t}(s),
\]
and Assumption \LT{} yields the upper bound 
\[ 
\bar D\leq\inf_{s\in [0,1]}e(s).
\]
Let $\d\hat{\mathbb P}_{mt}(\phi)=e^{-t\phi}\d{\mathbb P}_{mt}(\phi)$ and
$\d\hat{\mathbb P}_t(\phi)=\e^{-t\phi}\d {\mathbb P}_{t}(\phi)$. By Equ.~\eqref{QPrelat} and 
Proposition \ref{prologue-bounds} (2) one has
\[
D(\omega_{mt},\omega_{m(-t)})=D(\omega_{m2t},\omega_{m})
\ge\int\frac{\d\hat{\mathbb P}_{m2t}(\phi)}{1+\e^{-2t\phi}}.
\]
By Proposition \ref{paris-end}, $\hat{\mathbb P}_{m2t}\to\hat{\mathbb P}_{2t}$ weakly
and hence
$$
\liminf_{m\to\infty}\log D(\omega_{mt},\omega_{m(-t)})\ge
\lim_{m\to\infty}\log\int\frac{\d\hat{\mathbb P}_{m2t}(\phi)}{1+\e^{-2t\phi}}=
\log\int\frac{\d\hat{\mathbb P}_{2t}(\phi)}{1+\e^{-2t\phi}}
\ge\log\left(\frac12\,\hat{\mathbb P}_{2t}(]0,\infty[)\right),
$$
where we used the Chebyshev inequality in the last step. Since 
$$
e(s)=\lim_{t\rightarrow \infty}\frac{1}{t}\log \int\e^{s t\phi}\d \hat {\mathbb P}_{t}(\phi),
$$
Assumption \LT{} and Proposition \ref{GE-THM} imply
\[
\liminf_{t\rightarrow \infty}\frac{1}{2t}
\log\hat  {\mathbb  P}_{2t}(]0, \infty[)\geq -\varphi(0)=\inf_{s\in[0,1]}e(s),
\]
and hence
\[
\ubar D\geq \inf_{s\in[0,1]}e(s).
\]
\qed

The results described in this section shed some light on the relation between hypothesis testing and 
non-equilibrium statistical mechanics. If the systems ${\cal Q}_m$ converge, as $m\rightarrow \infty$,  
to a limiting infinitely extended open quantum system, then this relation can be elaborated further, 
see Section~\ref{sec-hypo} for details. 

\section{Algebraic preliminaries}
\label{sec-algebrapre}
\subsection{Notation}

Let $\fM$ be a $W^\ast$-algebra with unit $1$, dual $\fM^\ast$, and predual
$\fM_\ast\subset \fM^\ast$. The elements of $\fM_\ast$ are called normal functionals on $\fM$.
Hermitian, positive and faithful functionals as well as states on $\fM$ are defined as in the finite 
dimensional case (Section \ref{sec-setup}). $\fM_\ast^+$ denotes the set of positive elements of 
$\fM_\ast$. For $\nu,\omega\in\fM_\ast$, we write $\nu\ge\omega$ whenever 
$\nu-\omega\in\fM_\ast^+$.

An element $P\in \fM$ is called a projection if $P^2=P=P^\ast$. 
If $P_1, P_2$ are projections, then $P_1\leq P_2$ stands for $P_1P_2=P_1$ and 
$P_1\perp P_2$ for $P_1P_2=0$. The support of $\omega\in \fM_\ast^+$ is the projection 
${\rm s}_\omega$ defined by 
\[
\s_\omega=\inf\{ P\in \fM\,|\, \hbox{$P$ is a projection and $\omega(1 -P)=0$}\}.
\]
For any $A\in \fM$ and $P\geq \s_\omega$,  $\omega(A)=\omega(AP)=\omega(PA)$ and 
$$
\|\omega\|=\sup_{A\in\fM,\|A\|=1}|\omega(A)|=\omega(1)=\omega(\s_\omega).
$$
$\omega$ is faithful iff $\s_\omega=1$.

Any hermitian $\omega\in\fM_\ast$ has a unique Jordan decomposition $\omega=\omega_+-\omega_-$
where $\omega_{\pm}\in \fM_\ast^+$ and $\s_{\omega_+}\perp \s_{\omega_{-}}$.  
In particular, $\|\omega\|=\|\omega_+\| +\|\omega_-\|=\omega_+(1) +\omega_-(1)$ (see \cite{Tak}).

If $\nu, \omega \in \fM_\ast^+$  and  $\s_\nu\leq \s_\omega$ we shall say that $\nu$ is normal 
w.r.t.\;$\omega$, denoted $\nu\ll\omega$. If $\s_\nu=\s_\omega$ then $\nu$ and $\omega$ 
are called equivalent, denoted $\nu\sim\omega$.

Let $\cH$ be a Hilbert space and $\cB(\cH)$ the $C^\ast$-algebra of all bounded linear operators 
on $\cH$. If $\cA\subset\cB(\cH)$, then $\cA^\ast=\{A^\ast\,|\,A\in\cA\}$ is the adjoint of $\cA$ and
$\cA^\prime=\{B\in\cB(\cH)\,|\,BA=AB\text{ for all }A\in\cA\}$ denotes its commutant.
$\fM\subset \cB(\cH)$ is a $W^\ast$-algebra iff $\fM=\cA^{\prime\prime}$ for some 
$\cA=\cA^\ast\subset \cB(\cH)$. Such a $W^\ast$-algebra is called von Neumann algebra.
  
\subsection{Modular structure}
A $W^\ast$-algebra in standard form is a quadruple $(\fM,\cH,J,\cH^+)$ where $\cH$ is a Hilbert 
space, $\fM\subset \cB(\cH)$ is a $W^\ast$-algebra, $J$ is an anti-unitary involution on $\cH$ and 
$\cH^+$ is a cone in $\cH$ such that:
\begin{enumerate}[{\rm (1)}]
\item $\cH^+$ is self-dual, {\sl i.e.,} $\cH^+=\{\Psi\in\cH\,|\,(\Phi|\Psi)\ge0\text{ for all }\Phi\in\cH^+\}$;
\item $J\fM J=\fM^\prime$;
\item $JAJ =A^\ast $ for $A\in \fM\cap \fM^\prime$;
\item $J\Psi=\Psi$ for $\Psi\in \cH^+$;
\item $AJA\cH^+\subset \cH^+$ for $A\in \fM$.
\end{enumerate}
The quadruple $(\pi,\cH,J,\cH^+)$ is a standard representation of the $W^\ast$-algebra $\fM$ if
$\pi: \fM\rightarrow \cB(\cH)$ is a faithful representation and $(\pi(\fM),\cH,J,\cH^+)$ is in standard
form. A standard representation always exists. Moreover, if $(\pi_1,\cH_1,J_1,\cH_1^+)$ and 
$(\pi_2,\cH_2,J_2,\cH_2^+)$ are two standard representations of $\fM$ then there exists a unique 
unitary operator $U:\cH_1\to\cH_2$ such that $U\pi_1(A)U^\ast =\pi_2(A)$ for all 
$A\in \fM$, $UJ_1U^\ast=J_2$, and $U\cH_1^+=\cH_2^+$.

In what follows, without loss of generality, we will assume that all $W^\ast$-algebras are in standard 
form. For later reference we recall the following classical result (see, {\sl e.g.}, \cite{St}).
\bet
Let $\fM$ be a $W^\ast$-algebra in standard form. For any $\omega \in \fM_\ast^+$ 
there exists a unique $\Omega_\omega\in\cH^+$ such that
\[
\omega(A)=(\Omega_\omega|A\Omega_\omega),
\]
for all $A\in \fM$. The map $\fM_\ast^+\ni \omega\mapsto\Omega_\omega\in \cH^+$ is a bijection
and 
\begin{equation}
\|\Omega_\omega-\Omega_\nu\|^2
\leq \|\omega-\nu\|
\leq\|\Omega_\omega-\Omega_\nu\|\,\|\Omega_\omega +\Omega_\nu\|.
\label{araki-est}
\end{equation}
\eet 
{\bf\noindent Remark 1.} The upper bound in (\ref{araki-est}) is trivial to prove and the interesting part 
is the lower bound. At the end of Section~\ref{sec-q-prelim} we will give a new simple proof of this
lower bound.

{\bf\noindent Remark 2.} The support ${\rm s}_\omega$ projects on the closure of 
$\fM'\Omega_\omega$. It follows that ${\rm s}_\omega'=J\s_\omega J\in\fM'$ is the
orthogonal projection on the closure of $\fM\Omega_\omega$.

We now recall the definition of Araki's relative modular operator \cite{Ar1, Ar2}.  For 
$\nu, \omega \in {\fM}_\ast^+$ define $S_{\nu|\omega}$ on the domain 
$\fM\Omega_\omega + (\fM\Omega_\omega)^\perp$ by 
\[
S_{\nu|\omega}(A\Omega_\omega +\Theta)=\s_\omega A^\ast\Omega_\nu,
\]
where $\Theta \in (\fM\Omega_\omega)^\perp$. $S_{\nu|\omega}$ is a densely defined anti-linear 
operator. It is closable and we denote its closure by the same symbol. The positive operator
\[
\Delta_{\nu|\omega}=S_{\nu|\omega}^\ast S_{\nu|\omega}
\]
is called {\sl relative modular operator.} We denote $\Delta_{\omega}=\Delta_{\omega|\omega}$.
The  basic properties of the relative modular operator are (see \cite{Ar1, Ar2})
\bep\label{araki-prop-mod}
\begin{enumerate}[{\rm (1)}] 
\item $\Delta_{\lambda_1\nu|\lambda_2\omega}=\frac{\lambda_1}{\lambda_2}\Delta_{\nu|\omega}$,
for any $\lambda_1, \lambda_2 >0$. 
\item $\Ker \Delta_{\nu|\omega}=\Ker \s_\omega'\s_\nu$.
\item $S_{\nu|\omega}=J\Delta_{\nu|\omega}^{1/2}$ is the polar decomposition of $S_{\nu|\omega}$.
\item $\Delta_{\nu|\omega}^{1/2}\Omega_\omega
=\Delta_{\nu|\omega}^{1/2}\s_\nu\Omega_\omega=\s_\omega'\Omega_\nu$.
\item $J\Delta_{\omega|\nu}J\Delta_{\nu|\omega}
=\Delta_{\nu|\omega}J \Delta_{\omega|\nu}J =\s_\omega'\s_\nu$. 
\end{enumerate}
\eep

We note in particular that if $\nu$ and $\omega$ are faithful, then $\Delta_{\nu|\omega}>0$,
$\Delta_{\nu|\omega}^{-1}=J\Delta_{\omega|\nu}J$ and
$\Delta_{\nu|\omega}^{1/2}\Omega_\omega=\Omega_\nu$.

\subsection{Relative entropies}

For $\nu,\omega\in\fM_\ast^+$ we shall denote by $\mu_{\omega|\nu}$ the spectral measure 
for $-\log\Delta_{\omega|\nu}$ and $\Omega_\nu$. The relative entropy of $\nu$ w.r.t.\;$\omega$ 
is defined by 
\begin{equation}
{\rm Ent}(\nu|\omega)=
\begin{cases}
\displaystyle(\Omega_\nu|\log\Delta_{\omega|\nu}\Omega_\nu)
=-\int x\,\d\mu_{\omega|\nu}(x)& \text{if $\nu\ll\omega$},\\[7pt]
-\infty&\text{otherwise.}
\end{cases}
\label{quantum-rel-ent}
\end{equation}

\bep\label{prop-relative-q}
Suppose that $\nu \ll \omega$.
\begin{enumerate}[{\rm (1)}] 
\item For  $\lambda_1, \lambda_2>0$,
\[
\Ent(\lambda_1\nu|\lambda_2\omega)=\lambda_1\left(\Ent(\nu|\omega)
+\nu(1)\log \frac{\lambda_2}{\lambda_1}\right).
\]
\item 
\[
\Ent(\nu|\omega)\leq \nu(1)\log\frac{\omega(\s_\nu)}{\nu(1)}.
\]
In particular, if $\omega(\s_\nu)\leq \nu(1)$, then $\Ent(\nu|\omega)\leq 0$. 
\item 
If $\mathfrak{N}\subset\fM$ is a $W^\ast$-subalgebra then 
\[
\Ent(\nu|\omega)\leq \Ent (\nu_{|\mathfrak{N}}|\omega_{|\mathfrak{N}}).
\]
\end{enumerate}
\eep
Part (1) follows from Proposition \ref{araki-prop-mod} (1). The proof of (2) in the special case
$\omega(\s_\nu)=\nu(1)$ follows from the inequality $\log x\le x-1$. The general case is obtained by applying Part (1).
Part (3) is the celebrated monotonicity of the relative entropy \cite{LiR, Uh}.
Modern proofs of (3) can be found in \cite{OP} or  \cite{DJP}.

Let $\nu,\omega \in \fM_\ast^+$. R\'enyi's relative entropy of $\nu$ w.r.t.\;$\omega$ is defined by 
\[
{\rm Ent}_s(\nu|\omega)
= \log (\Omega_\omega|\Delta_{\nu|\omega}^s\Omega_\omega)
=\log\int\e^{-sx}\d\mu_{\nu|\omega}(x).
\]
We list below some of  its properties. Their proof is simple and can be found in \cite{JOPP2}.
\bep\label{prop-renyi}
\begin{enumerate}[{\rm (1)}] 
\item If $\nu\ll\omega$ then the function $\rr\ni s\mapsto\Ent_s(\nu|\omega)\in]-\infty, \infty]$
is convex. It is finite and continuous on $[0,1]$ and real analytic on $]0,1[$.

In the remaining statements we assume that $\nu\sim\omega$.
\item $\Ent_s(\nu|\omega)=\Ent_{1-s}(\omega|\nu)$.
\item $\Ent_s(\nu|\omega)\geq s\,\Ent(\omega|\nu)$.
\item 
${\displaystyle
\lim_{s \uparrow 1}\frac{1}{s-1}\Ent_s(\nu|\omega)
=-\Ent(\nu|\omega).
}$
\end{enumerate}
\eep

We finish with some estimates that will be used in the proof of the quantum Chernoff bound.
\bep\label{araki-multi}
\begin{enumerate}[{\rm (1)}]
\item Let $\nu_1, \nu_2, \omega\in \fM_\ast^+$ and $s\in[0,1]$. If $\nu_2\leq \nu_1$, then
$\Dom(\Delta_{\nu_1|\omega}^{s/2})\subset\Dom(\Delta_{\nu_2|\omega}^{s/2})$ and
\[
\|\Delta_{\nu_2|\omega}^{s/2}\Psi\|\leq\|\Delta_{\nu_1|\omega}^{s/2}\Psi\|, 
\]
for any $\Psi\in{\rm Dom}(\Delta_{\nu_1|\omega}^{s/2})$.
\item Let $\nu,\omega_1,\omega_2\in \fM_\ast^+$ and $s\in[0,1]$. If $\omega_2\leq\omega_1$, then
$$
\|\Delta_{\nu|\omega_2}^{s/2} A\Omega_{\omega_2}\|
\leq\|\Delta_{\nu|\omega_1}^{s/2} A\Omega_{\omega_1}\|, 
$$
for any $A\in\fM$.
\item Let $\nu_1,\nu_2,\omega_1,\omega_2\in \fM_\ast^+$ and $s\in[0,1]$. If $\nu_2\leq \nu_1$ 
and $\omega_2\leq\omega_1$, then 
\[
\|\Delta_{\nu_2|\omega_2}^{s/2}A\Omega_{\omega_2}\|
\leq\|\Delta_{\nu_1|\omega_1}^{s/2}A\Omega_{\omega_1}\|,
\]
for any $A\in\fM$.
\item Let $\nu_1, \nu_2, \omega_1, \omega_2\in \fM_\ast^+$ be faithful and $s\in[0,1]$. 
If $\nu_2\leq \nu_1$ and $\omega_2\leq  \omega_1$, then 
 \begin{equation}
(\Omega_{\omega_2}|(\Delta_{\nu_1|\omega_2}^s
-\Delta_{\nu_2|\omega_2}^s)\Omega_{\omega_2})
\leq(\Omega_{\omega_1}|(\Delta_{\nu_1|\omega_1}^s
-\Delta_{\nu_2|\omega_1}^s)\Omega_{\omega_1}).
\label{ogata-in}
\end{equation}
\end{enumerate}
\eep

{\bf\noindent Remark 1.} Part (4) was recently established  in \cite{Og3} with a proof based on Connes 
Radon-Nikodym cocycles. We will give below an alternative proof which emphasizes the connection of 
the above estimate with Araki-Masuda's theory of non-commutative $L^p$-spaces \cite{AM}.

{\bf\noindent Remark 2.} By an additional approximation argument the assumption that
$\nu_i,\omega_i$ are faithful can be removed, see \cite{Og3} for details.

{\bf\noindent Remark 3.} In the special case $s=1/2$,  Part (4) follows easily from 
well-known properties of the  cone $\cH^+$. Indeed, for $s=1/2$, the inequality (\ref{ogata-in}) is 
equivalent to the inequality 
\begin{equation}
(\Omega_{\omega_1}-\Omega_{\omega_2}|\Omega_{\nu_1}-\Omega_{\nu_2})\geq 0.
\label{ogata-in1}
\end{equation}
By the ordering property of the cone $\cH^+$, 
$\Omega_{\omega_1}-\Omega_{\omega_2}, \Omega_{\nu_1}-\Omega_{\nu_2}\in \cH^+$ 
and, since $\cH^+$ is a self-dual cone, (\ref{ogata-in1}) follows. 

The rest of this section is devoted to the proof of Proposition \ref{araki-multi}. We start with some 
preliminaries. 
\bel\label{Araki}
Let $\phi,\nu,\omega\in\fM_\ast^+$ and suppose that $\nu$ and $\omega$ are faithful. 
Then
\[
J \Delta_{\nu|\omega}^{\frac{1}{2}-s} A \Delta_{\phi|\omega}^s\Omega_\omega
=\Delta_{\phi|\nu}^s A^\ast \Omega_\nu
\]
holds for $s\in[0,1/2]$ and $A\in \fM$.
\eel
\proof
We just sketch the well-known argument (see, {\sl e.g.}, \cite{AM}). We recall that for 
$\phi_1,\phi_2\in\fM_\ast^+$ the Araki-Connes cocycle is defined by 
\[
[D\phi_1: D\phi_2]_t =\Delta_{\phi_1|\phi_2}^{ \i t}\Delta_{\phi_2}^{-\i t}.
\]
For a detailed discussion of this class of operators we refer the reader to Appendices  B and C of 
\cite{AM}. We only need that the cocycles are elements of $\fM$ and that the chain rule
\[
[D\phi_1: D\phi_2]_{t}[D\phi_2: D\phi_3]_t=[D\phi_1: D\phi_3]_t
\] 
holds whenever  $\s_{\phi_2}\geq \s_{\phi_1}$ or $\s_{\phi_2}\geq \s_{\phi_3}$. For $t$ real, one has
\[
\begin{split}
J\Delta_{\nu|\omega}^{\frac{1}{2}-\i t}A\Delta_{\phi|\omega}^{\i t}\Omega_\omega
&=\Delta_{\omega|\nu}^{-\i t}J\Delta_{\nu|\omega}^{\frac{1}{2}}A[D\phi:D\omega]_t\Omega_\omega\\[3mm]
&=\Delta_{\omega|\nu}^{-\i t}[D\phi:D\omega]_t^\ast A^\ast\Omega_\nu\\[3mm]
&=\Delta_{\nu}^{-\i t}[D\omega:D\nu]_t^\ast[D\phi:D\omega]_t^\ast A^\ast \Omega_\nu\\[3mm]
&=\Delta_{\nu}^{-\i t}([D\phi:D\omega]_t[D\omega:D\nu]_t)^\ast A^\ast \Omega_\nu\\[3mm]
&=\Delta_{\nu}^{-\i t}[D\phi:D\nu]_t^\ast A^\ast \Omega_\nu\\[3mm]
&= \Delta_{\phi|\nu}^{-\i t} A^\ast \Omega_\nu.
\end{split}
\]
For $B\in\fM$, the  functions
$$
f(z)=(A\Delta_{\phi|\omega}^{\bar z}\Omega_\omega|
\Delta_{\nu|\omega}^{\frac12-z}B\Omega_\omega),
\qquad
g(z)=(JB\Omega_\omega|\Delta_{\phi|\nu}^zA^\ast\Omega_\nu)
$$
are analytic in the open strip $0<\Re z<\frac{1}{2}$ and bounded and  continuous 
on its closure. Since, for $t$ real,
$$
f(-\i t)=(JB\Omega_\omega|
J\Delta_{\nu|\omega}^{\frac{1}{2}-\i t}A\Delta_{\phi|\omega}^{\i t}\Omega_\omega),
\qquad
g(-\i t)=(JB\Omega_\omega|\Delta_{\phi|\nu}^{-\i t}A^\ast\Omega_\nu),
$$
one concludes from the above calculation that $f(z)=g(z)$ holds in the
closed strip $0\le\Re z\le\frac{1}{2}$. In particular
$$
(A\Delta_{\phi|\omega}^{s}\Omega_\omega|
\Delta_{\nu|\omega}^{\frac12-s}B\Omega_\omega)
=(JB\Omega_\omega|\Delta_{\phi|\nu}^sA^\ast\Omega_\nu),
$$
for $s\in[0,1/2]$. Since $\fM\Omega_\omega$ is a core for $\Delta_{\nu|\omega}^{\frac12-s}$,
we conclude that $A\Delta_{\phi|\omega}^{s}\Omega_\omega$ is in the domain of
$\Delta_{\nu|\omega}^{\frac12-s}$ and that
$$
(JB\Omega_\omega|J\Delta_{\nu|\omega}^{\frac12-s}A\Delta_{\phi|\omega}^{s}\Omega_\omega)=
(\Delta_{\nu|\omega}^{\frac12-s}A\Delta_{\phi|\omega}^{s}\Omega_\omega|
B\Omega_\omega)
=(JB\Omega_\omega|\Delta_{\phi|\nu}^sA^\ast\Omega_\nu),
$$
from which the result follows.
\qed

\bel\label{Araki2}
Let $\nu, \omega \in \fM_\ast^+$ and suppose that $\omega$ is faithful. Let $U\in \fM$ be a partial
isometry such that $U^\ast U=\s_\nu$ and let $\nu_U(\,\cdot\,)=\nu(U^\ast\cdot\,U)$. Then 
\[
\Delta_{\nu_U|\omega}=U\Delta_{\nu|\omega}U^\ast.
\]
\eel 
\proof The vector representative of $\nu_U$ in $\cH^+$ is $UJUJ\Omega_\nu$. Hence, for $ A\in \fM$, 
\[
J\Delta_{\nu_U|\omega}^{\frac{1}{2}}A\Omega_\omega= A^\ast U JUJ\Omega_\nu
=JUJ A^\ast U\Omega_\nu = JU \Delta_{\nu|\omega}^{\frac{1}{2}}U^\ast A\Omega_\omega.
\]
 \qed

\medskip
{\bf Proof of Proposition \ref{araki-multi}.} Part (1) is well known (see Lemma C.3 in \cite{AM}).
To prove Part (2), we use Proposition \ref{araki-prop-mod} (2) and (5) to obtain
$$
\Delta_{\nu|\omega_i}^{s/2}A\Omega_{\omega_i}
=\Delta_{\nu|\omega_i}^{s/2}\s_\nu A\Omega_{\omega_i}
=\Delta_{\nu|\omega_i}^{s/2}J\Delta_{\omega_i|\nu}^{1/2}A^\ast \Omega_\nu
=J\Delta_{\omega_i|\nu}^{1/2-s/2}A^\ast \Omega_\nu,
$$
and the statement follows from Part (1). One proves Part (3) by successive applications
of  Parts (1) and (2).

To prove part (4), we first notice that it suffices to prove the statement for $s\in[0,1/2]$.
Indeed, if this case is established,  then the case $s\in[1/2,1]$ follows by exchanging the roles of 
the $\nu$ and $\omega$ and using  the identities 
\[ 
\|\Delta_{\omega_i|\nu_j}^{s/2}\Omega_{\nu_j}\|
=\|\Delta_{\nu_j|\omega_i}^{(1-s)/2}\Omega_{\omega_i}\|.
\]
One can further assume that $s\in]0,1/2[$ (the  case $s=0$ is trivial  and the case $s=1/2$ 
follows from $s<1/2$ by continuity).

Consider the Araki-Masuda space $L^p(\fM, \omega_i)$
(constructed w.r.t.\;the reference vector $\Omega_{\omega_i}$) for $p=1/s\in]2,\infty[$. Let 
\[
\cH^{\alpha}_{\Omega_{\omega_i}}=\{ \Delta_{\omega_i}^\alpha \fM_+\Omega_{\omega_i}\}^{\rm cl}
\]
be the usual $\alpha$-cone in $\cH$ (note that $\cH_{\Omega_{\omega_i}}^{1/4}=\cH^+)$. The 
positive part of $L^p(\fM, \omega_i)$ is defined by 
\[
L^p_+(\fM, \omega_i) =L^p(\fM, \omega_i)\cap\cH^{1/2p}_{\Omega_{\omega_i}},
\] 
and by the Lemma 4.3 in \cite{AM}, 
\[
L^p_+(\fM, \omega_i)=\{\Delta_{\nu|\omega_i}^{1/p}\Omega_{\omega_i}\,|\,\nu\in\fM_\ast^+\}.
\]
The polar decomposition in $L^p(\fM,\omega_i)$ (Theorem 3 in \cite{AM}) implies there exist unique
$\phi_i\in \fM_\ast^+$ and unique partial isometries $U_i\in\fM$ satisfying $U_i^\ast U_i =\s_{\phi_i}$ 
such that 
\begin{equation}
\Delta_{\nu_1|\omega_i}^s\Omega_{\omega_i}-\Delta_{\nu_2|\omega_i}^s\Omega_{\omega_i}
=U_i\Delta_{\phi_i|\omega_i}^s \Omega_{\omega_i}.
\label{rela}
\end{equation}
After applying $J\Delta_{\omega_i}^{1/2-s}$ to both sides of this identity we deduce from 
Lemmata~\ref{Araki} and \ref{Araki2} that
\[
\Delta_{\nu_1|\omega_i}^s \Omega_{\omega_i}- \Delta_{\nu_2|\omega_i}^s\Omega_{\omega_i}
=U_i^\ast\Delta_{\phi_{i\,U_i}|\omega_i}^s \Omega_{\omega_i}.
\]
By the uniqueness of the polar decomposition, $U_i^\ast=U_i$ and $\phi_{i\,U_i}=\phi_i$. It follows that 
$U_i$ is a self-adjoint partial isometry and so  its spectrum is contained in the set $\{-1,0, 1\}$. 
Hence,  $U_i= P_{i}^+-P_{i}^-$ and $\s_{\phi_i}=P_i^++P_i^-$, where $P_{i}^\pm$ are the spectral 
projections corresponding to the eigenvalues $\pm1$. It then follows from (\ref{rela}) that
\[
P_{i}^-\Delta_{\nu_1|\omega_i}^s\Omega_{\omega_i}
-P_{i}^-\Delta_{\nu_2|\omega_i}^s\Omega_{\omega_i}
=-P_{i}^-\Delta_{\phi_i|\omega_i}^s \Omega_{\omega_i}.
\]
Again applying $J\Delta_{\omega_i}^{1/2-s}$ to both sides we get
\[
\Delta_{\nu_1|\omega_i}^sP_{i}^- \Omega_{\omega_i}
-\Delta_{\nu_2|\omega_i}^sP_{i}^-\Omega_{\omega_i}
=-\Delta_{\phi_i|\omega_i}^s P_{i}^-\Omega_{\omega_i},
\]
and so 
\[
\|\Delta_{\nu_1|\omega_i}^{s/2}P_{i}^- \Omega_{\omega_i}\|^2
-\|\Delta_{\nu_2|\omega_i}^{s/2}P_{i}^- \Omega_{\omega_i}\|^2=
-\|\Delta_{\phi_i|\omega_i}^{s/2} P_{i}^-\Omega_{\omega_i}\|^2\leq 0.
\]
Since $\nu_1\geq \nu_2$, Proposition \ref{araki-multi} (1)  implies that the left hand side is positive and
we deduce
\[
\Delta_{\phi_i|\omega_i}^{s/2} P_{i}^-\Omega_{\omega_i}=0.
\]
Since $J\Delta_{\omega_i}^{1/2-s}P_{i}^-\Delta_{\phi_{i}|\omega_i}^s\Omega_{\omega_i}
=\Delta_{\phi_i|\omega_i}^s P_{i}^-\Omega_{\omega_i}=0$, we obtain
$P_{i}^-\Delta_{\phi_{i}|\omega_i}^s\Omega_{\omega_i}=0$ and therefore
$$
\Delta_{\phi_{i}|\omega_i}^s\Omega_{\omega_i}
=\s_{\phi_i}\Delta_{\phi_{i}|\omega_i}^s\Omega_{\omega_i}
=P_i^+\Delta_{\phi_{i}|\omega_i}^s\Omega_{\omega_i}
=U_i\Delta_{\phi_{i}|\omega_i}^s\Omega_{\omega_i}.
$$ 
Hence, Equ.~\eqref{rela} becomes
\begin{equation}
\Delta_{\nu_1|\omega_i}^s \Omega_{\omega_i}- \Delta_{\nu_2|\omega_i}^s\Omega_{\omega_i}
=\Delta_{\phi_i|\omega_i}^s \Omega_{\omega_i}.
\label{relatwo}
\end{equation}
Acting on both sides of the above relation with $J\Delta_{\omega_1|\omega_2}^{1/2-s}$
and applying Lemma \ref{Araki} we get
$$
\Delta_{\nu_1|\omega_1}^s \Omega_{\omega_1}- \Delta_{\nu_2|\omega_1}^s\Omega_{\omega_1}
=\Delta_{\phi_2|\omega_1}^s \Omega_{\omega_1}.
$$
Comparing the last relation with the Equ.~\eqref{relatwo} for $i=1$ yields
$$
\Delta_{\phi_1|\omega_1}^s \Omega_{\omega_1}=\Delta_{\phi_2|\omega_1}^s \Omega_{\omega_1},
$$
and so, by the uniqueness of the polar decomposition,  $\phi_1=\phi_2=\phi$. Finally,
Proposition \ref{araki-multi} (2) allows us to conclude
\[
(\Omega_{\omega_2}|(\Delta_{\nu_1|\omega_2}^s
-\Delta_{\nu_2|\omega_2}^s)\Omega_{\omega_2})
=(\Omega_{\omega_2}|\Delta_{\phi|\omega_2}^s \Omega_{\omega_2})
\leq (\Omega_{\omega_1}|\Delta_{\phi|\omega_1}^s \Omega_{\omega_1})
=(\Omega_{\omega_1}|(\Delta_{\nu_1|\omega_1}^s -\Delta_{\nu_2|\omega_1}^s)\Omega_{\omega_1}).
\]
\qed

For additional information about quantum relative entropies we refer the reader to \cite{OP}.

\subsection{Classical systems}
\label{sec-classical}
Let $(M, {\cal F}, P)$ be a probability space, where $M$ is a set, ${\cal F}$ a $\sigma$-algebra in $M$,
and $P$ a probability measure on $(M, {\cal F})$. The classical probabilistic setup fits into the 
algebraic framework as follows. Let ${\cal M}$ be the vector space of all complex measures on 
$(M, {\cal F})$ which are absolutely continuous w.r.t.\;$P$. If $\|\nu\|$ is the total variation of 
$\nu \in {\cal M}$, then $\|\cdot\|$ is a norm on ${\cal M}$ and ${\cal M}$  is a Banach space 
isomorphic to $L^1(M, \d P)$. $\fM=L^\infty(M, \d P)={\cal M}^\ast$, and the standard representation 
$(\pi, \cH, J, \cH^+)$ is identified as $\cH= L^2(M, \d P)$, $\pi(f) g = fg$, $J(f)=\bar f$, and 
$\cH^+ =\{ f\in \cH\,|\, f\geq 0\}$. $\fM_\ast={\cal M}$ ($\nu(f)=\int f\d\nu$), $\fM_\ast^+$ consists 
of positive measures in ${\cal M}$ and for $\omega \in \fM_\ast^+$, 
\[
\Omega_\omega = \left(\frac{\d\omega}{\d P}\right)^{1/2}.
\]
Let $\nu, \omega\in \fM_\ast^+$ and denote by $\nu=\nu_{\omega,{\rm ac}}+\nu_{\omega,{\rm sing}}$, 
with $\nu_{\omega,{\rm ac}}\ll\omega$ and $\nu_{\omega,{\rm sing}}\perp \omega$, the  Lebesgue 
decomposition of $\nu$ w.r.t.\;$\omega$. Then 
\[
\Delta_{\nu|\omega}(f)= \frac{\d \nu_{\omega,{\rm ac}}}{\d\omega} f.
\]
In particular, $\Delta_{\omega|\omega}=1$ and $\log \Delta_{\omega|\omega}=0$. 
If $\nu\ll \omega$ then the relative entropy and the R\'enyi relative entropy take  the familiar forms
\[
\Ent(\nu|\omega)=-\int_M \log \frac{\d\nu}{\d\omega}\d\nu
=-\int_M \frac{\d\nu}{\d\omega}\log \frac{\d\nu}{\d\omega}\d\omega,
\qquad
\Ent_s(\nu|\omega)=\log \int_M \left(\frac{\d\nu}{\d\omega}\right)^s \d\omega.
\]

\subsection{Finite systems}
\label{sec-finitesys}
Let $\cK$ be a finite dimensional Hilbert space and $\fM=\cO_\cK$, the $\ast$-algebra of all linear 
operators on $\cK$ equipped with the usual operator norm. $\fM^\ast=\fM_\ast$ is identified with 
$\cO_\cK$ equipped with the trace norm $\|\omega\|=\Tr |\omega|$, and $\fM_\ast^+$ identified 
with positive linear operators on $\cK$. 

The standard representation of $\fM$ is constructed as follows. $\cH=\cO_\cK$ equipped with the 
inner product $(A|B)={\rm Tr}\,A^\ast B$, $\pi(A)B=AB$,  $J(A)=A^\ast$, and
$\cH^+=\{A\in O_\cK\,|\, A\geq 0\}$. If $\nu,\omega\in\fM_\ast^+$ then 
$\Omega_\omega=\omega^{1/2}$ and 
\[
\Delta_{\nu|\omega}(A)=\nu A \omega^{-1}{\rm s}_\omega.
\] 
If in addition $\nu \ll \omega$ we recover the formulas 
\[ \Ent_s(\nu|\omega)= \log \Tr \,\nu^s \omega^{1-s}\s_\omega,
\qquad
\Ent(\nu|\omega)=\Tr \,\nu(\log \omega -\log \nu).
\]
 
 \section{Hypothesis testing in $W^\ast$-algebras}
 \label{sec-walgebra}
 
\subsection{Preliminaries}
\label{sec-q-prelim}

Let  $\cQ$ be a quantum system described by  a $W^\ast$-algebra $\fM$ in standard form 
on a Hilbert space $\cH$. Let $(\nu, \omega)$  be a pair of faithful normal states on  $\fM$ 
and suppose that one of the following two  hypotheses holds:

\begin{quote} Hypothesis I :  ${\cal Q}$ is in the state $\omega$; 
\end{quote} 
\begin{quote} Hypothesis II  :  ${\cal Q}$ is in the state $\nu$. 
\end{quote} 
A priori, Hypothesis I is realized with probability $p\in]0,1[$ and hypothesis II with probability $1-p$.
A {\em test} is a projection $T\in\fM$ and the result of a measurement of the corresponding 
observable is a number in $\sp(T)=\{0,1\}$. If the outcome is $1$, one accepts hypothesis I, 
otherwise one  accepts hypothesis II. The total error probability is
$$
(1-p)\nu(T) + p\,\omega(1 -T).
$$
We again  absorb the scalar factors into the functionals  and  consider the quantities 
\[
D(\nu, \omega, T)=\nu(T) +  \omega (1 - T)
\]
and 
\[
D (\nu, \omega)=\inf _{T}D(\nu,\omega,T)
\]
for a  pair of faithful normal functionals $\nu,\omega\in\fM_\ast^+$.

\bet\label{q-NP-Lemma}
\begin{enumerate}[{\rm (1)}]
\item
\[
D(\nu, \omega)=D(\nu, \omega, \s_{(\omega-\nu)_+})=\frac{1}{2}(\omega(1) +\nu(1)- \|\omega-\nu\|).
\]
\item  For $s\in[0,1]$, 
\begin{equation}
D(\nu, \omega)\leq(\Omega_\omega|\Delta_{\nu|\omega}^s\Omega_\omega)
=\int\e^{-sx}\d\mu_{\nu|\omega}(x).
\label{Cheb1}
\end{equation}
\item 
\begin{equation}
D(\nu, \omega)\geq 
(\Omega_\omega|\Delta_{\nu|\omega}(1+\Delta_{\nu|\omega})^{-1}\Omega_\omega)
=\int\frac{\d\mu_{\nu|\omega}(x)}{1+\e^x}.
\label{genubnd}
\end{equation}
\end{enumerate}
\eet

\proof  (1) For any test $T$, one has
\[
\begin{split}
\nu(T) + \omega (1-T)
&=\omega(1) -(\omega-\nu)(T)\geq \omega(1) -(\omega-\nu)_+(T)\\[3pt]
&\geq \omega(1)-(\omega-\nu)_+(1)=\frac{1}{2}(\omega(1) +\nu(1)- \|\omega-\nu\|).
\end{split}
\]
On the other hand, 
\[
D(\nu, \omega, \s_{(\omega-\nu)_+})=\omega(1)-(\omega-\nu)_+(1)
=\frac{1}{2}(\omega(1) +\nu(1) - \|\omega-\nu\|),
\]
and (1) follows. 

\noindent (2) Writing the Jordan decomposition of $\phi=\omega-\nu$ as $\phi=\phi_+-\phi_-$,
and using
\[
2\phi_+(1)=\|\phi\| +\phi(1),
\]
we rewrite Inequality~\eqref{Cheb1} in the equivalent form
\begin{equation}
(\Omega_\omega|(\Delta_{\omega|\omega}^s-\Delta_{\nu|\omega}^s)\Omega_\omega)\leq\phi_+(1).
\label{st-so1}
\end{equation}
Now, since $\omega=\nu+\phi\leq\nu+\phi_+$ and $\nu\le\nu+\phi_+$, applying successively
Proposition \ref{araki-multi} (1), (4) and (2) yields
\begin{align*}
(\Omega_\omega|(\Delta_{\omega|\omega}^s-\Delta_{\nu|\omega}^s)\Omega_\omega)
&\leq (\Omega_\omega|(\Delta_{\nu+\phi_+|\omega}^s-\Delta_{\nu|\omega}^s)\Omega_\omega)\\
&\leq(\Omega_{\nu+\phi_+}|(\Delta_{\nu+\phi_+|\nu+\phi_+}^s
-\Delta_{\nu|\nu+\phi_+}^s)\Omega_{\nu+\phi_+})\\
&\leq(\Omega_{\nu+\phi_+}|\Delta_{\nu+\phi_+|\nu+\phi_+}^s\Omega_{\nu+\phi_+})
-(\Omega_{\nu}|\Delta_{\nu|\nu}^s\Omega_{\nu})\\
&=(\nu+\phi_+)(1)-\nu(1)=\phi_+(1).
\end{align*}

(3) We start with an observation. Let $S$ and $P\geq 0$ be bounded self-adjoint operators on 
$\cH$ and $\lambda\ge0$. Then 
\begin{equation}
(1 -S)P(1-S)+\lambda SPS=\frac{\lambda}{1+\lambda}P
+ \frac{1}{1+\lambda}(1-(1+\lambda)S)P(1-(1+\lambda)S)\geq \frac{\lambda}{1+\lambda}P.
\label{sps}
\end{equation}
Set  $S=\s_{\phi_+}$. Since
\[
\nu(S)=(S\Omega_\nu| S\Omega_\nu)= (\Delta_{\nu|\omega}^{1/2}S\Omega_\omega|
\Delta_{\nu|\omega}^{1/2}S\Omega_\omega),
\]
we have 
\[
D(\nu, \omega)=\nu(S)+ \omega(1-S)
=(\Omega_\omega| (1-S + S\Delta_{\nu|\omega}S)\Omega_\omega).
\]
Let $P(\cdot)$ be  the spectral resolution of $\Delta_{\nu|\omega}$. Then, 
\[
\begin{split}
D(\nu, \omega)
&=\int_0^\infty\left(\Omega_\omega\left| \left[(1-S)\d P(\lambda)(1-S)
+ \lambda S\d P(\lambda) S\right]\right.\Omega_\omega\right)\\[3mm]
&\geq\int_0^\infty\frac{\lambda}{1+\lambda}
\left(\Omega_\omega\left|\d P(\lambda)\right.\Omega_\omega\right)
=(\Omega_\omega|\Delta_{\nu|\omega}(1+\Delta_{\nu|\omega})^{-1}\Omega_\omega),
\end{split}
\]
where in the second step we used the estimate (\ref{sps}). \qed 

We finish this section with some remarks.

{\bf\noindent Remark 1.} Using the scaling law of Proposition \ref{araki-prop-mod} (1) and the
easy fact that $\Omega_{\e^\theta\omega}=\e^{\theta/2}\Omega_\omega$, one deduces from
Part (3) of Proposition \ref{q-NP-Lemma} that
$$
D(\nu,\e^\theta\omega)\ge\int\frac{\e^\theta\d\mu_{\nu|\omega}(x)}{1+\e^{x+\theta}},
$$
for any $\nu,\omega\in\fM_\ast^+$ and $\theta\in\rr$.
The Chebyshev inequality yields that, for any $\vartheta\in\rr$,
\begin{equation}
D(\nu,\e^\theta\omega)\ge\frac{\e^\theta}{1+\e^{\vartheta+\theta}}
\,\mu_{\nu|\omega}(]-\infty,\vartheta]).
\label{aclasbnd}
\end{equation}
In particular, for $\theta=\vartheta=0$, one has
\begin{equation}
D(\nu,\omega)\ge\frac12\mu_{\nu|\omega}(]-\infty,0]).
\label{aclasbndzero}
\end{equation}

{\bf\noindent Remark 2.} As we have already mentioned, the essential difference between the 
classical and the non-commutative setting lies in the proof of Parts (2) and (3) of Theorem 
\ref{q-NP-Lemma}. In the classical setup these results are  elementary and are proven as follows. 
Suppose that  $\nu$ and $\omega$ are two equivalent  measures in $\fM_\ast^+$. Then by Theorem 
\ref{q-NP-Lemma} (1) (which is a simple result with the same proof in the classical and the 
non-commutative cases),
\begin{equation}
D(\nu, \omega)=\frac{1}{2}(\omega(1)+\nu(1)-\|\omega-\nu\|)
=\int\min \left(1, \frac{\d\nu}{\d \omega}\right)\d\omega.
\label{min-tri}
\end{equation}
One easily deduces the classical version of the lower bound \eqref{genubnd}
$$
D(\nu,\omega)\geq \int\frac{\d\mu_{\nu|\omega}(x)}{1+\e^x},
$$
where $\mu_{\nu|\omega}$ is the probability distribution of  the random variable 
$-\log\d\nu/\d\omega$ w.r.t.\;$\omega$. Note however that the lower bound
\[
D(\nu|\omega)\geq \mu_{\nu|\omega}(]-\infty,0]), 
\]
which also follows directly from \eqref{min-tri}, yields a stronger version of the
estimate \eqref{aclasbndzero}. In fact, this stronger bound does
not hold in the quantum case, it is easy to find a $2\times 2$ matrix example violating
this stronger inequality.

Since for $s\in[0,1]$ and $a, b\geq 0$, $\min (a, b)\leq a^{1-s}b^s$, 
(\ref{min-tri}) also implies  (2) in the classical case. (This  argument easily generalizes to the 
case where $\nu$ and $\omega$ are not necessarily equivalent measures). 

{\bf\noindent Remark 3.} After (2) and (3) of Theorem \ref{q-NP-Lemma} are established, there is
no essential difference in the study of the mathematical structure of hypothesis testing in the 
classical and the non-commutative setting.  

A similar  remark applies to  approximately finite hypothesis testing  treated in 
Section~\ref{sec-approxfin}. Proposition \ref{prologue-bounds} (2) has a very similar proof to 
Theorem \ref{q-NP-Lemma} (3). However, the proof of Proposition 
\ref{prologue-bounds} (1) is much simpler than the proof of Theorem \ref{q-NP-Lemma} (2). Once this
result is established, however, there is very little difference between proofs in the approximately 
finite setting and the general setting. 

{\bf\noindent Remark 4.} Theorem \ref{q-NP-Lemma} (2) in the special case $s=1/2$ is equivalent to 
the estimate 
\begin{equation}
\|\Omega_\omega-\Omega_\nu\|^2 \leq \|\omega -\nu\|,
\label{araki-est1}
\end{equation}
which is the non-trivial part of the Araki estimate
(\ref{araki-est}). The proof of Theorem \ref{q-NP-Lemma} (2) is
actually much simpler in this special case (see Remark 3 after
Proposition \ref{araki-multi}) and hence yields a novel simple proof of the
classical estimate (\ref{araki-est1}).

\subsection{Error exponents}
\label{sec-upper-eff}

Let ${\cal I}\subset\rr_+$ be an unbounded index set and for each $t\in{\cal I}$ let $\fM_t$ be a 
$W^\ast$-algebra in standard form on the Hilbert space $\cH_t$, and  $(\nu_t, \omega_t)$ a pair 
of faithful normal functionals on $\fM_t$. Let $\mathcal{I}\ni t\mapsto w_t>0$ be a weight
function such that $\lim_{t\to\infty}w_t=\infty$.

The associated Chernoff type error exponents are defined by 
\[
\bar D=\limsup_{t\rightarrow \infty}\frac{1}{w_t} \log D(\nu_t, \omega_t),
\qquad 
\ubar D=\liminf_{t\rightarrow \infty}\frac{1}{w_t} \log D(\nu_t, \omega_t).
\]
An immediate consequence of Theorem \ref{q-NP-Lemma} (2) is

\bep\label{chernoff-tri}
\[
\bar D \leq \inf_{s\in [0,1]} \limsup_{t\rightarrow \infty}\frac{1}{w_t}\Ent_s(\nu_{t}|\omega_t).
\]
\eep

Suppose that $\nu_t, \omega_t$ are states. For $r\in\rr$ the Hoeffding type exponents are defined
by 
\begin{align*}
\bar B(r)&=\inf_{\{T_t\}}\left\{\limsup_{t \rightarrow \infty}\frac{1}{w_t}\log \omega_t(1-T_t)
\,\, \bigg|\,\, \limsup_{t\rightarrow \infty}\frac{1}{w_t}\log\nu_t(T_t)<-r\right\},\\[3mm]
\ubar B(r)&=\inf_{\{T_t\}}\left\{\liminf_{t \rightarrow \infty}\frac{1}{w_t}\log \omega_t(1-T_t)
\,\,\bigg|\,\, \limsup_{t\rightarrow \infty}\frac{1}{w_t}\log\nu_t(T_t)<-r\right\},\\[3mm]
B(r)&=\inf_{\{T_t\}}\left\{\lim_{t \rightarrow \infty}\frac{1}{w_t}\log \omega_t(1 -T_t)
\,\, \bigg|\,\, \limsup_{t\rightarrow \infty}\frac{1}{w_t}\log\nu_t(T_t)<-r\right\},
\end{align*}
where in the last case the infimum is taken over all families of tests $\{T_t\}_{t \in {\cal I}}$ 
for which  $\lim_{t \rightarrow \infty}\frac{1}{w_t}\log \omega_t(1 -T_t)$ exists. 
\bep\label{basic-expo}
\begin{enumerate}[{\rm (1)}]
\item The Hoeffding exponents  are increasing functions of $r$, and 
$\ubar B(r)\leq \bar B(r)\leq B(r)\leq 0.$
\item  $\ubar B(r)=\bar B(r)= B(r)=-\infty$ if $r < 0$.
\item The functions $\ubar B(r), \bar B(r), B(r)$ are  upper-semicontinuous and right continuous. 
\end{enumerate}
\eep
\proof The only part that requires a proof is (3). We will establish (3) for $\ubar B(r)$, the other cases 
are identical. For a given family of tests $\{T_t\}$ set 
\[
h_{\{T_t\}}(r)=
\begin{cases}
\displaystyle\liminf_{t\rightarrow\infty}\frac{1}{w_t}\log\omega_t(1-T_t)
&\text{if }\displaystyle\limsup_{t\rightarrow \infty}\frac{1}{w_t}\log \nu_t(T_t)<-r\\[8pt]
\infty &\text{otherwise.}
\end{cases}
\]
Since the  function $r\mapsto h_{\{T_t\}}(r)$ is  upper-semicontinuous so is 
\[
\ubar B(r)=\inf_{\{T_t\}} h_{\{T_t\}}(r).
\]
Since $\ubar B(r)$ is increasing and upper-semicontinuous, it is also right continuous.
\qed

Suppose that $\nu_t, \omega_t$ are states. For $\epsilon \in ]0,1[$ set 
\begin{equation}
\begin{split}
&\bar B_\epsilon=\inf_{\{T_t\}}\left\{\limsup_{t \rightarrow \infty}\frac{1}{w_t}\log \omega_t(1-T_t)
\,\, \bigg|\,\,\nu_t(T_t)\leq \epsilon\right\},\\[3mm]
&\ubar B_\epsilon=\inf_{\{T_t\}}\left\{\liminf_{t \rightarrow \infty}\frac{1}{w_t}\log \omega_t(1-T_t)
\,\,\bigg|\,\,\nu_t(T_t)\leq \epsilon\right\},\\[3mm]
&B_\epsilon=\inf_{\{T_t\}}\left\{\lim_{t \rightarrow \infty}\frac{1}{w_t}\log \omega_t(1-T_t)
\,\,\bigg|\,\, \nu_t(T_t)\leq \epsilon\right\},
\end{split}
\label{stein-1}
\end{equation}
where in the last case the infimum is taken over all families of tests $\{T_t\}_{t \in {\cal I}}$ for which  
$\lim_{t \rightarrow \infty}\frac{1}{w_t}\log \omega_t(1-T_t)$ exists.  Note that if
\[
\beta_t(\epsilon)=\inf_{T:\nu(T)\leq\epsilon}\omega_t(1 -T),
\]
then 
\[
\liminf_{t\rightarrow \infty} \frac{1}{w_t}\log \beta_t(\epsilon)
=\ubar B_\epsilon, \qquad \limsup_{t\rightarrow \infty}
\frac{1}{w_t}\log \beta_t(\epsilon)=\bar B_\epsilon.
\]
If $\ubar B_\epsilon= \bar B_\epsilon$ for some $\epsilon$, then also 
\[
\lim_{t \rightarrow \infty} \frac{1}{w_t}\log \beta_t(\epsilon)=B_\epsilon.
\]

 We also define 
\begin{equation}
\begin{split}
\bar B&=\inf_{\{T_t\}}\left\{\limsup_{t \rightarrow \infty}\frac{1}{w_t}\log \omega_t(1-T_t)
\,\,\bigg|\,\,\lim_{t\rightarrow \infty}\nu_t(T_t)=0\right\},\\[3mm]
\ubar B&=\inf_{\{T_t\}}\left\{\liminf_{t \rightarrow \infty}\frac{1}{w_t}\log \omega_t(1-T_t)
\,\,\bigg|\,\,\lim_{t\rightarrow \infty}\nu_t(T_t)=0\right\},\\[3mm]
B&=\inf_{\{T_t\}}\left\{\lim_{t \rightarrow \infty}\frac{1}{w_t}\log \omega_t(1-T_t)
\,\,\bigg|\,\,\lim_{t\rightarrow \infty}\nu_t(T_t)=0\right\},
\end{split}
\label{stein-2}
\end{equation}
where in  the last case the infimum is taken over all families of tests $\{T_t\}_{t \in {\cal I}}$ 
for which $\lim_{t \rightarrow \infty}\frac{1}{w_t}\log \omega_t(1-T_t)$ exists. 

We shall call the numbers (\ref{stein-1}) and (\ref{stein-2}) the  Stein type exponents. 
Clearly, for any $\epsilon\in]0,1[$, one has the relations
\begin{equation}
\begin{array}{ccccc}
\ubar B_\epsilon&\le&\bar B_\epsilon&\le&B_\epsilon\\[-5pt]
\vertleq&&\vertleq&&\vertleq\\[7pt]
\ubar B&\le&\bar B&\le&B
\end{array}
\label{SteinRelat}
\end{equation}

The following general lower bounds holds \cite{HP}.
\bep\label{pain2}
For any $\epsilon\in]0,1[$, one has
\[
\liminf_{t\rightarrow \infty}\frac{1}{w_t}\Ent(\nu_t|\omega_t)
\leq \min\{\ubar B, (1-\epsilon)\ubar B_\epsilon\}.
\]
\eep
\proof 
Let $T_t$ be a test. The monotonicity of the relative entropy (Proposition \ref{prop-relative-q} (3)) 
applied to the 2-dimensional abelian algebra $\mathfrak{N}=\{1,T_t\}''$ implies  
\[
\begin{split}
\Ent(\nu_t|\omega_t)\leq\Ent(\nu_{t|\mathfrak{N}}|\omega_{t|\mathfrak{N}})&=
-\nu_t(T_t)\log \nu_t(T_t) -(1-\nu_t(T_t))\log(1-\nu_t(T_t)) \\[3mm]
 &\,\,\,+ \nu_t(T_t)\log\omega_t(T_t)+ (1-\nu_t(T_t))\log (1-\omega_t(T_t))\\[3mm]
&\leq 2\max_{x\in [0,1]}(-x\ln x) + (1-\nu_t(T_t))\log\omega_t(1-T_t),
\end{split}
\]
from which the statement follows. 
\qed

The further study  of error  exponents is based on 

\newcommand{\Wone}{{\hyperref[W1]{{\rm(W1)}}}}
\begin{quote}\label{W1}
{\bf Assumption (W1).} For $s\in[0,1]$, the limit 
\[
e(s)= \lim_{t \rightarrow \infty}\frac{1}{w_t}\Ent_s(\nu_t|\omega_t)
\]
exists, the function $e(s)$ is continuous on $[0,1]$, differentiable on $]0,1[$, and $D^+e(0)<D^-e(1)$. 
\end{quote}

Note that $e(s)$ is convex on $[0,1]$. If $\nu_t, \omega_t$ are states, then $e(0)=e(1)=0$ and 
$e(s)\leq 0$ for $s\in [0,1]$.

\subsection{Chernoff bound}

\bet  Suppose that \Wone{} holds. Then 
\[
\ubar D=\bar D=  \inf_{s\in [0,1]} e(s).
\]
\label{chernoff-thm}
\eet 

We omit the proof since it is identical to the proof of Theorem \ref{fin-chernoff}.

\subsection{Hoeffding bound}
\label{sec-HB}

In this section we shall make use of the result described in Section~\ref{sec-FLT}. In particular, 
the reader should  recall  the function $\psi(r)$  defined in terms of $e(s)$ in Proposition \ref{var-b-1}. 
\bet\label{hoeffding}
Suppose that \Wone{} holds. Then for all $r$, 
\begin{equation}
\ubar B(r)=\bar B(r)=B(r)=\psi(r).
\label{hoeff-for}
\end{equation}
\eet
{\bf\noindent Remark.} In words, the Hoeffding bound asserts that for any family of tests 
$\{T_t\}_{t\in {\cal I}}$ satisfying 
\begin{equation}
\limsup_{t \rightarrow\infty}\frac{1}{w_t}\log \nu_t(T_t)<-r,
\label{words1}
\end{equation}
one has 
\[
\liminf_{t \rightarrow\infty}\frac{1}{w_t}\log \omega_t(1-T_t)\geq\psi(r),
\] 
and that there exists a family of tests $\{T_t\}$  such that (\ref{words1}) holds and 
\begin{equation}
\lim_{t \rightarrow\infty}\frac{1}{w_t}\log \omega_t(1-T_t)=\psi(r).
\label{lim-te1}
\end{equation}

\proof The proof follows standard arguments (see, {\sl e.g.}, \cite{HMO1}). Since for $r<0$ all the 
terms in \eqref{hoeff-for} are $-\infty$, we only need to consider $r\geq 0$. We recall the definitions
\[
\varphi(\theta)=\sup_{s\in [0,1]}(\theta s -e(s)),
\qquad 
\hat \varphi(\theta)=\varphi(\theta)-\theta,
\]
and set 
\begin{equation}
T_t(\theta)=\s_{(\e^{\theta w_t}\omega_t -\nu_t)_+}.
\label{special-test}
\end{equation} 
Since $\Ent_s(\nu_t|\e^{\theta w_t}\omega_t)=\theta(1-s)w_t + \Ent_s(\nu_t|\omega_t)$, Theorem 
\ref{q-NP-Lemma} (2) implies
\begin{equation} 
D(\nu_t, \e^{\theta w_t}\omega_t)
=\nu(T_t(\theta))+ \e^{\theta w_t}\omega_t(1- T_t(\theta))
\leq \e^{\theta(1-s)w_t}\e^{\Ent_s(\nu_t|\omega_t)},
\label{upper-b-bound-1}
\end{equation}
and hence
\[
\nu(T_t(\theta))\leq \e^{\theta(1-s)w_t}\e^{\Ent_s(\nu_t|\omega_t)}, 
\qquad \omega_t(1- T_t(\theta))\leq \e^{-\theta sw_t}\e^{\Ent_s(\nu_t|\omega_t)},
\]
for $s\in[0,1]$. It follows that 
\begin{equation}
\begin{split}
&\limsup _{t\rightarrow \infty}\frac{1}{w_t}\log D(\nu_t, \e^{\theta w_t}\omega_t)
\leq  \inf_{s\in [0,1]}(e(s)-\theta(s-1))=-\hat\varphi(\theta),\\[3mm]
&\limsup _{t\rightarrow \infty}\frac{1}{w_t}\log \nu_t(T_t(\theta))
\leq  \inf_{s\in [0,1]}( e(s)-\theta(s-1))=-\hat\varphi(\theta),\\[3mm]
&\limsup_{t\rightarrow \infty}\frac{1}{w_t}\log \omega_t(1 -T_t(\theta))
\leq \inf_{s\in [0,1]}  (e(s)-\theta s)=-\varphi(\theta).
\end{split}
\label{chest}
\end{equation}
The estimate \eqref{aclasbnd} yields
$$
D(\nu_t,\e^{\theta w_t}\omega_t)\geq\frac12\,\e^{\theta w_t}\,\mu_{\nu_t|\omega_t}(]-\infty,-\theta w_t]).
$$
Hence, by Theorem \ref{GE-THM}, for $\theta\in]D^+e(0), D^-e(1)[$, 
\[
\liminf_{t\rightarrow \infty}\frac{1}{w_t}\log D(\nu_t,\e^{\theta w_t} \omega_t)\geq \theta -\varphi(\theta)=-\hat \varphi(\theta).
\]
Combining this estimate with (\ref{chest}) we derive that for $\theta \in ]D^+e(0), D^{-}e(1)[$, 
\begin{equation}
\lim_{t\rightarrow \infty}\frac{1}{w_t}\log D(\nu_t, \e^{\theta t}\omega_t)=-\hat \varphi(\theta).
\label{Trylonmidnight}
\end{equation}
Let
\[
D^+e(0)<\theta <\theta^\prime<D^-e(1),
\]
and let $\{T_t\}$ be tests such that 
\[
\limsup_{t\rightarrow \infty}\frac{1}{w_t}\log\nu_t(T_t)<-\hat\varphi(\theta).
\]
We recall that $\hat\varphi$ is strictly decreasing on $]D^+e(0), D^-e(1)[$ and so 
$-\hat\varphi(\theta)<-\hat\varphi(\theta^\prime)$. Using \eqref{Trylonmidnight}, we derive
\[
\begin{split}
-\hat \varphi(\theta^\prime)=\liminf_{t\to\infty}\frac1{w_t}\log D(\nu_t,\e^{\theta'w_t}\omega_t)&
\leq \liminf_{t\to\infty}\frac1{w_t}\log D(\nu_t,\e^{\theta'w_t}\omega_t,T_t)\\
&=\liminf_{t \rightarrow \infty}\frac{1}{w_t}\log (\nu_t(T_t)+\e^{\theta^\prime w_t}\omega_t(1-T_t))\\[3mm]
&\leq \max \left\{ \liminf_{t\rightarrow \infty}\frac{1}{w_t}\log \nu_t(T_t), 
\theta^\prime +\liminf_{t\rightarrow \infty}\frac{1}{w_t}\log \omega_t(1-T_t)\right\}\\[3mm]
&\leq \max \left\{-\hat\varphi(\theta), \theta^\prime +\liminf_{t\rightarrow \infty}
\frac{1}{w_t}\log \omega_t(1-T_t)\right\},
\end{split}
\]
from which we can conclude that
\[
\liminf_{t\rightarrow \infty}\frac{1}{w_t}\log \omega_t(1-T_t)
\geq - \hat \varphi(\theta^\prime)-\theta^\prime=-\varphi(\theta^\prime).
\]
Since $\varphi$ is continuous, taking $\theta^\prime \downarrow \theta$ we derive
\begin{equation}
\liminf_{t\rightarrow \infty}\frac{1}{w_t}\log \omega_t(1-T_t)\geq 
-\varphi(\theta),
\label{thur-hom}
\end{equation}
and so for $\theta \in ]D^+e(0), D^-e(1)[$, 
\[
\ubar B(\hat \varphi(\theta))\geq -\varphi(\theta).
\]
Combining (\ref{thur-hom}) and (\ref{chest}) yields
\[
\lim_{t\rightarrow \infty}\frac{1}{w_t}\log \omega_t(1-T_t(\theta))= - \varphi(\theta),
\]
and so if $\theta \in ]D^+e(0), D^-e(1)[$ then
\[
B(\hat \varphi(\theta)-\epsilon)\leq - \varphi(\theta)
\]
holds for any $\epsilon >0$.

If $\hat\varphi(\theta)$ is a point of continuity of $B$, taking $\epsilon\downarrow 0$ we see that
\begin{equation}
\ubar B(\hat \varphi(\theta))=B(\hat \varphi(\theta))=-\varphi(\theta).
\label{sem}
\end{equation}
The functions $B$ and $\ubar B$ are  finite, increasing and right continuous on the open interval 
$\hat\varphi(]D^+e(0), D^-e(1)[)$.
Since $\hat\varphi$ is continuous and strictly decreasing on $]D^+e(0), D^-e(1)[$,
there exists a countable set ${\cal N}$ such that $\hat\varphi({\cal N})$ is the set of discontinuity 
points of $B$ and $\ubar B$ on $\hat \varphi(]D^+e(0), D^-e(1)[)$. 
Relation~\eqref{sem} holds for all $\theta\in]D^+e(0), D^-e(1)[\setminus {\cal N}$. 
The right continuity implies that \eqref{sem} holds for all $\theta \in ]D^+e(0), D^-e(1)[$.

Suppose that $D^+e(0)>-\infty$. Since $\hat \varphi(\theta)=-\theta$, $ \varphi(\theta)=0$ for 
$\theta \leq  D^+e(0)$ (Theorem \ref{prop-fenchel} (6)), continuity of $\varphi$ and $\hat\varphi$
implies
$$
\lim_{\theta\downarrow D^+e(0)}\hat\varphi(\theta)=-D^+e(0),
\qquad
\lim_{\theta\downarrow D^+e(0)}\varphi(\theta)=0.
$$
Invoking the upper semicontinuity of $B$, we derive from \eqref{sem}
$$
B(-D^-e(0))\geq\limsup_{\theta\downarrow D^+e(0)}B(\hat\varphi(\theta))
=-\lim_{\theta\downarrow D^+e(0)}\varphi(\theta)=0,
$$
and $B$ being non-positive, we conclude that $B(-D^-e(0))=0$. Since $B$ is increasing,
we must have
$$
B(\hat\varphi(\theta))=0=-\varphi(\theta)
$$
for $\theta\leq D^+e(0)$. The same argument shows that 
$\ubar B(\hat\varphi(\theta))=-\varphi(\theta)$ for $\theta\leq D^+e(0)$, and hence
\eqref{sem} holds for all $\theta< D^-e(1)$. Using Equ. \eqref{def-br} 
we see that  for $r>0$, 
\begin{equation} \ubar B(r)=B(r)=\psi(r).
\label{fri-sun}
\end{equation}
Again, since  the functions $\ubar B$, $B$ and $\psi$ are right continuous, taking $r\downarrow0$ in
\eqref{fri-sun} we get that \eqref{fri-sun} also holds for $r=0$. Since $\ubar B(r)\leq \bar B(r)\leq B(r)$, 
the statement follows.
\qed

The above proof gives that if   $r>0$ and $\varphi(\theta)=r$,  then (\ref{lim-te1}) holds for the tests $T_t(\theta)$.

\subsection{Stein's Lemma}
\label{sec-SL}
In accordance with the terminology used in non-equilibrium statistical mechanics, we shall call 
\[
\Sigma^+=D^-e(1)
\]
{\em entropy production} of the hypothesis testing. By Proposition \ref{prop-renyi} (2)--(3)
$$
\Ent_{1-s}(\nu_t|\omega_t)=\Ent_s(\omega_t|\nu_t)\geq s\,\Ent(\nu_t|\omega_t),
$$
and hence, for $s\in]0,1[$,
$$
\frac{e(1-s)}s=\lim_{t\to\infty}\frac1{w_t}\frac{\Ent_{1-s}(\nu_t|\omega_t)}{s}
\ge\limsup_{t\to\infty}\frac1{w_t}\Ent(\nu_t|\omega_t).
$$
Taking the limit $s\downarrow0$, we get
\[
-\Sigma^+\geq \limsup_{t\rightarrow\infty} \frac{1}{w_t}\Ent(\nu_t|\omega_t).
\]

\bep\label{pain1}
Suppose that \Wone{} holds. Then  all  Stein type exponents are $\leq -\Sigma^+$. 
\eep

\proof
By Relation \eqref{SteinRelat}, the largest Stein exponent is $B$ and since 
$\limsup_{t\to\infty}\frac1{w_t}\log\nu_t(T_t)<0$ implies 
$\lim_{t\to\infty}\nu_t(T_t)=0$ one has $B\le B(0)$. Theorem \ref{hoeffding} and Proposition 
\ref{var-b-1} (2) yield
$$
B\leq B(0)=\psi(0)=-\Sigma^+.
$$
\qed

To derive lower bounds we need to strengthen \Wone{} and assume

\newcommand{\Wtwo}{{\hyperref[W2]{{\rm(W2)}}}}
\begin{quote}\label{W2}
{\bf Assumption (W2).}  \Wone{} holds, $\Sigma^+$ is finite, and for all $s\in \rr$, 
\begin{equation}
\lim_{t\rightarrow \infty}(\Omega_{\nu_t}|\Delta_{\omega_t|\nu_t}^{\i s/w_t}\Omega_{\nu_t})
=\e^{-\i s\Sigma^+}.
\label{weak-law}
\end{equation}
\end{quote}

The following result is known as Stein's Lemma.

\bet\label{Stein-lemma}
Suppose that \Wtwo{} holds. Then for all  $\epsilon \in ]0,1[$,
\[
\ubar B = \bar B  =B= \ubar B_\epsilon = \bar B_\epsilon= B_\epsilon =-\Sigma^+.
\]
\eet

\proof  By Proposition \ref{pain1} and Relation \eqref{SteinRelat}, it suffices to prove that 
$\ubar B_\epsilon \geq -\Sigma^+$.
Assumption \Wtwo{} and the Levy continuity theorem imply that the spectral measure $\tilde\mu_t$ of 
$-\frac1{w_t}\log \Delta_{\omega_t|\nu_t}$ for $\Omega_{\nu_t}$ converges weakly to the Dirac 
measure at $\Sigma^+$. Let $\{T_t\}$ be a family of tests such that $\nu_t(T_t)\le\epsilon\in]0,1[$
for all $t\in\mathcal I$. For $\theta>\vartheta>\Sigma^+$, the estimate \eqref{aclasbnd} yields
$$
\omega_t(1-T_t)\ge D(\e^{-\theta w_t}\nu_t,\omega_t)-\e^{-\theta w_t}\nu_t(T_t)\ge
\e^{-\theta w_t}\left(\frac{\tilde\mu_t(]-\infty,\vartheta[)}{1+\e^{(\vartheta-\theta) w_t}}-\epsilon\right).
$$
Since $\tilde\mu_t(]-\infty,\vartheta[)/(1+\e^{(\vartheta-\theta) w_t})\to1$ as $t\to\infty$, we conclude 
that
$$
\liminf_{t\to\infty}\frac1{w_t}\log\omega_t(1-T_t)\ge -\theta
$$
holds for any $\theta>\Sigma^+$. Taking $\theta\downarrow\Sigma^+$ yields the result.
\qed

In many models one can show that the following holds: 

\newcommand{\Wthree}{{\hyperref[W3]{{\rm(W3)}}}}
\begin{quote}\label{W3}
{\bf Assumption (W3).}  For some $\delta >0$ and $s\in [0, 1+\delta[$ the limit 
\[ e(s)=\lim_{t \rightarrow \infty}\frac{1}{w_t}\Ent_s(\nu_t|\omega_t)\]
exists,  the function $e(s)$ is continuous on $[0, 1+\delta[$, differentiable on $]0, 1+\delta[$, and 
$e^\prime(1)=\Sigma^+>0$.   
\end{quote}

We note that  \Wthree{} $\Rightarrow$ \Wtwo{}. Indeed, Proposition \ref{prop-renyi} (4) and the
convexity of $e(s)$ imply
\begin{equation} 
\Sigma^+=e^\prime(1)=\lim_{t\to\infty}\left.\frac{\d\ }{\d s}\frac1{w_t}\Ent_s(\nu_t|\omega_t)\right|_{s=1}
=-\lim_{t\rightarrow \infty}\frac{1}{w_t}\Ent (\nu_t|\omega_t).
\label{ep-test}
\end{equation} 
Propositions \ref{LDP-basic}  and \ref{prop-renyi} (2) imply (\ref{weak-law}).


\section{Entropy production and full counting statistics for $W^\ast$-dynamical systems}
\label{sec-NECSM}
\subsection{Setup}
Our starting point is a quantum dynamical system $\cQ=(\fM, \tau, \omega)$. There, 
$\fM$ is a $W^\ast$-algebra in standard form on a Hilbert space $\cH$,
$\tau=\{\tau^t\,|\, t\in \rr\}$ is a $W^\ast$-dynamics, {\sl i.e.,} a weakly  continuous group of 
$\ast$-automorphisms of $\fM$, and $\omega$ is a faithful normal state on $\fM$.
The self-adjoint elements of $\fM$ describe observables of the physical system under consideration 
which evolve in time as $A_t=\tau^t(A)$. 
The state  $\omega$ describes the initial thermodynamic  state of the system. 
The state of the system at time $t$ is $\omega_t=\omega \circ \tau^t$. 
Obviously, $\omega_t(A)=\omega(A_t)$.
 
A time-reversal of $\cQ$ is an anti-linear $\ast$-automorphism $\Theta:\fM\rightarrow \fM$
such that
\[
\Theta\circ \Theta= {\rm id}, \qquad \tau^t \circ \Theta=\Theta \circ \tau^{-t}.
\label{tri}
\]
The state $\omega$ is time-reversal invariant if $\omega(\Theta(A)) =\omega(A^\ast)$ for 
all $A\in \fM$ and in this case we say that the system $\cQ$ is TRI.

Let $\beta >0$. We recall that $\omega$ is a $(\tau,\beta)$-KMS state if for all $A, B \in \fM$ 
the function 
\[
F_{A, B}(t)=\omega(AB_t)
\]
has an analytic continuation to the strip $0 < \Im z <\beta$ which is bounded and continuous 
on its closure, and which satisfies the KMS-boundary condition
\[
F_{A,B}(t+\i \beta)=\omega(B_tA),
\]
for $t\in\rr$.
A $(\tau, \beta)$-KMS state is $\tau$-invariant and faithful. In algebraic quantum statistical mechanics
a $(\tau, \beta)$-KMS state describes a physical state of thermal equilibrium at inverse temperature 
$\beta$. For additional information about KMS states we refer the reader to \cite{BR2}.

\subsection{Entropy production observable}
\label{sec-epo}

Consider the Araki-Connes cocycles 
\[
[D\omega_t : D\omega]_u=\Delta_{\omega_t|\omega}^{\i u}\Delta_{\omega}^{-\i u},
\]
with $u,t\in\rr$. They are unitaries in $\fM$ satisfying
\[
[D\omega_{t_1+t_2} : D\omega]_u
=\tau^{-t_2}([D\omega_{t_1} : D\omega]_u)[D\omega_{t_2} : D\omega]_u.
\]

\newcommand{\Epone}{{\hyperref[Ep1]{{\rm(Ep1)}}}}
\begin{quote}\label{Ep1}
{\bf Assumption (Ep1).} For all $t\in \rr$ the map 
$u \mapsto [D\omega_t : D\omega]_u\in \fM$ is weakly differentiable at $u=0$.
\end{quote}

The derivatives 
\[
\ell_{\omega_t|\omega}=\left.\frac{1}{\i} \frac{\d\ }{\d u}[D\omega_t : D\omega]_u\right|_{u=0}
\]
are self-adjoint elements of $\fM$ satisfying 
\begin{equation}
\ell_{\omega_{t_1+t_2}|\omega}=\tau^{-t_2}(\ell_{\omega_{t_1}|\omega}) +\ell_{\omega_{t_2}|\omega}.
\label{cocycle}
\end{equation}

Note also that 
\[
\log \Delta_{\omega_t|\omega}=\log \Delta_\omega +\ell_{\omega_t|\omega}.
\]
In the terminology of Araki \cite{Ar3}, $\ell_{\omega_t|\omega}$ is the relative Hamiltonian of
$\omega_t$ w.r.t.\;$\omega$. The entropy cocycle is defined by $c^t =\tau^t(\ell_{\omega_t|\omega})$ 
and satisfies 
\[
c^{t_1+t_2}= c^{t_2} + \tau^{t_2}(c^{t_1}).
\]
In the finite dimensional setting $c^t =S_t - S$, where $S=-\log \omega$ is the entropy observable.

We also need 

\newcommand{\Eptwo}{{\hyperref[Ep2]{{\rm(Ep2)}}}}
\begin{quote}\label{Ep2}
{\bf Assumption (Ep2).} The map $t \mapsto \ell_{\omega_t|\omega}\in \fM$ is weakly differentiable 
at $t=0$.
\end{quote}

The {\em entropy production observable} is defined by 
\[
\sigma=\left.\frac{\d\ }{\d t}\ell_{\omega_t|\omega}\right|_{t=0} =\left.\frac{\d\ }{\d t} c^t\right|_{t=0},
\]
and is the exact non-commutative analog of the phase space contraction rate in classical 
non-equilibrium statistical mechanics \cite{JPR}. The cocycle relation (\ref{cocycle}) yields 
\[
\ell_{\omega_t|\omega}=\int_0^t \sigma_{-u}\d u, \qquad c^t =\int_0^t \sigma_u \d u,
\]
where the integrals converge weakly ({\sl i.e.,} for all $\nu \in \fM_\ast$, 
$\nu(\ell_{\omega_t|\omega})=\int_0^t \nu(\sigma_{-u})\d u$, see Section~2.5.3 in \cite{BR1}).

The observable of {\em mean entropy production rate over the time interval} $[0, t]$ is 
\[
\Sigma^t =\frac{c^t}{t}=\frac{1}{t}\int_0^t \sigma_u \d u.
\]

\bep Suppose that \Epone{} and \Eptwo{}  hold. Then: 
\begin{enumerate}[{\rm (1)}]  
\item $\omega(\sigma)=0$.
\item If $\cQ$ is {\rm TRI}, then $\Theta(\sigma)=-\sigma$. 
\item The entropy balance equation
\[
\omega(\Sigma^t)=-\frac1t\Ent(\omega_t|\omega),
\]
holds. In particular, $\omega(\Sigma^t)\geq 0$ for $t>0$. 
\end{enumerate}
\eep

Let us again illustrate the above  definitions on the example of open quantum systems. 
Consider thermal reservoirs $\cR_j$ described by quantum dynamical systems 
$(\fM_j, \tau_j, \omega_j)$ in standard form on the Hilbert spaces $\cH_j$,  $j=1, \cdots, n$. 
For each $j$, let $\vartheta_j=\{ \vartheta_j^t\,|\, t\in \rr\}$  be a given gauge group, {\sl i.e.,} a 
$W^\ast$-dynamics on $\fM_j$ commuting with $\tau_j$ 
($\vartheta_j^t \circ \tau_j^t=\tau_j^t \circ \vartheta_j^t$ for all $t$). 
We assume that $\cR_j$ is in thermal equilibrium at inverse temperature $\beta_j$ and chemical 
potential $\mu_j$, {\sl i.e.,} that $\omega_j$ is a $\beta_j$-KMS state for the $W^\ast$-dynamics 
$\tau_j^t \circ \vartheta_j^{-\mu_jt}$. We denote by $\delta_j/\xi_j$ the generator of $\tau_j/\vartheta_j$,
{\sl i.e.,} $\tau^t=\e^{t\delta_j}/\vartheta^t=\e^{t\xi_j}$.
The complete reservoir system $\cR=\cR_1 +\cdots +\cR_n$ is  
described by $(\fM_\cR, \tau_\cR, \omega_\cR)$, where 
\[
\fM_\cR=\otimes_{j=1}^n \fM_j, \qquad \tau_\cR=\otimes_{j=1}^n\tau_j, \qquad \omega_\cR=
\otimes_{j=1}^n\omega_j.
\]
Let $\cS$ be a finite quantum system described by the Hilbert space $\cH_\cS$, the Hamiltonian 
$H_\cS$ and the state $\omega_\cS(A)=\Tr A/\dim \cH_\cS$. We set $\fM_\cS=\cO_{\cH_{\cS}}$, 
$\tau_\cS^t(A)=\e^{\i t H_\cS}A\e^{-\i t H_{\cS}}$.
The joint but decoupled system $\cS+\cR$ is described by $(\fM, \tau_0, \omega)$, where 
\[
\fM=\fM_{\cS}\otimes \fM_{\cR}, \qquad \tau_0=\tau_{\cS}\otimes \tau_\cR, 
\qquad \omega=\omega_\cS\otimes \omega_\cR.
\]
The coupling of $\cS$ with $\cR_j$ is described by a self-adjoint element $V_j\in\fM_\cS\otimes \fM_j$
and the full interaction is $V=\sum_{j=1}^n V_j$. Let $\tau$ be the perturbed $W^\ast$ dynamics 
generated by $\delta=\delta_0 +\i [V,\,\cdot\,]$, where $\delta_0$ is the generator of $\tau_0$. 
The coupled open quantum system $\cQ=\cS+\cR$ is described by $(\fM, \tau, \omega)$.

Suppose that $V$ is in the domain of the generators $\delta_j/\xi_j$ for all $j$ (we abbreviate 
$\delta_j \otimes {\rm id}, {\rm id} \otimes \delta_j$ by $\delta_j$, etc). 
The observables 
\begin{equation} \Phi_j =\delta_j(V), \qquad {\cal J}_j =\xi_j(V),
\label{flux-gener}
\end{equation}
describe the energy/charge currents out of $\cR_j$. Under the above assumptions,
\Epone{} and \Eptwo{} hold and 
\[
\sigma =-\sum_{j=1}^n \beta_j(\Phi_j -\mu_j {\cal J}_j).
\]
For the proofs and additional information we refer the reader to \cite{JOPP2}.

\subsection{Full counting statistics} 
\label{sec-fcs}
We continue with the framework of the previous sections, adapting to this setting the
construction of Section \ref{FCS-finite}. Let $\mu_{\omega_t|\omega}$ be the
spectral measure for $-\log \Delta_{\omega_t|\omega}$ and $\Omega_\omega$.
The R\'enyi entropic functional of the system $\cQ$ is
\[
e_t(s)=
\Ent_s(\omega_t|\omega)=(\Omega_\omega|\Delta_{\omega_t|\omega}^s\Omega_\omega)
=\log \int\e^{-sx} \d\mu_{\omega_t|\omega}(x).
\] 
Following Eq. \eqref{QPrelat}, we define the full counting statistics of $\cQ$ as
the family $\{{\mathbb P}_t\}_{t>0}$ of Borel probability measures on $\rr$ given by 
\[
\d {\mathbb P}_t(\phi) = \d\mu_{\omega_{-t}|\omega}(t\phi).
\]
Since
\[
e_{-t}(s)= \Ent_s(\omega_{-t}|\omega)=\Ent_s(\omega|\omega_t)
=\Ent_{1-s}(\omega_t|\omega)=e_{t}(1-s),
\]
one has
\[
e_t(1-s)= \log \int\e^{-st \phi}\d{\mathbb P}_t(\phi),
\]
and if the system $\cQ$ is TRI, then 
\[
\Ent_{s}(\omega_t|\omega)= \Ent_{s}(\omega_{-t}|\omega),
\]
and so the transient Evans-Searles  fluctuation relation 
\[
e_t(s)=e_t(1-s)
\]
holds (this relation was  proven for the first time in \cite{TM1}). Equivalent formulations of 
this fluctuation relation are the identities (1)--(4) on page 
\pageref{ESequiv}.

To describe the fluctuations of ${\mathbb P}_t$ as $t\rightarrow \infty$ we need

\newcommand{\LTone}{{\hyperref[LT1]{{\rm(LT1)}}}}
\begin{quote}\label{LT1}
{\bf Assumption  (LT1).}  There is an open interval ${\mathbb I}\supset [0,1]$ such that for
$s\in {\mathbb I}$ the limit 
\[
e(s)=\lim_{t\rightarrow \infty}\frac{1}{t}e_t(s)
\]
exists, is finite, and the function ${\mathbb I}\ni s \mapsto e(s)$ is differentiable.
\end{quote}

For our purposes we may assume w.l.o.g.\;that ${\mathbb I}$ is symmetric around $s=1/2$. 
The function $e(s)$ is convex, $e(0)=e(1)=0$, $e(s)\leq 0$ for $s\in [0,1]$, and $e(s)\geq 0$ 
for $s\not\in [0,1]$. The entropy production of $\cQ$ is
\[
\Sigma^+=e^\prime(1)=-\lim_{t\rightarrow \infty}\frac{1}{t}\Ent(\omega_t|\omega).
\]
We also set
\[
\bar \Sigma^+=-e^\prime(0)=-\lim_{t\rightarrow \infty}\frac{1}{t}\Ent(\omega|\omega_t),
\]
\[
\ubar \theta =\inf_{s\in {\mathbb I}}e^\prime(s), 
\qquad \bar \theta =\sup_{s\in {\mathbb I}}e^\prime(s),
\]
and
\[
\varphi(\theta)=\sup_{s\in {\mathbb I}}(-\theta s - e(s)).
\]
If $\cQ$ is TRI, then the Evans-Searles fluctuation relations 
\[
e(s)=e(1-s), \qquad \varphi(-\theta)=\varphi(\theta)+\theta,
\]
hold and $\bar \Sigma^+=\Sigma^+$, $\bar \theta =-\ubar \theta$.

The assumption \LTone{} and the G\"artner-Ellis theorem imply that
${\mathbb P}_t\to\delta_{\Sigma^+}$ weakly and ${\mathbb E}_t(\phi)\to\Sigma^+$ as  
$t\rightarrow  \infty$. Moreover,  for any open set ${\mathbb J}\subset (\ubar \theta, \bar\theta)$, 
\[
\lim_{t\rightarrow \infty}\frac{1}{t}\log {\mathbb P}_t({\mathbb J})
=-\inf _{\theta \in {\mathbb J}}\varphi(\theta).
\]
If $e(s)$ is analytic in a neighborhood of zero, then 
$$
\lim_{t\to\infty}t^{k-1}C^{(k)}_t=\partial_s^k e(s)|_{s=1},
$$
where $C^{(k)}_t$ denotes the $k$-th cumulant of ${\mathbb P}_t$.

The physical  interpretation of  ${\mathbb P}_t$ is in terms of the full counting statistics of an 
approximating sequence of finite quantum systems. In  a given model the choice of this approximating 
sequence is dictated by physical considerations. A typical mathematical setup for these approximations
is the following.

\begin{quote}{\bf (Ap1)} The approximating sequence of quantum systems $\cQ_m$, 
$m\in {\mathbb N}$, is described by a sequence of  finite dimensional Hilbert spaces ${\cH}_m$,
algebras $\cO_{\cH_m}$, Hamiltonians $H_m$ and states $\omega_m>0$. 
\end{quote}

\begin{quote} {\bf (Ap2)} For all $m$ there is a faithful representation 
$\pi_m:\cO_{\cH_m}\rightarrow \fM$ such that 
$\pi_m(\cO_{\cH_m})\subset \pi_{m+1}(\cO_{\cH_{m+1}})$.
\end{quote}

\begin{quote} {\bf (Ap3)} For $A\in\cup_{m}\pi_m(\cO_{\cH_m})$ and $t\in \rr$, 
\[
\lim_{m\rightarrow \infty}\omega_{mt}(\pi_m^{-1}(A))=\omega_t(A).
\]
\end{quote}

\begin{quote} {\bf (Ap4)} For all $s$ in some open set containing $[0,1]$ and for all $t$,
\[
\lim_{m\rightarrow \infty}\Ent_s(\omega_{mt}|\omega_m)=\Ent_s(\omega_t|\omega).
\]
\end{quote}

(Ap4) ensures that the full counting statistics ${\mathbb P}_{mt}\to{\mathbb P}_t$ weakly
as $m\rightarrow \infty$, thus giving a direct physical interpretation to ${\mathbb P}_t$, the full counting 
statistics of the infinitely extended system. The same approximation scheme is needed to give a 
physical interpretation the energy/charge current observables (\ref{flux-gener}).

The approximation setup (Ap1)--(Ap4) is well suited for spin systems and fermionic systems but 
needs to be adjusted for bosonic systems.

Again, for the proofs and additional information we refer the reader to \cite{JOPP2}

\subsection{Hypothesis testing of the arrow of time}
\label{sec-hypo}

The quantum hypothesis testing of the family  $\{(\omega_t, \omega_{-t})\}_{t>0}$ with
the weight $w_t=2t$ or, equivalently, of the family $\{(\omega, \omega_{t})\}_{t>0}$ with
the weight $w_t=t$ yields

\bet Suppose that \LTone{} holds and that $\Sigma^+>0$. Then: 
\begin{enumerate}[{\rm (1)}] 
\item
\[
\lim_{t\rightarrow \infty}\frac{1}{2t}\log (2-\|\omega_t-\omega_{-t}\|)=\inf_{s\in [0,1]}e(s).
\]
If $(\fM, \tau, \omega)$ is {\rm TRI}, then $\inf_{s\in [0,1]}=e(1/2)$.

\item For all $r\in \rr$, 
\[ \ubar B(r)=\bar B(r)=B(r)=\psi(r)=-\sup_{0\leq s<1}\frac{-sr -e(s)}{1-s}.
\]
\item For any $\epsilon \in ]0,1[$, 
\[\ubar B=\bar B=B=\ubar B_\epsilon=\bar B_\epsilon =B_\epsilon=-\Sigma^+.
\]
\end{enumerate}
\label{noneq-hp}
\eet
{\bf Remark.} Clearly, the statements of this theorem hold under more general conditions then 
\LTone{}. We have stated it in the present form for a transparent comparison with the large deviation
principle for full counting statistics. 

\bigskip
Theorem \ref{noneq-hp} quantifies the separation between the past and the future as time 
$t\uparrow\infty$. Since 
\[ 
\frac{1}{2}(2-\|\omega_t-\omega_{-t}\|)= 
\omega_t(1-\s_{(\omega_{t}-\omega_{-t})_+}) +\omega_{-t}(\s_{(\omega_t-\omega_{-t})_+}),
\]
Part (1) implies 
\begin{equation}
\limsup_{t \rightarrow \infty}\frac{1}{2t}\log \omega_t(1-\s_{(\omega_{t}-\omega_{-t})_+})
\leq \inf_{s\in [0,1]}e(s),
\label{sep-1}
\end{equation}
\begin{equation}
\limsup_{t \rightarrow \infty}\frac{1}{2t}\log \omega_{-t}(\s_{(\omega_{t}-\omega_{-t})_+})
\leq \inf_{s\in [0,1]}e(s).
\label{sep-2}
\end{equation}
Therefore, as $t \uparrow \infty$, the states $\omega_t$ are  concentrating exponentially fast along 
the projections $\s_{(\omega_t-\omega_{-t})_+}$ while the states $\omega_{-t}$ are concentrating 
exponentially fast along the orthogonal complement $1-\s_{(\omega_t-\omega_{-t})_+}$. Note 
that one of the inequalities (\ref{sep-1}), (\ref{sep-2}), must be an equality. If $(\fM, \tau, \omega)$ 
is TRI, then 
$\omega_t(1-\s_{(\omega_{t}-\omega_{-t})_+})=\omega_{-t}(\s_{(\omega_{t}-\omega_{-t})_+})$, 
and in this case 
 \[
 \lim_{t \rightarrow \infty}\frac{1}{2t}\log \omega_t(1-\s_{(\omega_{t}-\omega_{-t})_+})=
 \lim_{t \rightarrow \infty}\frac{1}{2t}\log \omega_{-t}(\s_{(\omega_{t}-\omega_{-t})_+})=e(1/2).
 \]

Let  $r>0$ and let  $\{T_t\}$ be a family of projections in $\fM$ such that 
\begin{equation}
\limsup_{t\rightarrow \infty}\frac{1}{2t}\log \omega_{-t}(T_t) <-r.
\label{sat-sick}
\end{equation}
Part (2) of Theorem \ref{noneq-hp} asserts that
\[
\liminf_{t\rightarrow \infty}\frac{1}{2t}\log\omega_t(1-T_t)\geq  \psi(r),
\]
and that there exists such a family $\{T_t\}$ for which
\[
\lim_{t\rightarrow \infty}\frac{1}{2t}\log\omega_t(1-T_t)= \psi(r).
\]
Hence, if $\omega_t$ is concentrating exponentially fast along $1-T_t$ with the rate $<-r$, then
$\omega_{-t}$ is concentrating along $T_t$ with the optimal exponential rate $\psi(r)$.

Part (3) of Theorem \ref{noneq-hp} asserts that for any family of projections $\{T_t\}$ such that 
\begin{equation}
\sup_t \omega_{-t}(T_t)<1,
\label{lv-horse}
\end{equation}
one has 
\[
\liminf_{t\rightarrow \infty}\frac{1}{2t}\log \omega_t(1-T_t)\geq -\Sigma^+,
\]
and that for any $\delta >0$ one can find a family $\{T_t^{(\delta)}\}$ satisfying
(\ref{lv-horse}) and 
\[
\lim_{t\rightarrow \infty}\frac{1}{2t}\log \omega_t(1-T_t^{(\delta)})\leq  -\Sigma^+ +\delta.
\]
Hence, if no restrictions are made on $T_t$ w.r.t.\;$\omega_{-t}$ except  (\ref{lv-horse}) (which 
is needed to avoid trivial results), the optimal  exponential rate of concentration of $\omega_t$ as 
$t\uparrow \infty$ is precisely $-\Sigma^+$.  

\section{Examples}
\label{sec-openqs}

\subsection{Spin-fermion model}
The spin-fermion model is an open quantum system describing the interaction of a
spin $1/2$ with finitely many free Fermi gas reservoirs $\cR_1,\ldots,\cR_n$. 

We shall use the notation introduced in Section~\ref{sec-CAR}. Let $\ch_j$ and $h_j$ be the single 
fermion Hilbert space and Hamiltonian of $\cR_j$.  The Hilbert space and Hamiltonian 
of $\cR_j$ are $\cH_j =\Gamma_{\rm f}(\ch_j)$ and $H_j =\d\Gamma(h_j)$ (the second quantization
of $h_j$). The algebra of observables of $\cR_j$ is $\cO_j =\CAR(\fh_j)$  and its dynamics is 
\[ 
\tau_j^t(A)=\e^{\i t H_j}A\e^{-\i t H_j}.
\]
In particular, $\tau_j^t( a_j^\#(f))=a_j^\#(\e^{\i t h_j}f)$, where $a_j^\#(\cdot)$ stands for  the 
annihilation/creation operator in $\cO_j$. The pair $(\cO_j, \tau_j)$ is a $C^\ast$-dynamical system, 
{\sl i.e.,} $\tau_j$ is a strongly continuous group of $\ast$-automorphisms of $\cO_j$. Initially, 
the reservoir $\cR_j$ is in thermal equilibrium at inverse temperature $\beta_j>0$ and chemical 
potential $\mu_j$, {\sl i.e.,} its initial state $\omega_j$ is the quasi-free state with the Fermi-Dirac 
density 
\[
T_{\beta_j, \mu_j}=\left(1 +\e^{\beta_j (h_j-\mu_j)}\right)^{-1}.
\]  
 
The full reservoir system $\cR=\cR_1+\cdots+\cR_n$ is described by 
\[
\cO_{\cR}=\cO_1\otimes\cdots\otimes\cO_n, \qquad 
\tau_\cR=\tau_1\otimes\cdots\otimes\tau_n, \qquad
\omega_\cR=\omega_1\otimes\cdots\otimes \omega_n.
\]
We also set $\cH_\cR=\cH_1\otimes\cdots\otimes\cH_n$ and $H_\cR= H_1 +\cdots + H_n$.

The small system $\cS$ is described by the Hilbert space $\cH_\cS=\cc^2$ and 
Hamiltonian\footnote{$\sigma^{(1)}, \sigma^{(2)}, \sigma^{(3)}$ denote the usual Pauli matrices.} 
$H_\cS=\sigma^{(3)}$. 
The initial state of $\cS$ is $\omega_{\cS}>0$. The Hilbert space of the coupled system $\cS+\cR$
is $\cH=\cH_\cS\otimes\cH_\cR$, its algebra of observables is $\cO=\cO_{\cH_\cS}\otimes\cO_\cR$,
and its initial state is $\omega=\omega_\cS\otimes\omega_\cR$. The interaction of $\cS$ with $\cR_j$
is described by $V_j=\sigma^{(1)}\otimes\varphi_j(\alpha_j)$, where $\varphi_j(\cdot)$ is the field 
operator in $\cO_j$ and $\alpha_j\in\ch_j$ are given vectors (in the present context, they are often 
called form-factors). The full interaction is $V=\sum_{j=1}^n V_j$ and the Hamiltonian of the coupled 
system is 
\[
H_\lambda=H_\cS + H_\cR + \lambda V,
\]
where $\lambda\in\rr$ is a coupling constant. The dynamics of the coupled system is described by 
\[
\tau_\lambda^t(A)=\e^{\i t H_\lambda}A\e^{-\i t H_\lambda}.
\]
The pair $(\cO, \tau_\lambda)$ is a $C^\ast$-dynamical system and the triple 
\[
\cQ=(\cO,\tau_\lambda,\omega)
\]
is the spin-fermion model. Needless to say, we are considering the simplest non-trivial case, for 
various generalizations see Remark 3 below. Passing to the  GNS representation of $\cO$ 
associated to $\omega$ one verifies that  the spin-fermion model is an example of an abstract open 
quantum system described at the end of Section~\ref{sec-epo}.
 
Starting with the seminal papers \cite{Dav, DS, LS}, the spin-fermion and the closely related spin-boson 
model have been one of the basic paradigms of non-equilibrium quantum statistical mechanics.
Although many basic questions are still open, systematic efforts over the last decade have lead to a 
reasonably good understanding of these models in the weak coupling regime ($\lambda$ small).
To describe the results that concern us here, we make the following regularity assumptions.

\newcommand{\SFone}{{\hyperref[SF1]{{\rm(SF1)}}}}
\begin{quote}\label{SF1}
{\bf  Assumption (SF1).} $\mu_j=0$, $\ch_j=L^2(\rr_+,\d x;{\mathfrak H}_j)$ for some auxiliary 
Hilbert space ${\mathfrak H}_j$ and $h_j$ is the operator of multiplication by the variable $x\in\rr_+$.
\end{quote}
 
For example, if $\ch_j =L^2(\rr^d, \d x)$ and $\ch_j =-\Delta$, where $\Delta$ is the
usual Laplacian on $\rr^d$, passing first to the momentum representation and then to polar 
coordinates shows that \SFone{} holds after a unitary transformation. (This is Example 1 in Sect.~3.1 in \cite{JKP}.)
 
We extend the form factors $\alpha_j$  to $\rr$ by setting $\alpha_j(-x)=\alpha_j(x)$.

\newcommand{\SFtwo}{{\hyperref[SF2]{{\rm(SF2)}}}}
\begin{quote}\label{SF2}
{\bf Assumption (SF2).} There is $\delta >0$ such that the $\alpha_j$'s extend to 
analytic functions on the strip $|\Im z|<\delta$ satisfying 
\[
\sup_{|y|<\delta}\int_{\rr}\e^{-a x}\|\alpha_{j}(x+\i y)\|_{{\mathfrak H}_j}^2 \d x <\infty
\]
for all $a>0$.
\end{quote}

\newcommand{\SFthree}{{\hyperref[SF3]{{\rm(SF3)}}}}
\begin{quote}\label{SF3}
{\bf Assumption (SF3).} For all $j$, $\|\alpha_j(2)\|_{{\mathfrak H}_j}>0$ ($2=1-(-1)$ is the
Bohr frequency of the $2$-level system $\cS$).
\end{quote}

Let  $\ch_{mj}$, $m=1,2\ldots,$  be  an increasing sequence of finite dimensional subspaces of $\fh_j$
such that $\cup_m \ch_{mj}$ is dense in $\fh_j$. Let $h_{mj}\in\cO_{\ch_{mj}, {\rm self}}$ be such that 
$h_{mj}\to h_j$ as $m\rightarrow \infty$ in the strong resolvent sense ($h_{mj}$ being extended to
$0$ on $\ch_{mj}^\perp$). Let $\alpha_{mj}\in\fh_{mj}$ be such that 
$\e^{a h_m}\alpha_{mj}\to\e^{ah}\alpha_j$ for all $a \in \rr$ and all $j$. Consider the sequence 
$\cQ_m=(\cO_m, \tau_{m\lambda}, \omega_m)$ of finite dimensional spin-fermion systems
constructed over $(\ch_{jm}, h_{jm})$ with form factors $\alpha_{mj}$. $\cQ_m$ is a (very general) 
sequence of finite dimensional approximations of the infinitely extended spin-fermion model $\cQ$. 

Let ${\mathbb P}_{m\lambda t}$ / ${\mathbb P}_{\lambda t}$ be the full counting statistics of 
$\cQ_m$ / $\cQ$ and  $e_{m\lambda t}(s)$ / $e_{\lambda t}(s)$ the corresponding R\'enyi
entropies (we denote explicitly the dependence on the coupling strength $\lambda$). 
We adopt the shorthand notation
\[
{\mathbb I}(\Lambda,r)=\{(\lambda,s)\in\rr^2\,|\,|\lambda|<\Lambda, s\in]1/2-r,1/2+r[\}.
\]

\bet Suppose that \SFone--\SFthree{} hold. Then:
\begin{enumerate}[{\rm (1)}]
\item For all $s, \lambda \in \rr$ and $t>0$, 
\[
\lim_{m \rightarrow \infty} e_{m\lambda t}(s) = e_{\lambda t}(s)
\]
exists and is finite.  In particular ${\mathbb P}_{m\lambda t}\rightarrow {\mathbb P}_{\lambda t}$
weakly. The function $(\lambda, s)\mapsto e_{\lambda t}(s)$ is real analytic and all cumulants of 
${\mathbb P}_{m \lambda t}$ converge to the corresponding cumulants of ${\mathbb P}_{\lambda t}$.

\item Let $r>0$ be given. Then there is $\Lambda >0$ such that for 
$(\lambda, s)\in{\mathbb I}(\Lambda,r)$
\[
e_{\lambda}(s)=\lim_{t\rightarrow \infty}\frac{1}{t}e_\lambda (s)
\]
exists. Moreover the function 
\[
 {\mathbb I}(\Lambda,r)\ni (\lambda, s)\mapsto e_{\lambda}(s)
\]
is real analytic, does not depend on the choice of $\omega_\cS$, and satisfies the Evans-Searles 
symmetry
\begin{equation}
e_{\lambda}(s)=e_\lambda(1-s).
\label{es-spin-fermion}
\end{equation}
\item For $0<|\lambda|< \Lambda$, $\Sigma_\lambda^+=e_\lambda^{\prime}(1)>0$ unless 
$\beta_1=\cdots=\beta_n$. Moreover, 
\[
\Sigma_\lambda^+ =\lambda^2\mathfrak{S}^+ + O(\lambda^4), 
\]
where 
\[
\mathfrak{S}^+=\frac\pi2\sum_{i,j}
\frac{\|\alpha_i(2)\|^2_{{\mathfrak H}_i}\|\alpha_j(2)\|^2_{{\mathfrak H}_j}}
{\sum_k \|\alpha_k(2)\|_{{\mathfrak H}_k}^2}
\frac{(\beta_i-\beta_j)\sinh (\beta_i-\beta_j)}{\cosh \beta_i\cosh \beta_j}.
\]
In particular, unless all $\beta_j$'s are the same, 
$]\frac{1}{2}-r, \frac{1}{2} +r[ \,\ni s\mapsto e_\lambda(s)$ is strictly 
convex for $0 <|\lambda|<\Lambda$. If all $\beta_j$'s are equal, $e_\lambda(s)$ is identically equal 
to zero.
\end{enumerate}
\label{sf-maintheorem}
\eet

{\bf\noindent Remark 1.} Part (1) is easy to prove and one also easily verifies  that (Ap1)--(Ap3) of 
Section~\ref{sec-fcs} hold. \SFone--\SFthree{} are not needed for (1) and it suffices that $\alpha_j$
is in the domain of $\e^{a h_j}$ for all $a$ and $j$.  Part (2) is implicit in \cite{JOP3}.
For the spin boson model (2) is proven in \cite{Ro} (with only notational modifications this 
proof also applies to the spin-fermion model).  (3) is proven in \cite{JOP3}.

{\bf\noindent Remark 2.} The spin-fermion model is automatically TRI (see \cite{JOP3}) and this 
observation implies (\ref{es-spin-fermion}).

{\bf\noindent Remark 3.} These results extend to the more general model where $\cS$ is an arbitrary 
finite quantum system and $V_j$ is a finite sum of terms of the form $Q\otimes P$ where 
$Q\in\cO_{\cH_\cS, {\rm self}}$ and $P\in\cO_j$ is a self-adjoint polynomial in the field operators 
with form factors satisfying \SFtwo{}. The Fermi Golden Rule assumption \SFthree{}, which ensures 
that $\cS$ is effectively coupled to the reservoirs at the Bohr frequency (difference of eigenvalues of 
$H_\cS$) , is replaced with the assumption that, for all $j$,  the generator $K_j$ of the quantum Markov semigroup of 
the system $\cS+\cR_j$  that arises in the van Hove limit $t\uparrow \infty$, $\lambda \downarrow 0$, 
$\lambda^2t$ fixed, has zero as simple eigenvalue and no other real eigenvalue. The time-reversal 
invariance is no more automatic and has to be assumed separately. In Part (3), $\mathfrak{S}^+$ 
is the entropy production of the quantum dynamical semigroup generated by $\sum_j K_j$ 
(as described in \cite{LS}).

{\bf\noindent Remark 4.} With only minor modifications Theorem \ref{sf-maintheorem} extends to the
multi-parameter full counting statistics. The connection of this more general  result with quantum 
hypothesis testing is unclear at the moment.

\subsection{Electronic black box model}
\label{sec-ebb}

The electronic black box (EBB) model is a free Fermi gas consisting of a finite part $\cS$ -- the sample
-- coupled to several, say $n$, infinitely extended reservoirs -- the leads. This model has been a basic 
paradigm in the study of coherent transport in electronic systems in mesoscopic physics. We shall 
consider here a very simple variant of the EBB model referring the reader to \cite{JOPP1} for proofs. 

We denote by $\fh_\cS$ and $h_\cS$ the single fermion Hilbert space and Hamiltonian of $\cS$, 
where $\ch_\cS$ is assumed to be finite dimensional. $\fh_j$ and $h_j$ are the single fermion Hilbert 
space and Hamiltonian the reservoir  $\cR_j$. We shall assume that $\cR_j$ is a semi-infinite
lead described in the tight binding approximation: $\fh_j=\ell^2(\nn)$ and $h_j =-\frac{1}{2}\Delta$ 
where $\Delta u_x = u_{x+1} + u_{x-1}-2u_x$ is the discrete Laplacian with Dirichlet boundary 
condition $u_{-1}=0$. Using the discrete Fourier transform 
\[
\widehat u(k)= \sqrt{\frac{2}{\pi}}\sum_{x\in \nn} u_x \sin (k(x+1)),
\]
one identifies $\ch_j$ with $L^2([0, \pi], \d k)$ and $h_j$ becomes the operator of multiplication by $\varepsilon(k)=1-\cos k$.
The single particle Hilbert space and Hamiltonian of the joint but decoupled system are 
\[
\fh=\fh_\cS \oplus \ch_\cR, \qquad h_0=h_\cS \oplus h_\cR.
\]
where
\[
\ch_\cR= \oplus_{j=1}^n \ch_j, \qquad h_\cR=\oplus_{j=1}^n h_j.
\]
The junction Hamiltonian $v_j$ which allows for the flow of fermions between $\cS$ and $\cR_j$ is 
\[
v_j = |\chi_j\rangle \langle \delta_0^{(j)}| + |\delta_0^{(j)}\rangle \langle \chi_j|,
\]
where $\chi_j$'s are unit vectors in  $\ch_\cS$ and $\delta_0^{(j)}$ is the Dirac delta function at the 
site $x=0$ of the $j$-th lead. The single particle Hamiltonian of the coupled system is 
\[
h_\lambda =h_0 +\lambda \sum_{j=1}^n v_j,
\]
where $\lambda\in\rr$ is a coupling constant. The initial state of $\cS$ is the quasi-free state with
density $T_\cS$ and the initial state of the reservoir $\cR_j$ is the quasi-free state with
Fermi-Dirac density
\[
T_{\beta_j, \mu_j}=\left(1 +\e^{\beta_j(h_j-\mu_j)}\right)^{-1}.
\] 

The EBB model is described by the triple $\cQ=(\cO,\tau_\lambda,\omega)$, where
$\cO=\CAR(\ch)$, 
\[
\tau_\lambda^t(a^\#(f))=a^\#(\e^{\i t h_\lambda}f),
\]
and $\omega$ is the quasi-free state with density
\[
T=T_\cS\oplus(\oplus_{j=1}^n T_j).
\]
Note that for $A\in\cO$, 
\[
\tau_\lambda^t(A)=\e^{\i t H_\lambda}A\e^{-\i t H_\lambda},
\]
where 
\[
H_\lambda=\d\Gamma(h_\lambda)
=H_0  + \lambda \sum_{j=1}^n (a^\ast(\chi_j)a(\delta_0^{(j)}) + a^\ast(\delta_0^{(j)})a(\chi_j))
\]
is the Hamiltonian of the EBB model. The EBB model is TRI iff the sample system $\cS$ is TRI,
more precisely iff there is an orthogonal basis of $\fh_\cS$ in which both $h_\cS$ and $T_\cS$ have
a real matrix representation and the vectors $\chi_j$ have real components.

Finite dimensional approximations of the EBB model are naturally constructed by replacing the
semi-infinite leads $\cR_j$ with finite leads with single fermion Hilbert space $\ell^2([0, m])$ and 
Hamiltonian $-\frac{1}{2}\Delta_m$, where $\Delta_m$ is the discrete Laplacian with Dirichlet boundary 
condition. The corresponding EBB model is denoted $\cQ_m=(\cO_m, \tau_{m\lambda}, \omega_m)$.
${\mathbb P}_{mt}$ denotes the full counting statistics of ${\cal Q}_m$. 

The orthogonal projection $1_{\cR}$ onto $\fh_\cR$ coincide with the projection onto the absolutely
continuous part of $h_0$. Since $h_\lambda - h_0$ is finite rank, the wave operators 
\[
{\rm w}_{\pm}=\slim_{t\to\pm\infty}\e^{\i t h_\lambda}\e^{-\i t h_0}1_{\cR}
\]
exist. 
In what follows we suppose that the spectrum of $h_\lambda$ is purely absolutely continuous and 
hence that the wave operators are complete.  If $\sp(h_\cS)\subset ]0,2[=\sp(h_\cR)$, under very 
general conditions this assumption holds for non-zero and small enough $\lambda$ \cite{JKP} (for 
large values of $\lambda$, $h_\lambda$ will have eigenvalues outside $[0, 2]$). The scattering matrix 
${\mathfrak s}={\rm w}_+^\ast {\rm w}_-$ is unitary on $\ch_\cR$ and acts as the operator of multiplication 
by a unitary $n\times n$ matrix $[{\mathfrak s}_{ij}(k)]$. Let $\varsigma(k)$ be a diagonal matrix 
with entries 
\[
\varsigma_{jj}(k)=-\beta_j(\varepsilon (k)-\mu_j),\]
and let 
\[
T(k)= \left(1 + \e^{-\varsigma(k)}\right)^{-1}.
\]

\bet
\begin{enumerate}[{\rm (1)}]
\item For all $\lambda, s$, and $t>0$,  
\[
\lim_{m\to\infty}e_{mt}(s)= e_t(s).
\]
In particular, ${\mathbb P}_{mt}\to{\mathbb P}_t$ weakly. The function $s\mapsto e_t(s)$ is real
analytic and all the cumulants of ${\mathbb P}_{mt}$ converge to corresponding cumulants of 
${\mathbb P}_t$. 

In the remaining statements we assume that $h_\lambda$ has purely absolutely continuous spectrum.

\item For all $s$ the limit
\[
e(s)=\lim_{t \rightarrow \infty}\frac{1}{t}e_t(s)
\]
exists and 
\begin{equation}
e(s)=\int_0^\pi\log\det\left(
1+T(k)\left(\e^{-s \varsigma(k)}{\mathfrak s}(k)\e^{s \varsigma (k)}{\mathfrak s}^\ast(k)-1\right)
\right)\frac{\d\varepsilon (k)}{2\pi}.
\label{cergy-sat}
\end{equation}
The function $e(s)$ is real analytic and does not depend on $T_{\cS}$.
\item 
\begin{equation}
\Sigma^+=\sum_{i,j=1}^n\int_0^\pi t_{ji}(k)
(\varrho_j(k)-\varrho_i(k))\varsigma_{jj}(k)\frac{\d\varepsilon (k)}{2\pi}, 
\label{landauer}
\end{equation}
where $t_{ji}(k)=|s_{ji}(k)-\delta_{ji}|^2$ and 
\[
\varrho_{j}(k)= \left( 1 +\e^{\beta_j(\varepsilon (k)-\mu_j)}\right)^{-1}
\]
is the Fermi-Dirac density of the $j$-th reservoir. $\Sigma^+>0$ iff there exists $j,i\in \{1,\ldots,n\}$ 
and a set $S\in [0, \pi]$ of positive Lebesgue measure such that $j\not=i$, $s_{ji}(k)\not=0$ for 
$k\in S$, and $(\beta_j, \mu_j)\not=(\beta_i, \mu_i)$. $e(s)$ is strictly convex if $\Sigma^+>0$ and 
is identically equal to zero if $\Sigma^+=0$.
\end{enumerate}
\label{ebb-maintheorem}
\eet

{\bf\noindent Remark 1.} The sufficient and necessary condition for the strict positivity of the entropy
production described in (3) can be rephrased as follows: $\Sigma^+>0$ iff there is an open scattering 
channel between two reservoirs $\cR_j$, $\cR_i$, which are not in mutual thermal equilibrium.

{\bf\noindent Remark 2.} Remark 4 after Theorem \ref{sf-maintheorem} applies to Theorem 
\ref{ebb-maintheorem}.

{\bf\noindent Remark 3.} (\ref{landauer}) is the Landauer-B\"uttiker formula for the entropy production 
of the EBB model. One can also derive the Landauer-B\"uttiker formulas for individual fluxes. The 
energy and charge operators of $\cR_j$ are $H_j=\d\Gamma (h_j)$ and $N_j=\Gamma(1_{\cR_j})$.
The corresponding heat and charge flux observables are
\begin{align*}
\Phi_j&=-\left.\frac{\d\ }{\d t}\e^{\i t H_\lambda}H_j\e^{-\i t H_\lambda}\right|_{t=0}
=\i\lambda(a^\ast(h_j \delta_0^{(j)})a(\chi_j)-a^\ast(\chi_j)a(h_j\delta_0^{(j)})),\\[5pt]
{\cal J}_j&=-\left.\frac{\d\ }{\d t}\e^{\i t H_\lambda}N_j\e^{-\i t H_\lambda}\right|_{t=0}
=\i \lambda(a^\ast( \delta_0^{(j)})a(\chi_j)-a^\ast(\chi_j)a(\delta_0^{(j)})).
\end{align*}
The entropy production observable is
\[
\sigma=-\sum_{j=1}^n \beta_j(\Phi_j-\mu_j{\cal J}_j).
\]
Then 
\begin{equation}
\begin{split}
\langle \Phi_j\rangle_+&=\lim_{t\rightarrow \infty}\omega(\tau_\lambda^t(\Phi_j))=
\sum_{i=1}^n \int_0^\pi t_{ji}(k)(\varrho_j(k)-\varrho_i(k))\varepsilon(k)\frac{\d\varepsilon (k)}{2\pi},\\[3mm]
\langle {\cal J}_j\rangle_+&=\lim_{t\rightarrow \infty}\omega(\tau_\lambda^t({\cal J}_j))=
\sum_{i=1}^n \int_0^\pi t_{ji}(k)(\varrho_j(k)-\varrho_i(k))\frac{\d\varepsilon (k)}{2\pi}.
\end{split}
\end{equation}
These formulas can be derived either directly \cite{AJPP2} or via the multi-parameter extension of the 
formula (\ref{cergy-sat}) (see \cite{JOPP1, JOPP2}).

\subsection{The XY quantum spin chain}

The XY spin chain on a finite sublattice $\Lambda=[l_1,l_2]\subset\zz$ is described by the Hilbert 
space
\[
\cH_\Lambda =\bigotimes_{x\in \Lambda}\cc^2,
\]
and the Hamiltonian
\[
H_\Lambda=-\frac{J}{4}\sum_{l_1 \leq x<l_2}(\sigma_x^{(1)}\sigma_{x+1}^{(1)}
+\sigma_x^{(2)}\sigma_{x+1}^{(2)})-\frac\lambda2\sum_{l_1\leq x\leq l_2} \sigma_x^{(3)},
\]
where $\sigma_x^{(j)}=(\otimes_{y=l_1}^{x-1}1)\otimes\sigma^{(j)}\otimes(\otimes_{y=x+1}^{l_2}1)$, 
$J$ is the nearest neighbor coupling constant and $\lambda$ is the strength of the magnetic 
field in the (3) direction. The thermal equilibrium state at inverse temperature $\beta>0$ is 
\[
\omega_{\Lambda \beta}=\frac{\e^{-\beta H_\Lambda}}{\Tr\, \e^{-\beta H_\Lambda}}.
\]

Let $n>0$ be a fixed integer and for $m>n$ let $\cQ_m$ denote the XY spin chain on 
$\Lambda_m=[-m, m]$ with initial state 
\[
\omega_m=
\omega_{[-m, -n-1]\beta_L}\otimes \omega_{[-n, n]\beta}\otimes \omega_{[n+1, m]\beta_R}.
\]
$\cQ_m$ is a TRI open quantum system. The small subsystem $\cS$ is the XY spin chain on
$[-n, n]$ and the reservoirs are the  XY spin chains on $[-m,-n-1]$ and $[n+1,m]$.
We denote by ${\mathbb P}_{mt}$ the full counting statistics of $\cQ_m$ and by
$e_{mt}(s)$ its R\'enyi entropic functional. 

\bet
\begin{enumerate}[{\rm (1)}]
\item For all $s\in \rr$ and $t>0$ the limit 
\[
e_{t}(s)=\lim_{m\rightarrow \infty}e_{mt}(s)
\]
exists. In particular, ${\mathbb P}_{mt}$ converges weakly to a probability measure ${\mathbb P}_t$, 
the TD limit full counting statistics. The function $s\mapsto e_t(s)$ is real-analytic and all 
cumulants of ${\mathbb P}_{mt}$ converge to the corresponding cumulants of ${\mathbb P}_t$. 
\item For all $s\in\rr$ the limit
\[
e(s)=\lim_{t\rightarrow \infty}\frac{1}{t}e_t(s)
\]
exists and
$$
e(s)=\frac{1}{J\pi}\int_{u_-}^{u_+}\log\left(
1-\frac{\sinh(su\Delta\beta)\sinh((1-s)u\Delta\beta)}{\cosh(u\beta_L)\cosh(u\beta_R)}
\right)\d u,
$$
where $u_\pm=(\lambda\pm J)/2$ and $\Delta\beta=\beta_R-\beta_L$.
\item
\[
\Sigma^+=e^\prime(1)=\frac{1}{J\pi}\int_{u_-}^{u_+}
(u\beta_L-u\beta_R)\left(\tanh(u\beta_L)-\tanh(u\beta_R)\right)\d u
\]
and $\Sigma^+ >0$ iff $\beta_L\not=\beta_R$.
\end{enumerate}
\label{thm-XY}
\eet

{\bf\noindent Remark 1.} The Jordan-Wigner transformation maps the system $\cQ_m$ to a finite
EBB model. The scattering matrix of the corresponding infinitely extended EBB model can be explicitly 
computed and so Theorem \ref{thm-XY} is a corollary of Theorem \ref{ebb-maintheorem} 
(see \cite{JOPP1}). 

{\bf\noindent Remark 2.} One can show that the limiting functional $e(s)$ is analytic on the
strip $|\Im s|<\pi/(|\lambda|+|J|)\Delta\beta$.


\end{document}